\documentclass{emulateapj-rtx4}
\usepackage{amsmath,amssymb,color}
\newcommand{\M}{\mbox{${\cal M}$}}
 \newcommand{\msol}{\M$_\odot$}
 \newcommand{\hi}{\mbox{H{\sc i}}}
 \newcommand{\kms}{km s$^{-1}$}

\definecolor{grey}{rgb}{0.5,0.6,0.7}
\definecolor{darkgreen}{rgb}{0.0,0.4,0.0}

\newcommand{\densunit}{\begin{tiny}($10^{-3}$ \msol\ pc$^{-3}$)\end{tiny}}
\newcommand{\odensunit}{$10^{-3}$ \msol\ pc$^{-3}$}
 
\slugcomment{{\it The Astronomical Journal}, 142, 109}

\shorttitle{Improved modeling of the mass distribution of galaxies 
with the Einasto halo model} 
\shortauthors{Chemin et al.}

\begin{document}

\title{Improved modeling of the mass distribution of disk galaxies by the Einasto halo model}
\author{Laurent Chemin\altaffilmark{1,2}, W. J. G. de Blok\altaffilmark{3} and Gary A. Mamon\altaffilmark{4}}

\email{chemin@obs.u-bordeaux1.fr, edeblok@ast.uct.ac.za, gam@iap.fr}
\altaffiltext{1}{Universit\'e de Bordeaux, Observatoire Aquitain des Sciences de l'Univers, BP 89, 33271 Floirac Cedex, France}
\altaffiltext{2}{CNRS, Laboratoire d'Astrophysique de Bordeaux-UMR 5804, BP 89, 33271 Floirac Cedex, France}
\altaffiltext{3}{ACGC,  Department of Astronomy, University of Cape Town, Rondebosch 7700, South Africa}
\altaffiltext{4}{Institut d'Astrophysique de Paris (UMR 7095: CNRS \& UPMC), 98 bis Bd. Arago, 75014 Paris, France}

\begin{abstract}
 
The analysis of the rotation curves (RCs) of spiral galaxies provides an efficient
diagnostic
for studying the properties of dark matter halos and their relations
with the baryonic material. Since the cored pseudo-isothermal (Iso) model usually
provides a better description of observed RCs than does the cuspy NFW model,
there have been concerns that the $\Lambda$CDM primordial density fluctuation
spectrum may not be the correct one.  
We have modeled the RCs of galaxies   from The
HI Nearby Galaxy Survey (THINGS)   with the Einasto halo model, which has
 emerged as the best-fitting model of the halos arising in 
dissipationless cosmological $N$-body simulations.
We find that the RCs are
significantly better fit with the Einasto halo than with either Iso or NFW
halo models.
 In our best-fit  Einasto models, the radius of density slope $-2$ and the density at this radius are highly
correlated. 
The Einasto index, which controls the overall shape of the density profile,  
is near unity on average for intermediate and low mass halos. This is not in
agreement with the predictions from $\Lambda$CDM simulations. 
The indices of the most massive halos are in rough agreement with those cosmological simulations 
and appear correlated with the halo virial mass. 
 We find that a typical Einasto  density profile  declines more strongly in its outermost parts than any of the Iso or NFW models whereas  
 it is relatively shallow in its innermost regions. 
The core nature of those regions of halos thus extends the cusp-core controversy found for
 the NFW model with low surface density galaxies to the Einasto halo with more massive galaxies like those of 
 THINGS. 
 The Einasto concentrations decrease as a function of halo mass, 
 in agreement with trends seen in numerical simulations. However 
 they are generally smaller than values expected for simulated Einasto halos. 
  We thus find that the Einasto halo model provides, so far, the best match to the observed RCs,
 and can therefore be considered as a new standard  model for dark matter halos.

\end{abstract}


\section{Context}
\label{intro}
One of the major   tools in   galactic dynamics is  the
decomposition of the rotation curves (hereafter, RCs) of spiral and lenticular galaxies  
into baryonic and dark
components.  This mass decomposition produces constraints 
on the distribution of dark matter in these galaxies, which permits the  investigation of the possible relations
between baryons and dark matter, and most importantly to study the
  properties of galactic dark  matter  halos. 

Two spherical halo models are usually used for the mass decomposition of RCs.  On one
hand, cosmologists tend to favor 
the Navarro et al. (1996, hereafter, NFW) model
\begin{equation}
\rho_{\rm NFW}(r) = 4\,\rho_{-2}
\,{r_{-2}\over r}\,\left
    ({r_{-2}\over r+r_{-2}}\right)^2 \ ,
\label{eq:rhonfw}
\end{equation}
where $r_{-2}$ is the radius where the density profile has a (logarithmic)
slope of $-2$ (the ``isothermal" value) and $\rho_{-2}$ is the local density at that radius.
The NFW model was
originally thought \citep{nfw97} to provide a universal description of halos
of different mass produced in dissipationless (i.e. dark matter only)
cosmological simulations run with different cosmologies, in particular in the
Lambda Cold Dark Matter (hereafter, $\Lambda$CDM) paradigm.
It has a cuspy inner structure with an inner slope of $-1$ and
logarithmically diverges in
mass at large radii.
Note that steeper inner cusps (inner slopes as steep as $-3/2$)
have been produced in other dissipationless cosmological
simulations \citep{moo99,jin00,die04}.

On the other hand, the analysis of RCs has traditionally used the so-called
pseudo-isothermal (hereafter, Iso) model 
\begin{equation}
\rho_{\rm Iso}(r) = 4\,\rho(a)
\,{a^2\over r^2+a^2} \ ,
\label{eq:rhoiso}
\end{equation}
which is often 
called the core halo because of its finite central density (but linearly
diverging mass at large radii, where the asymptotic density slope is $-2$). This
halo has no cosmological background but is often seen to better fit
galactic RCs   than the NFW model, particularly
for dark matter dominated objects like low surface density galaxies
\citep[e.g.][]{deb02, kuz06, kuz08}, for which the baryonic
contribution can be largely neglected and the  mass  density
profile as derived  from  the RC ``directly''  traces 
that of the dark matter halo.  The disagreement between  an observed,
apparently finite, central  density and the expectation of a
steeper density profile from the $\Lambda$CDM cosmological simulations
 remains an  unsolved  problem in galactic dynamics \citep[but see][]{gov10}. In fact,
 this \emph{cusp-core controversy} has
 often been invoked as a major weakness of the current standard cosmological
 model of hierarchical growth of structures starting from a nearly
 homogeneous Universe, seeded with density fluctuations arising from a
 $\Lambda$CDM power spectrum. However, the impressive agreement between
 $\Lambda$CDM predictions and observations of the angular fluctuation
 spectrum of the Cosmic Microwave Background
 suggests that other explanations for the controversy must be
 sought, either in the dissipative and feedback physics of the baryons or in
 a more detailed analysis of halo density profiles from $\Lambda$CDM
 cosmological simulations.

Recently, \citet{nav04} have proposed another model that fits the density
profiles of 
halos in $\Lambda$CDM simulations even better than the NFW model. It was later realized
\citep{mer06} that the model advocated by \citet{nav04} had been previously
introduced for the distribution of stellar light and mass  in galaxies
 \citep{ein65,ein68,ein69}, with density profile
\begin{equation}
\rho_{\rm E}(r) = \rho_{-2}\,\exp\left\{-2n\,\left[\left ({r\over
    r_{-2}}\right)^{1/n}-1\right]\right\} \ ,
\label{eq:rhoeinasto}
\end{equation}
where, again, $r_{-2}$ is the radius where the density profile has a
slope of $-2$ and $\rho_{-2}$ is the local density at that radius.
While both the NFW and Iso models are described by two parameters, a
characteristic scale and a characteristic density at that radius, the Einasto
model involves a third parameter, $n$, the Einasto index, which describes the
shape of the density profile.
The Einasto model is the three-dimensional equivalent of the S\'ersic model
\citep{ser68} that provides an excellent fit to the surface brightness
profiles of elliptical galaxies (e.g. \citealp{kor09}).\footnote{Both the NFW
  model \citep{lok01} and halos in $\Lambda$CDM simulations \citep{mer05} are
  well-fit by an $n=3$ S\'ersic  model (S\'ersic 1968).}

 %%%%%%%%%%%%%%%%%%%%% compare chi2 %%%%%%%%%%%%%%%%%%%%%%%
 \begin{figure}[!t]
 \begin{center}
\includegraphics[width=0.85\columnwidth]{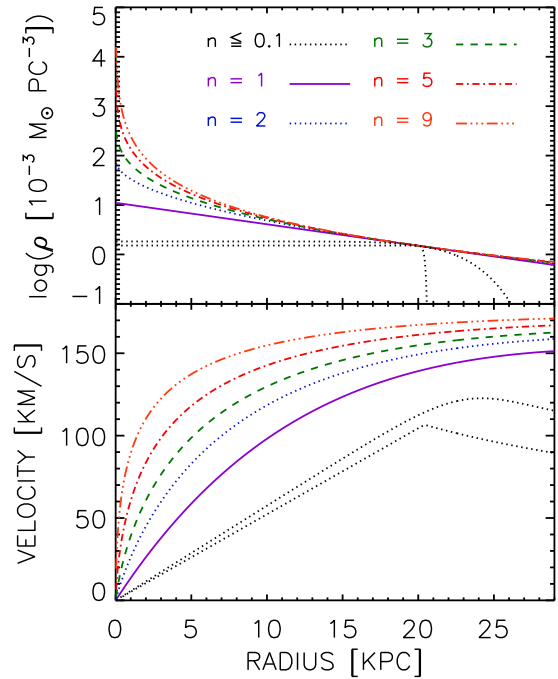}
 \caption{Density profile (\emph{top}) and rotation curve (\emph{bottom}) for
   Einasto models of different indices.  
   The   characteristic radius $r_{-2} = 20$ kpc and density $\rho_{-2} =
   1.5 \times 10^{-3}$ \msol\ pc$^{-3}$ are the same for all
   models. Only the Einasto index $n$ changes as indicated by values
   and lines of various colour and style. The  dotted black lines 
   correspond to models with $n=0.005$ (bottom curve) and $n=0.1$
   (upper curve).}
 \label{fig:rceinasto}
 \end{center}
 \end{figure}
%%%%%%%%%%%%%%%%%%%%%%%%%%%%%%%%%%%%%%%%%%%%%%%%%%%%%%%%%%%
  
The mass profile of the Einasto halo \citep{car05, mam05} is written
as
\begin{equation}
M_{\rm E}(r) = 4\pi n r_{-2}^{3} \rho_{-2}\ {\rm e}^{2n}\ (2n)^{-3n} \gamma
\left(3n,{r\over r_{-2}}\right),
\label{eq:Meinasto}
\end{equation}
where $\gamma(3n,x)=\int ^{x}_{0} {\rm e}^{-t}t^{3n-1} dt$ is the incomplete gamma function. 

As  illustrated in Fig.~\ref{fig:rceinasto} (top panel),  the 
net effect of increasing the halo shape index $n$ at fixed
characteristic density and radius is to increase and steepen the
density  profile  $\rho (r)$ in the central part of the halo.  This
then leads to increase the inner slope and amplitude of the
RC (bottom panel of Fig.~\ref{fig:rceinasto}). 
 Equation~\ref{eq:Meinasto} also implies that an Einasto halo has a finite mass, 
whose physical feature is advantageous with respect to the divergent NFW and Iso mass profiles. At fixed 
characteristic  Einasto scale density and radius, the mass profile 
converges more rapidly to its virial mass for small indices than for larger ones.

Because the inner RCs are expected to probe deeply inside the
halos down to scales of 1/300th of the virial radius, where Navarro and
co-workers have found small but significant departures between the density profiles of
$\Lambda$CDM halos and the NFW model, it has been suggested
(e.g. \citealp{sto06}) that the shallower inner slope of the
Einasto model should reconcile the observed RCs with
$\Lambda$CDM models.
However, no extensive
mass  decompositions  of galactic RCs  using Einasto
halos  have yet been carried out.  Only the RC of the Andromeda
galaxy has been modeled with the Einasto formula \citep{che09}, though
with no real  improvement  with respect to the usual core and NFW
halos as caused by the peculiar shape of the M31 RC.   Note also that \citet{gra06}  compare  the central
dark matter  densities  of  a  sample of low surface brightness
galaxies \citep{deb04} to a family of Einasto halos and  conclude 
that  these match  the data reasonably well.

The present article aims to provide the first mass decompositions of  
RCs with the Einasto model on an important sample of spiral galaxies.
 The questions we wish to address are the
following. Is the Einasto halo a good fit to the RCs of
galaxies? Is it a better description of observations than the usual
NFW and Iso halos? What does a typical galactic Einasto halo look
like?  

We use a subsample from The HI
Nearby Galaxy Survey (THINGS), as described in \citeauthor{deb08}
(\citeyear{deb08}, hereafter D08) and in \S\ref{sec:observations},
and compare the quality of
Einasto halo fits with those obtained using the NFW halo and
pseudo-isothermal sphere  (\S\ref{sec:massmodels}). We investigate
the statistical significance of possible improvements of the Einasto
model (\S\ref{sec:comparisoneinastoisonfw}), determine the
 parameter  space of Einasto halos for the THINGS sample
(\S\ref{sec:halofamily}) and compare the properties of galaxy-sized
halos generated in numerical simulations to those derived from our
sample (\S\ref{sec:compsimu}). We also describe a range of 
 Einasto indices (constrained two-parameter models) that provide better fits to RCs 
than the Iso and NFW models (\S\ref{sec:twoparam}). 
 This article is  intended to be 
the first  in  a series that  aims  at investigating how well the
Einasto formalism  works in galactic dynamics  and  at quantifying the degree of cuspiness of dark matter 
halos as a function of galaxy mass. 
The WMAP3 cosmology \citep{spe07} is adopted throughout this work.

\section{The THINGS Sample}
\label{sec:observations}
 
 The  \hi\ observations of galaxies used in this study are from The
\hi\ Nearby Galaxy Survey \citep[THINGS,][]{wal08}. This survey
consists  of high-resolution  21-cm observations of thirty-four nearby (closer than 15 Mpc)
 spiral and irregular galaxies (Sb to Im). 
The galaxies were observed with the Very Large Array
(VLA) in its B, C, and D configurations.  The \hi\ data and the complete details of
their reduction have been presented in previous articles from the
THINGS series \citep[see e.g.][]{wal08}.    We analyze here
the sample of seventeen rotationally dominated and undisturbed galaxies presented in D08, which is a 
subset of the whole sample from \cite{wal08}.

\section{Mass modeling}
\label{sec:massmodels}

\subsection{Generalities}
The reader is referred to D08 for the details of the kinematical analysis of the  
\hi\ velocity fields  by tilted ring models to extract the rotation curves.
D08 has also presented a complete dynamical analysis of those RCs with
the NFW and Iso models. 

 We have  modeled the mass distribution using different RCs from those published in D08, 
as explained in \S\ref{sec:independentvelocities}. 
We have performed Levenberg-Marquardt non-linear least-squares fits to the
  RCs, taking into account  the contributions of the 
 gas, a spherical stellar bulge, a stellar disk, and a spherical dark matter component.  

 These contributions were taken from  D08 for the stellar and gaseous RCs and  from 
equation~(\ref{eq:Meinasto}) for the halo.  Basically the  (atomic only)  gaseous disk RCs 
come  from THINGS \hi\ surface densities   and 
 the stellar RCs from  $3.6\,\mu m$ surface
brightness profiles, available for all our galaxies  from
complementary observations with the \textit{Spitzer} Nearby Galaxy Survey \citep[SINGS,][]{ken03}. 
The advantage of using these near-infrared luminosities, is that they are
little affected by internal or Galactic extinction. 
 We have not considered models with free mass-to-light ratios in
 combination  with Einasto halos. Such an analysis will be presented in a future paper (Chemin et al., in preparation).  Instead 
we have adopted the fixed  mass-to-light ratios $\Upsilon_{3.6\mu m}$ of D08, who derived $\Upsilon_{3.6\mu m}$ from the 2MASS $J-{K_s}$ colors, adopting 
$\Upsilon_{3.6\mu m}=0.92\, \Upsilon_{K_s}-0.05$ \citep{oh08}, and
$\log \Upsilon_{K_s} = 1.43\,(J-{K_s})-1.38$, extrapolated from \citet{bel01}. \citet{bel01}  derived
stellar mass-to-light ratios assuming  stellar populations with a bursty star
formation history with a diet-Salpeter initial mass function (IMF). As in D08 we also consider the \citet{kro01}
IMF, which decreases the  mass-to-light ratios (hence stellar masses) by 0.15 dex.
 A  $\rm sech^2(z/z_0)$ law \citep{vdk81} was used to derive the 
vertical distribution of the stellar disk, under the hypothesis that the disk scale-height $z_0$ is one fifth  
  the radial disk scale-length \citep{vdk81,kre02}.  

 None of our mass models take into account the contribution of molecular gas because the total gas surface densities are 
atomic gas dominated for the majority of our sample \citep{ler08}. However we briefly report  on the effects of molecular gas 
on Einasto halo parameters for the extreme case of NGC 6946  in \S\ref{sec:molecgas}. 

Following D08 the RCs of the galaxies NGC~2403 and NGC~3198 have
been decomposed  using either  one single stellar disk component
(labeled ``NGC2403d" and ``NGC3198d") or two stellar
components. Also,   for NGC7793 we present the analysis of the whole
RC, as well as that  of its rising part only
(``NGC7793s'' ). For each assumed IMF, fits have been performed
using fixed and free Einasto indices for sake of comparison with
cosmological simulations, as will be presented in
Section~\ref{sec:compsimu}. Further technical details of the fitting procedure are given in
\citet{che09}.

\subsection{The influence of RC sampling on mass models}
\label{sec:independentvelocities}
 
The RCs we fit here are  sampled   differently
from  those analyzed in D08,  where  the \hi\ rotation velocities
used for the dynamical analysis are not totally independent.  D08
used a sampling of two data points per spatial resolution element.   
As a consequence they  did not directly  compare the $\chi^2$ values
obtained for the core and  cusp  models but instead   studied the
 differences in  $\chi^2$.  
  
Since a direct comparison of the reduced $\chi^2$  values  for the
different two- and three-parameter models  is essential  to
 identify  the ``best''  halo model, one has first to investigate
 whether any correlation between spatially adjacent rotation
velocities could alter the results of the mass modeling.  For this
purpose, we have  derived  mass models of several galactic RCs   
from D08 and \citet{che09}   using  three  different 
samplings,  assuming  Einasto, NFW and Iso
halos.
  
A first modeling exercise  (S1)  was  done with RCs
sampled with two points per beam (as in D08), a second one (S2) with
one  point  per beam \citep[as in][]{che09} and a third one (S3)
with  one point every two beams.  The last two samplings provide
independent data points,  as opposed to  the first one.    The
velocity uncertainties  are  defined as in D08 or \citet{che09},
taking into account the (generally dominant) kinematical asymmetry observed
between the receding the approaching halves of each galactic disk.

We observe the following results  (which are strictly speaking only
valid for the range of samplings explored here and should not be
further generalized) : (i) the  parameter  space of  the  dark
matter halo  models  does not vary with the spatial sampling of
 their  RC, (ii)  the relative fit quality of the three
halo models does not vary with resolution,  and (iii) the $\chi^2$
 values   increase  from S1 to S2 and then  decrease for  S3,
 which  generally exhibits the smallest $\chi^2$.  To illustrate,
  the fits of the NFW, Iso and Einasto halos to the RC
of the Andromeda galaxy lead to a reduced $\chi^2 \sim 7$ for S1
(sampling of $\sim$200 pc), $\chi^2 \sim 21$ for S2  (a sampling of
$\sim$400 pc, as in Tab. 5 of \citet{che09}, for their ``hybrid''
model)  and $\chi^2 \sim 8$ for S3 (sampling of $\sim$800
pc).  For this galaxy,  the fit parameters of the NFW halo, the
characteristic velocity $v_{200}$ and the halo concentration $c$, are
$v_{200} = (145 \pm 2)$ \kms\ and $c = 21 \pm 1$ for S1, $v_{200} =
(146 \pm 4)$ \kms\ and $c = 20 \pm 2$ for S2, $v_{200} = (147 \pm 5)$
\kms\ and $c = 21 \pm 2$ for S3, indeed independent of the sampling.

 The non-independent  rotation velocities in   S1 are thus
responsible  for the  smaller $\chi^2$  compared to  S2, whereas
 the  larger sampling  interval  explains  the  smaller $\chi^2$
for model S3.   The risk of using large intervals (like S3) is, of
course, that one starts loosing small-scale details in the rotation
curve and decreases the velocity gradient in the inner parts of the
galaxies.

 %%%%%%%%%%%%%%%%%%%%% compare chi2 dependent-independent bins %%%%%%%%%%%%%%%%%%%%%%%
 \begin{figure}[!t]
 \begin{center}
\includegraphics[width=0.75\columnwidth]{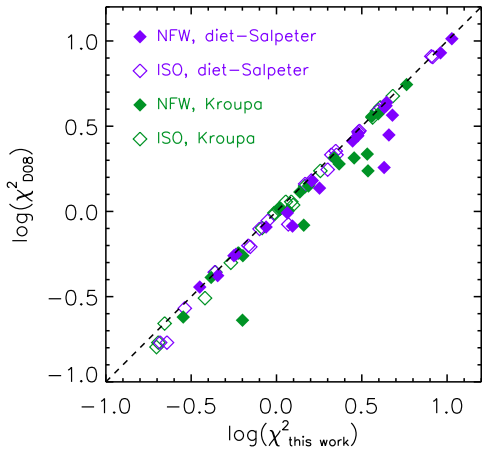}
 \caption{Comparison between reduced $\chi^2$ from mass models fit
   to THINGS RCs  from D08 and from this study, with the Iso (\emph{Open
     diamonds}) and NFW (\emph{closed diamonds}), using the 
Kroupa (\emph{green}) and diet-Salpeter (\emph{purple}) IMFs. }
 \label{fig:veloind}
 \end{center}
 \end{figure}
%%%%%%%%%%%%%%%%%%%%%%%%%%%%%%%%%%%%%%%%%%%%%%%%%%%%%%%%%%%

 We therefore  conclude that the optimum sampling of RCs
for mass   models of galaxies is   one which   avoids
 non-independent  velocities generating  artificially  small
$\chi^2$, and  which  preserves an appropriate  spatial 
resolution  that  does not smooth  out small-scale details in the 
RC, especially in the inner regions, which are the most sensitive to
the differences between the
Iso, NFW and Einasto models.  We have thus modeled the mass  distributions of the
THINGS galaxies  from RCs  with  the S2 sampling,
 i.e., with galactocentric radii separated by one synthesized beam
size.  
 
Note  that the conclusions given in D08 still hold: the halo
parameters we have fit are  similar to theirs, the
Iso model usually provides better results than the NFW model. 
Only the  values of the  reduced $\chi^2$ have increased when
 going  from  non-independent  to independent velocities, as shown
in Figure~\ref{fig:veloind}.  Finally, because the analysis of
individual differences between the NFW and Iso halos is beyond the
scope of this article, we refer the reader to Tab.~\ref{tab:resfit1} and~\ref{tab:resfit2}
 for the resulting  halo parameters and to D08  for further discussion.

\subsection{The Einasto model fits}
\label{results}

\begin{figure*}
 \begin{center}
\includegraphics[width=0.25\textwidth]{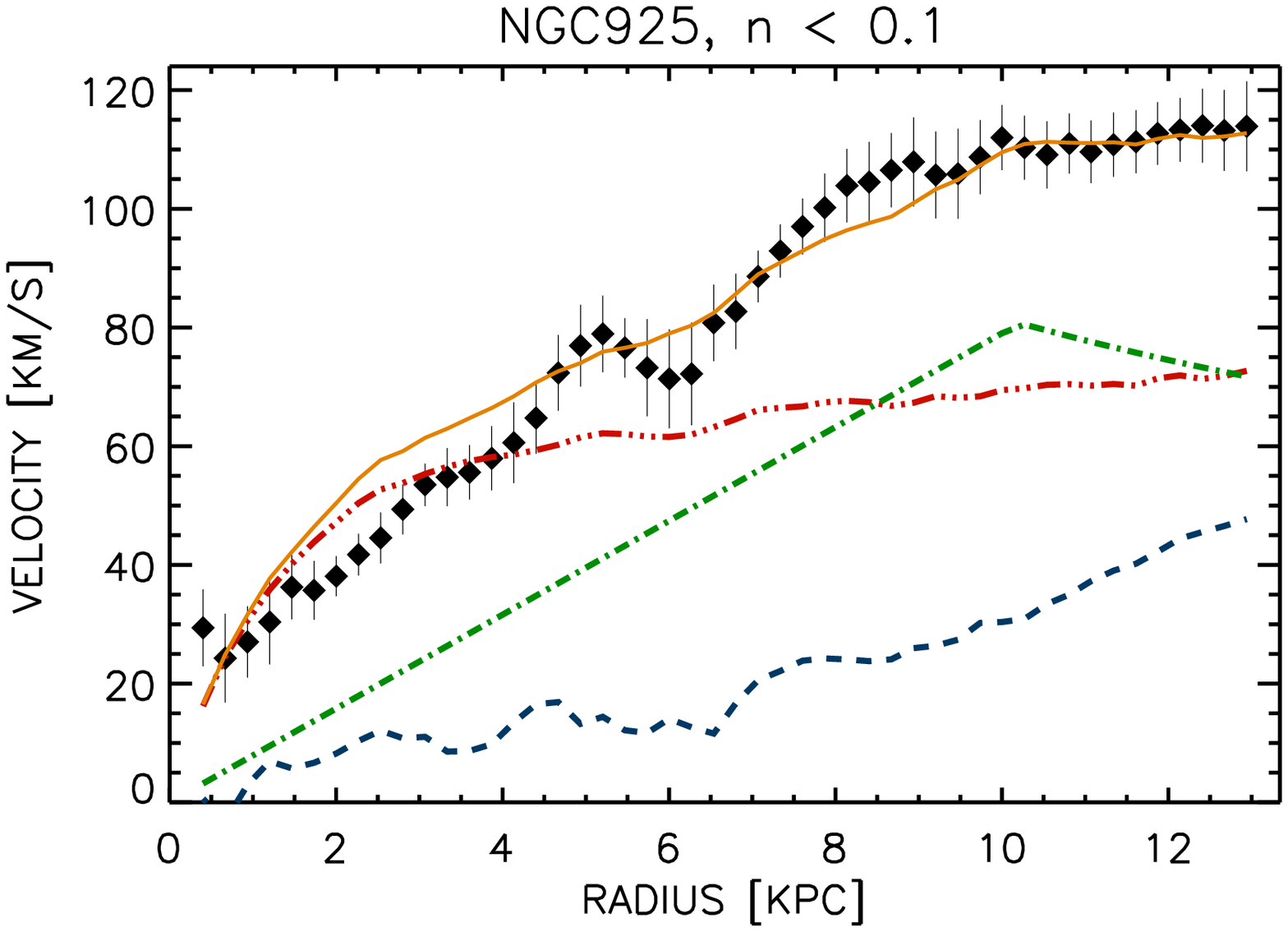}\includegraphics[width=0.25\textwidth]{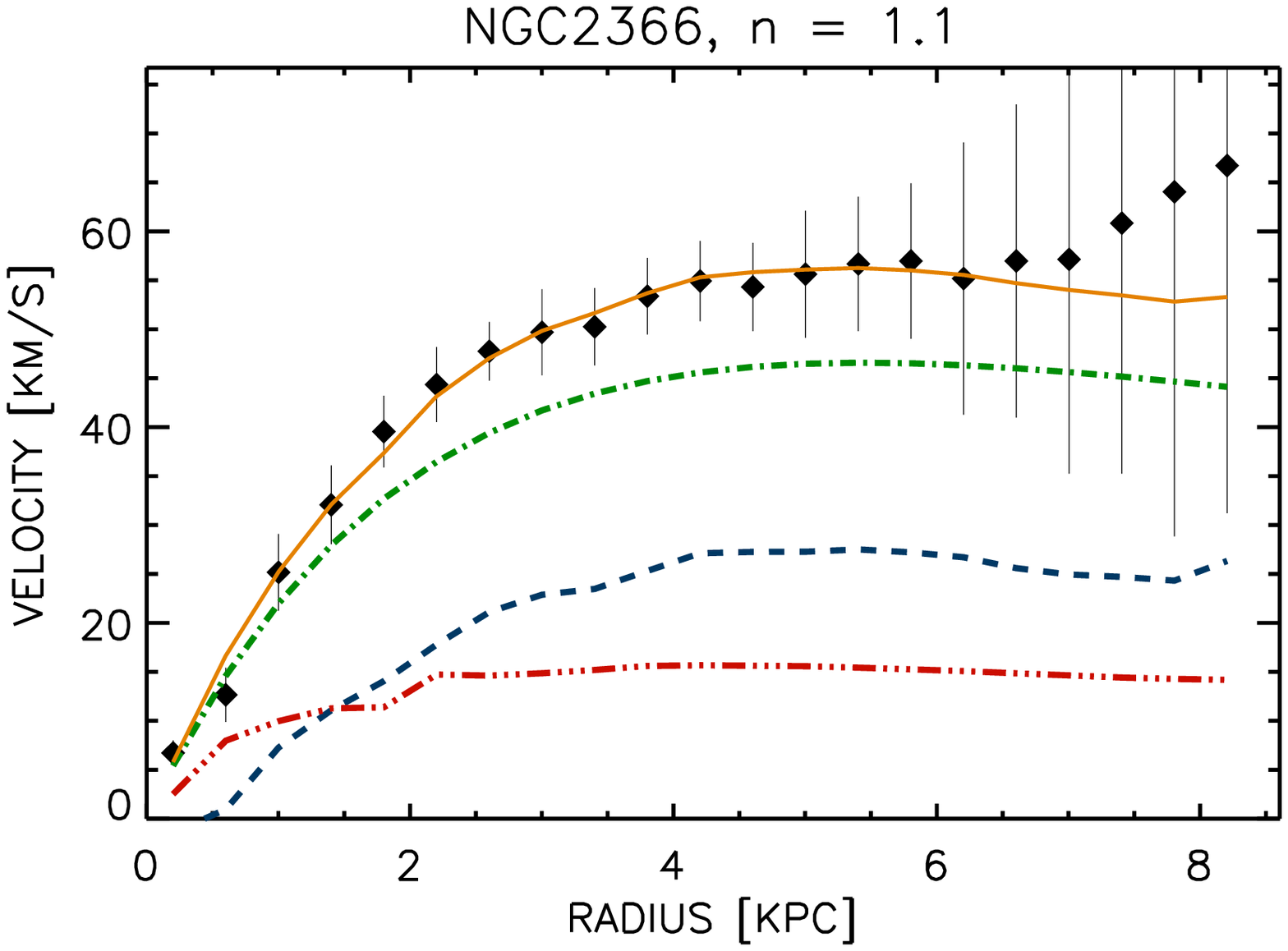}\includegraphics[width=0.25\textwidth]{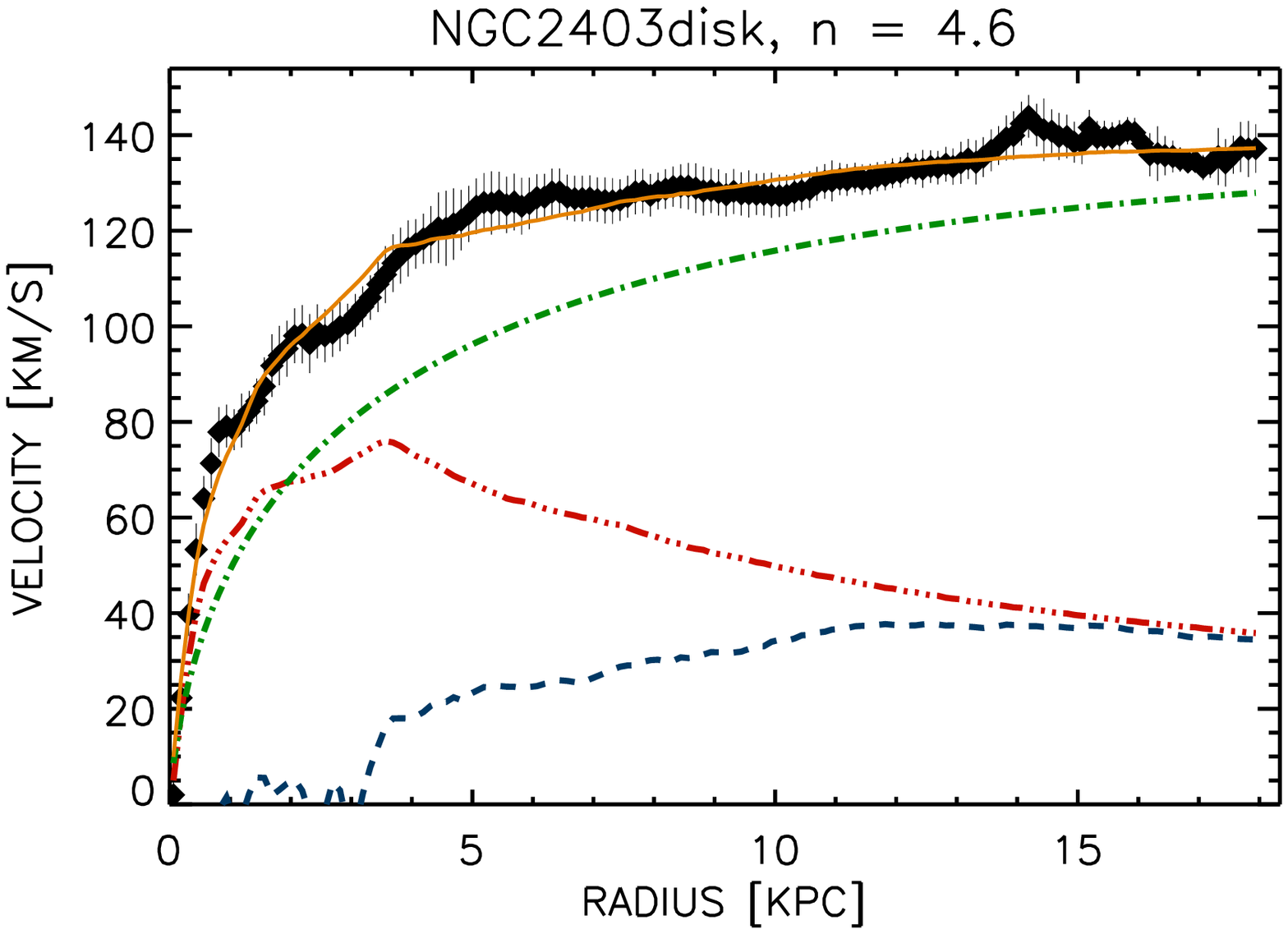}\includegraphics[width=0.25\textwidth]{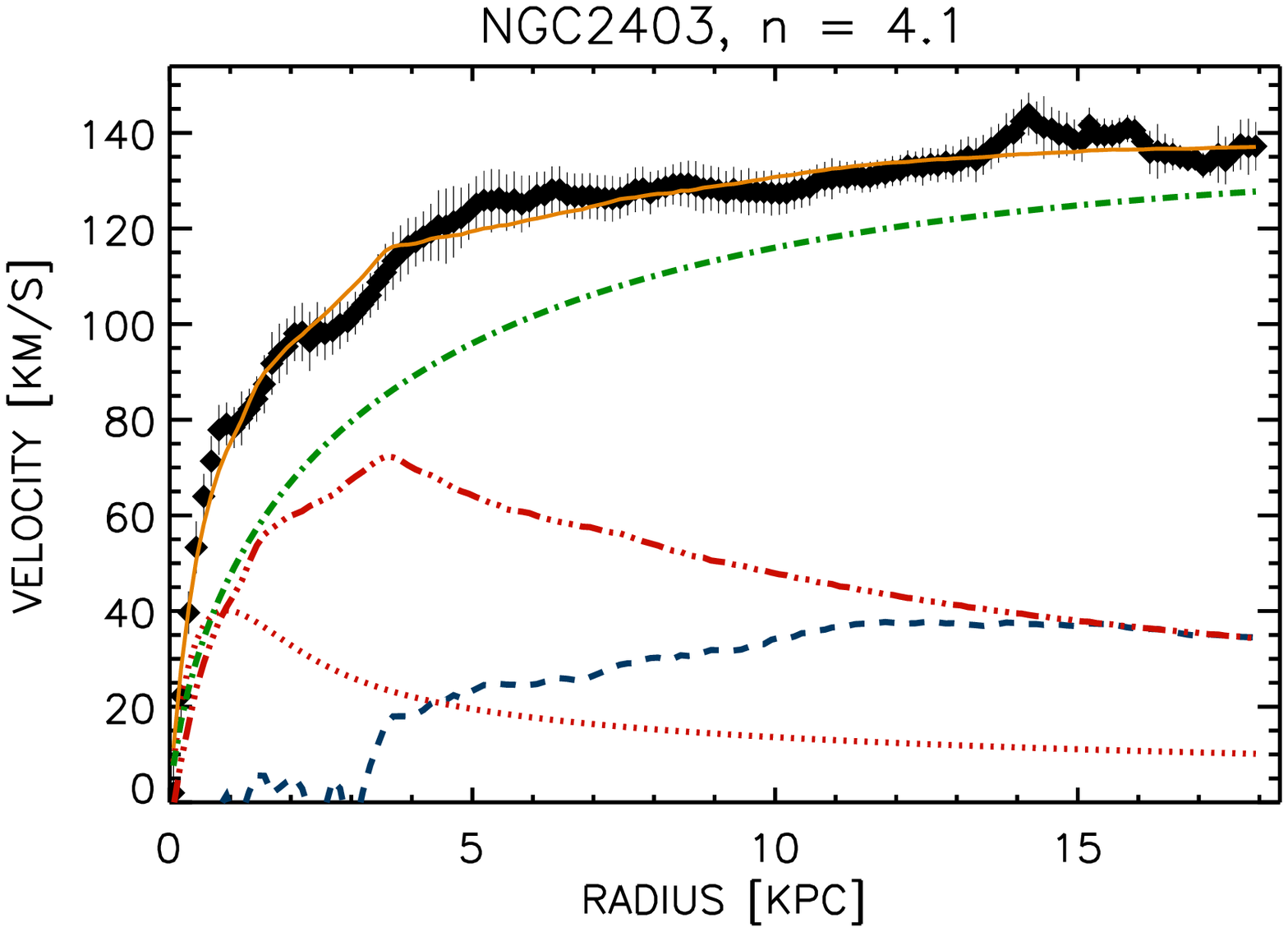}
\includegraphics[width=0.25\textwidth]{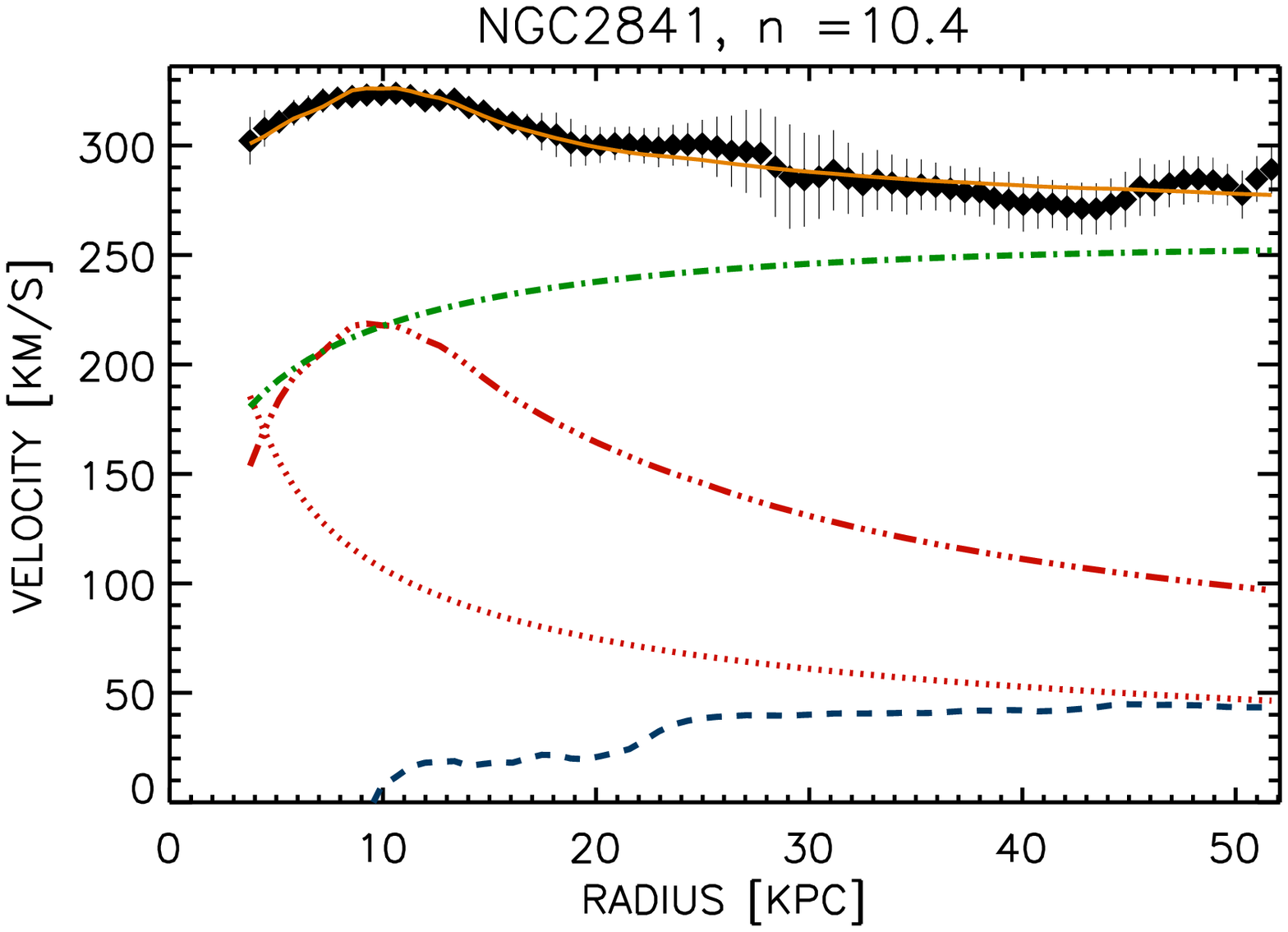}\includegraphics[width=0.25\textwidth]{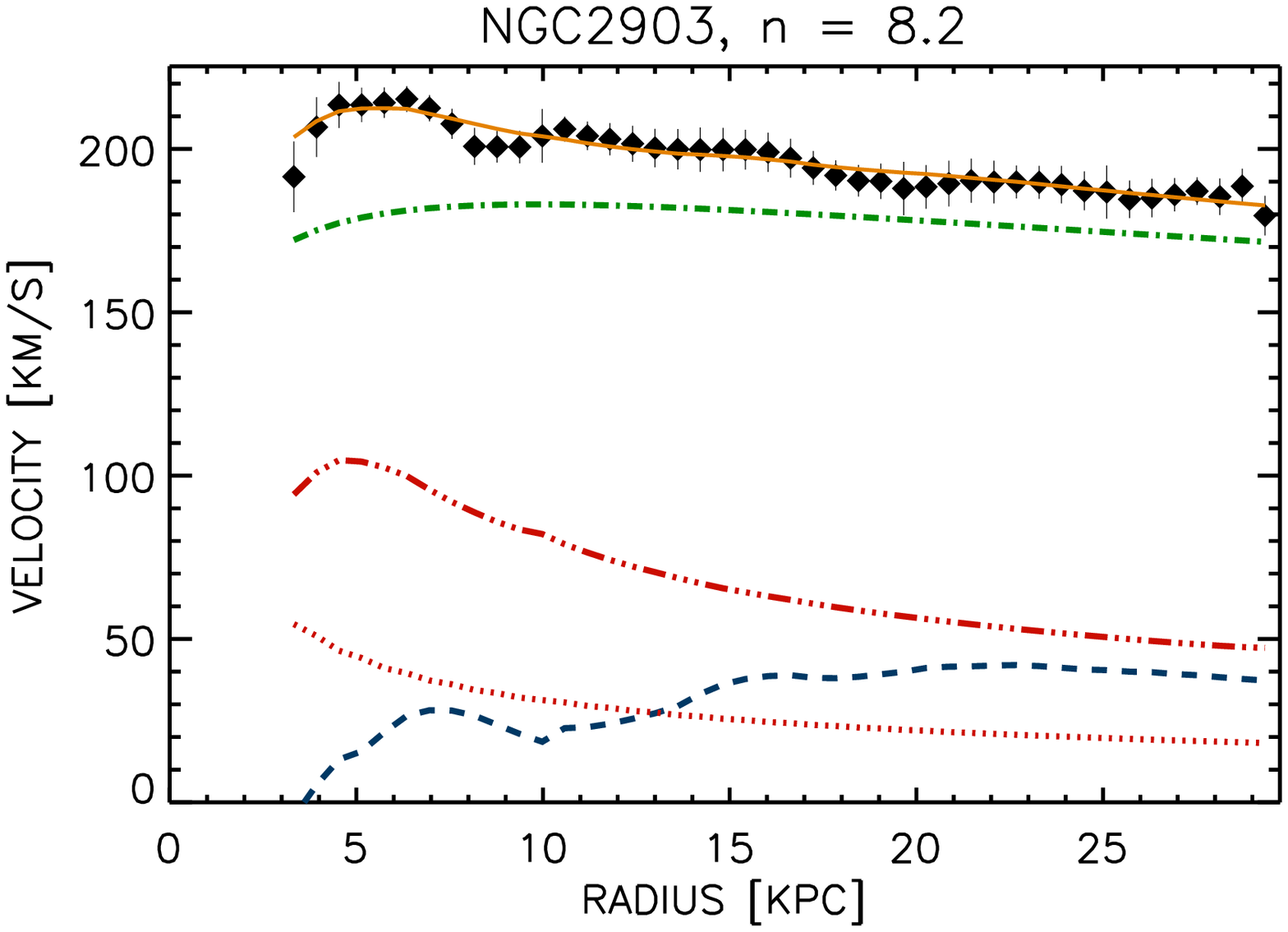}\includegraphics[width=0.25\textwidth]{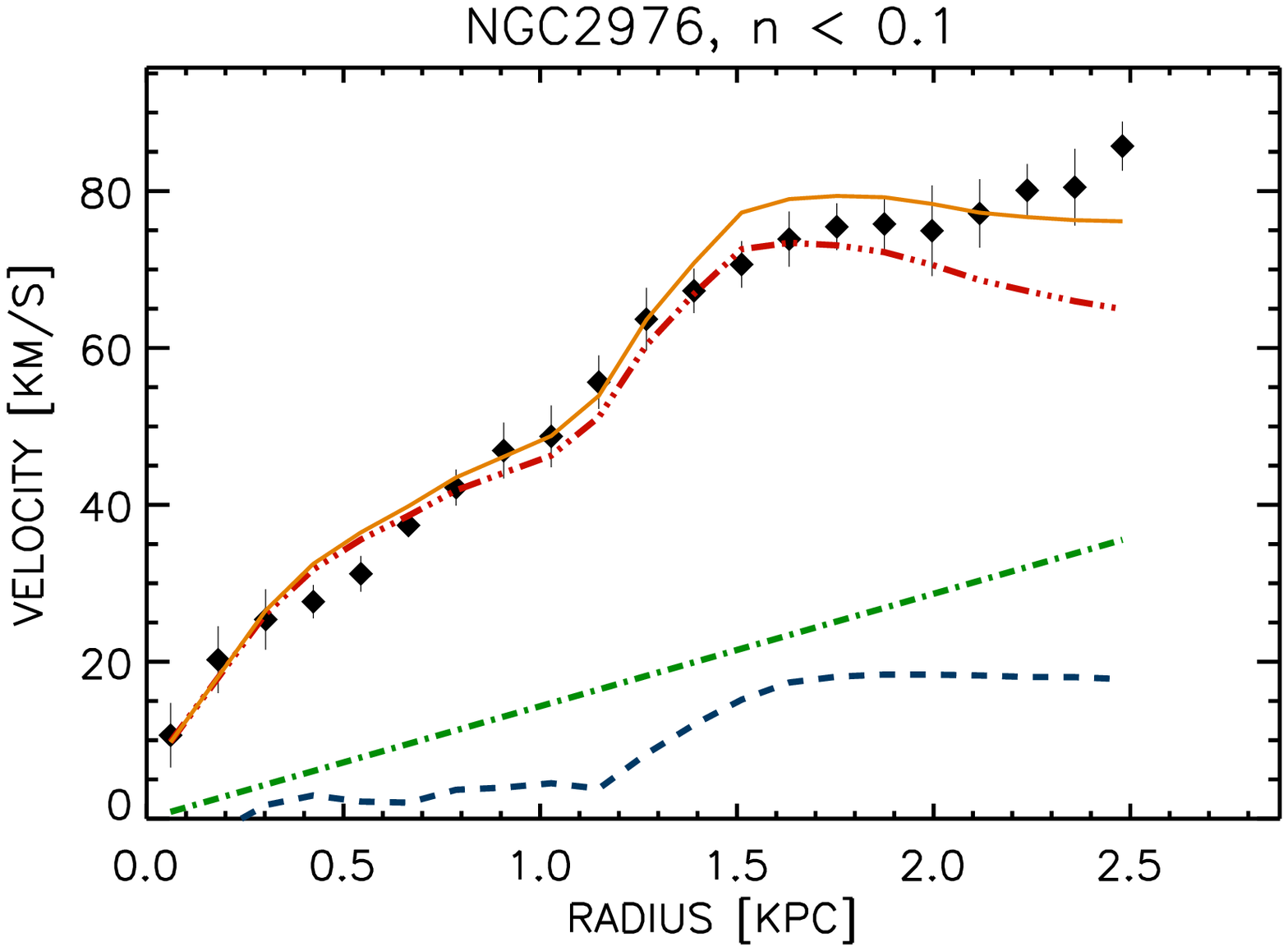}\includegraphics[width=0.25\textwidth]{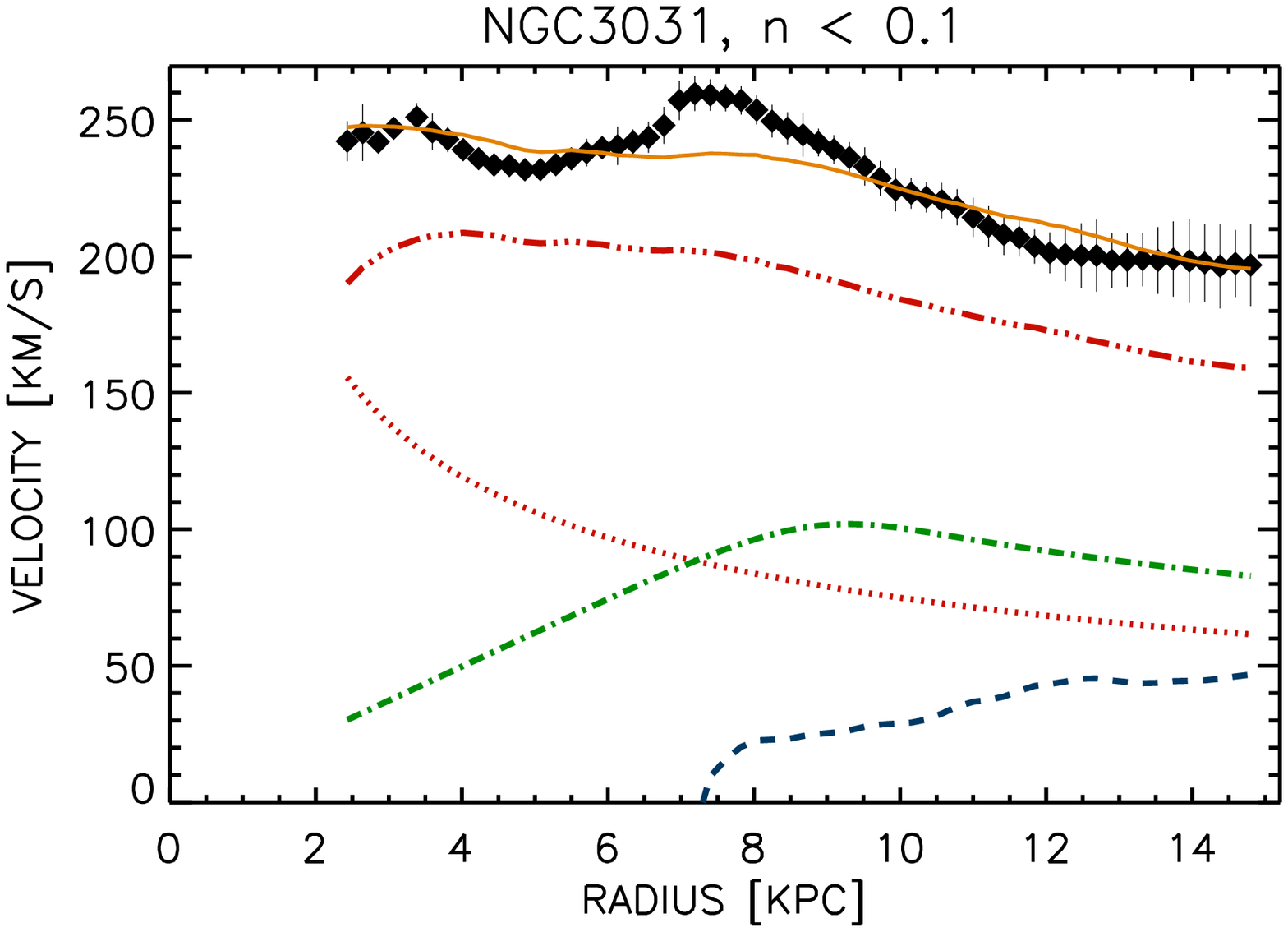}
\includegraphics[width=0.25\textwidth]{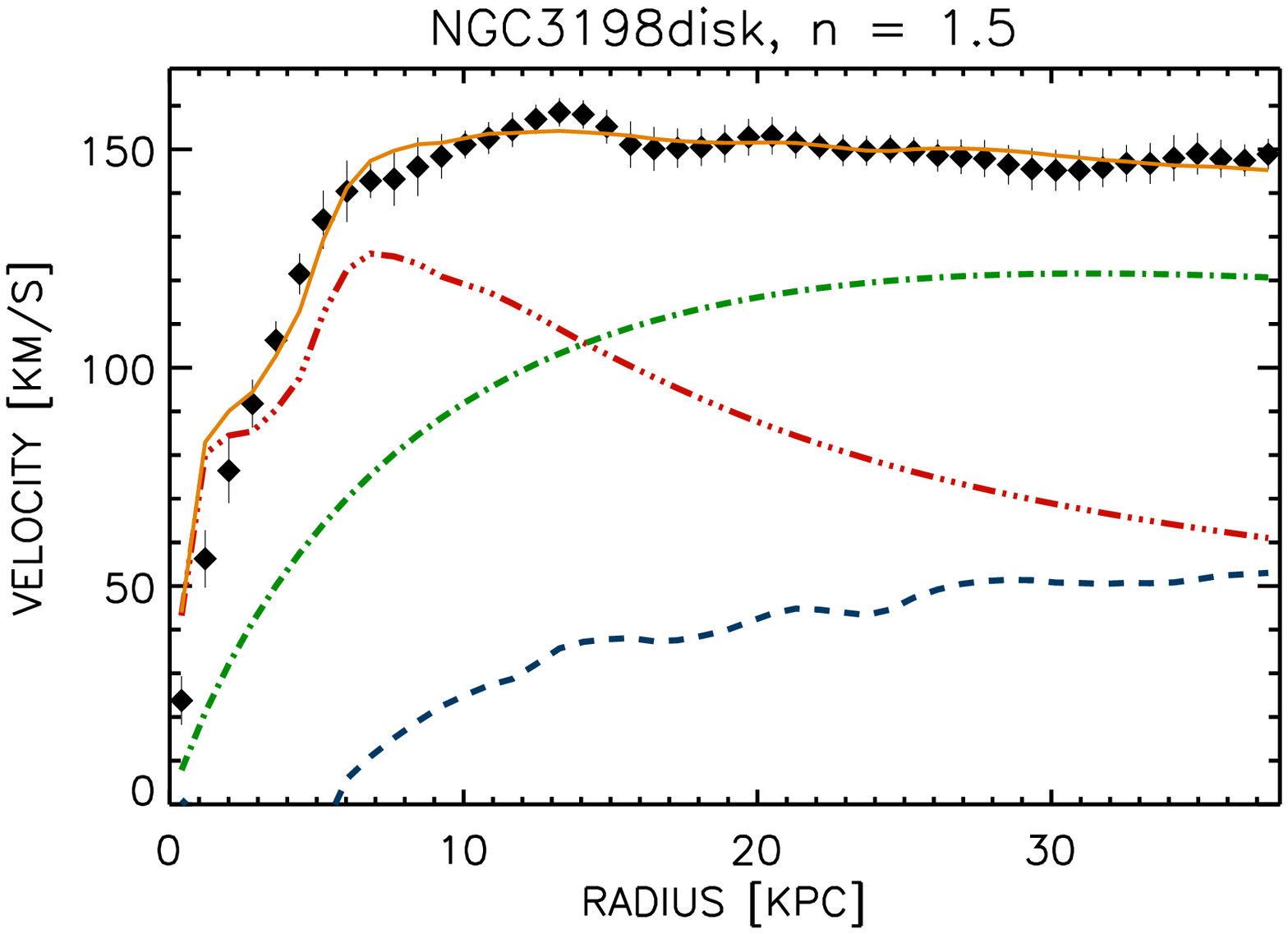}\includegraphics[width=0.25\textwidth]{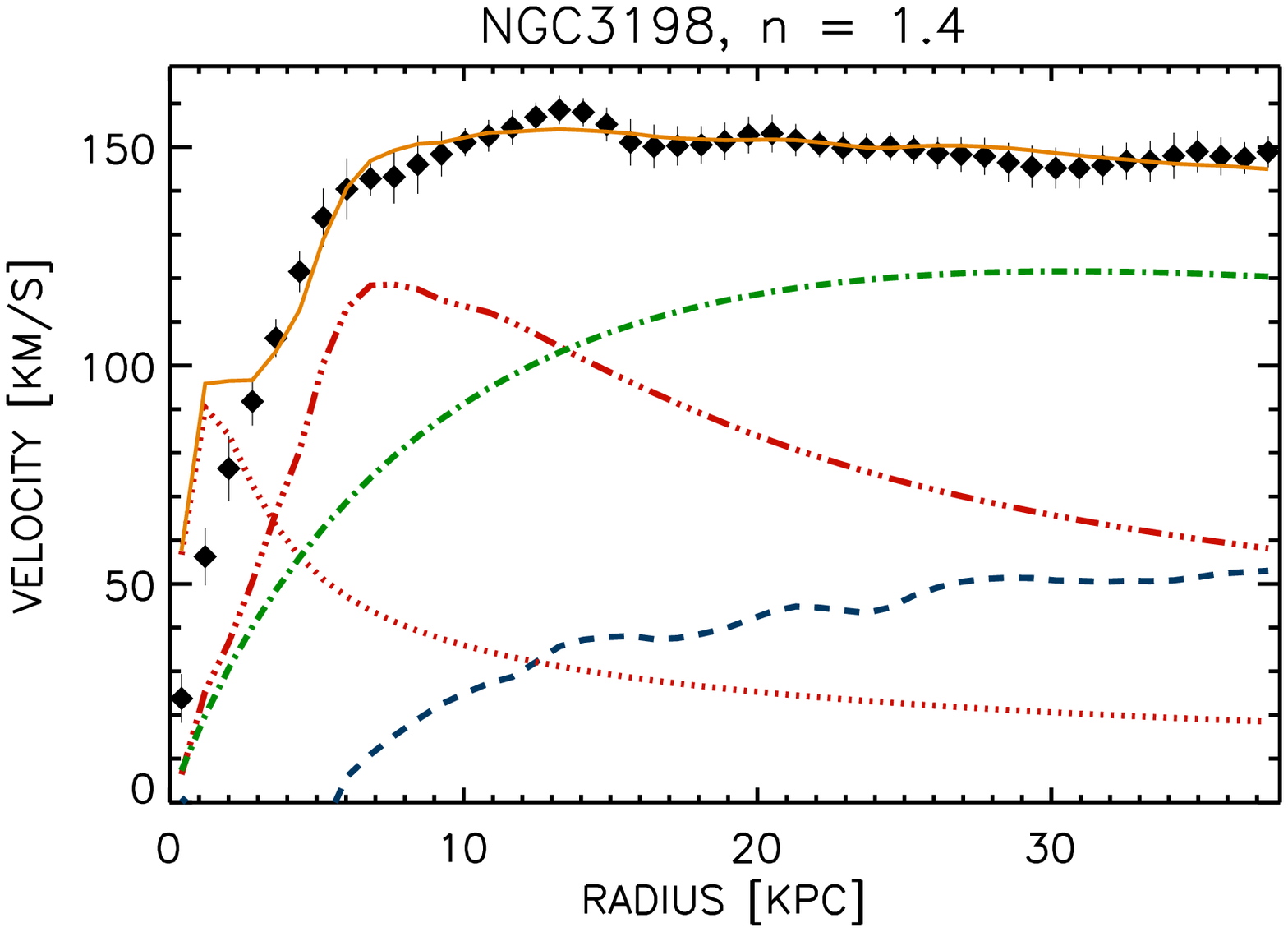}\includegraphics[width=0.25\textwidth]{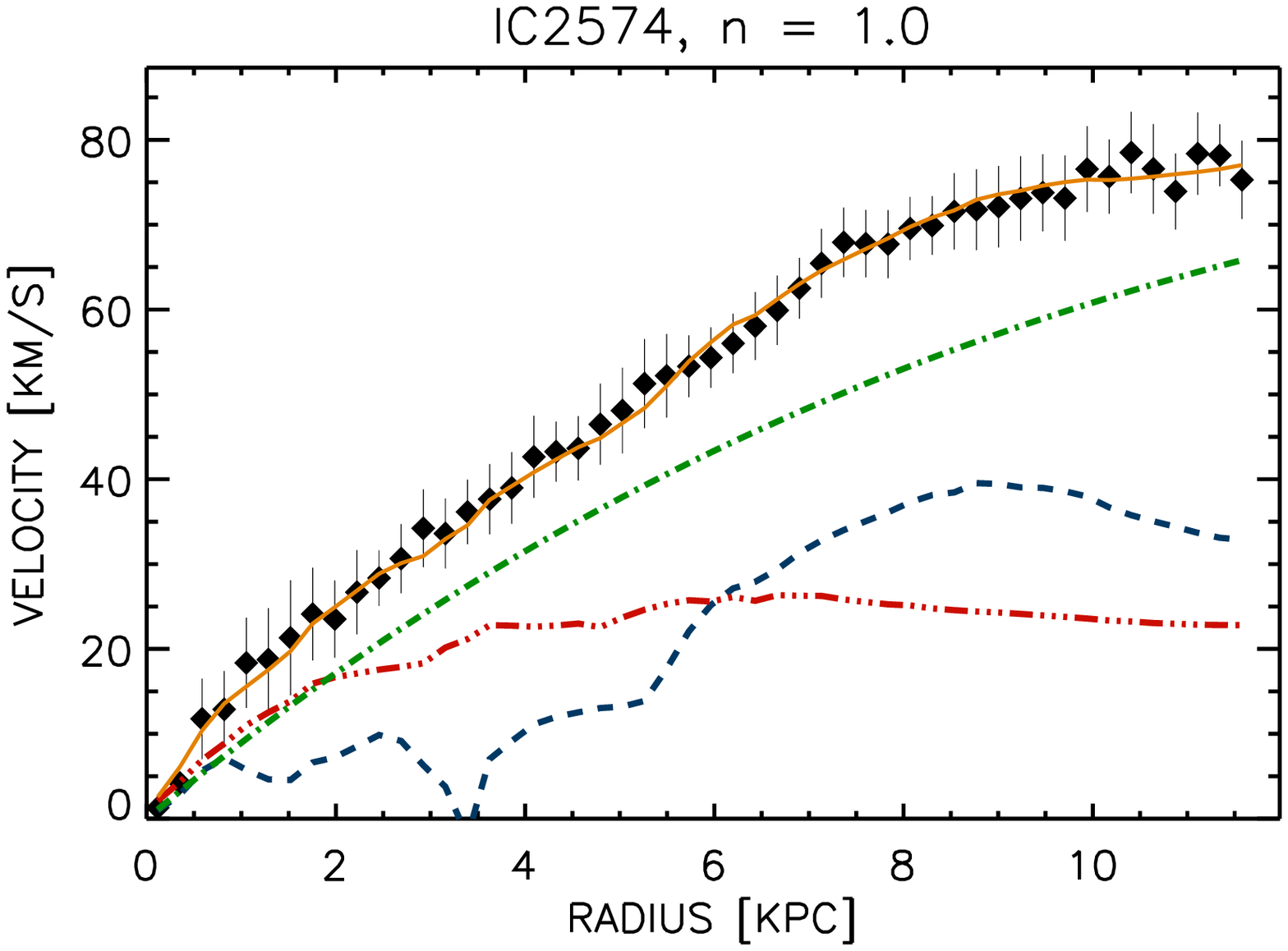}\includegraphics[width=0.25\textwidth]{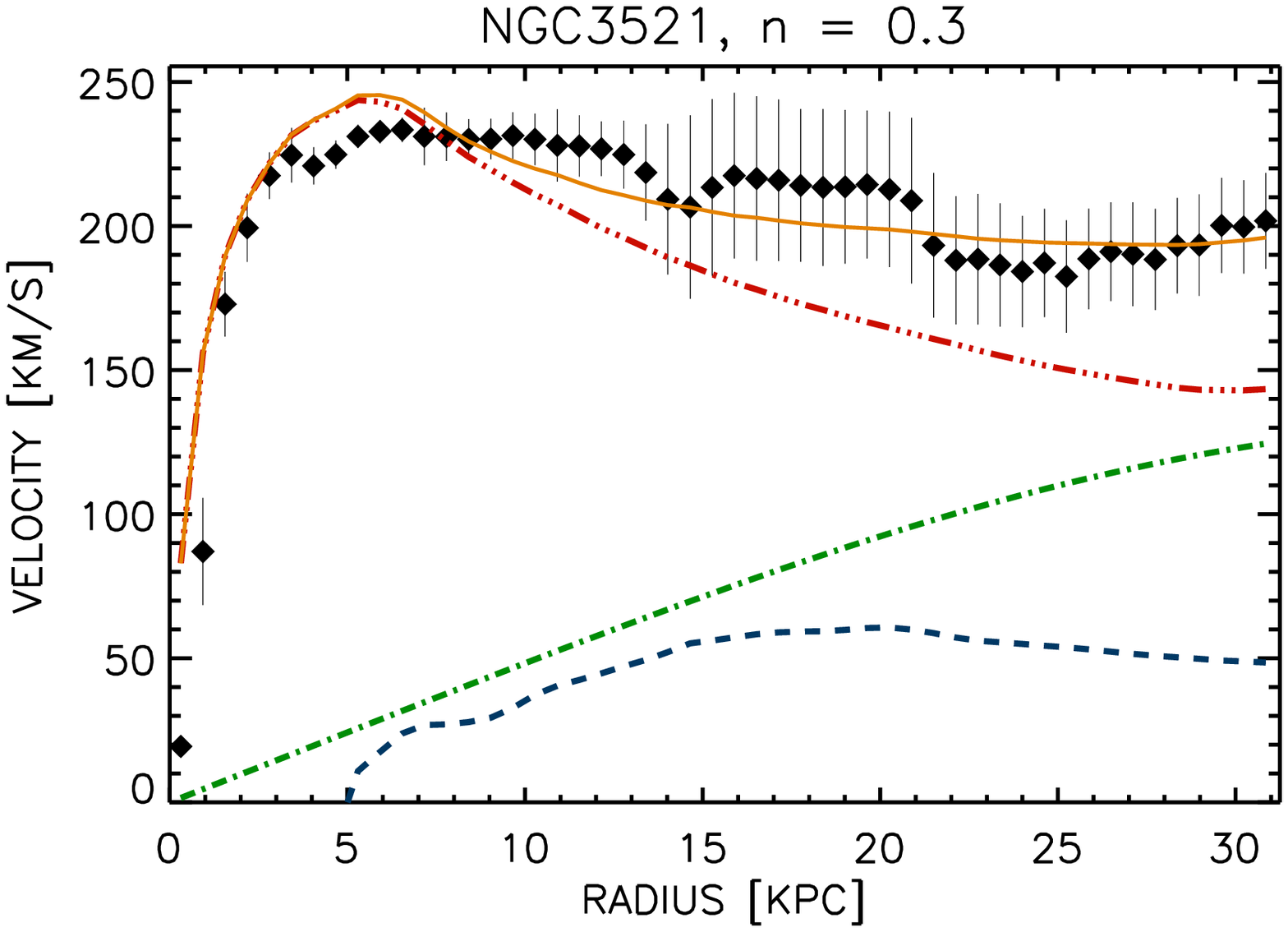}
\includegraphics[width=0.25\textwidth]{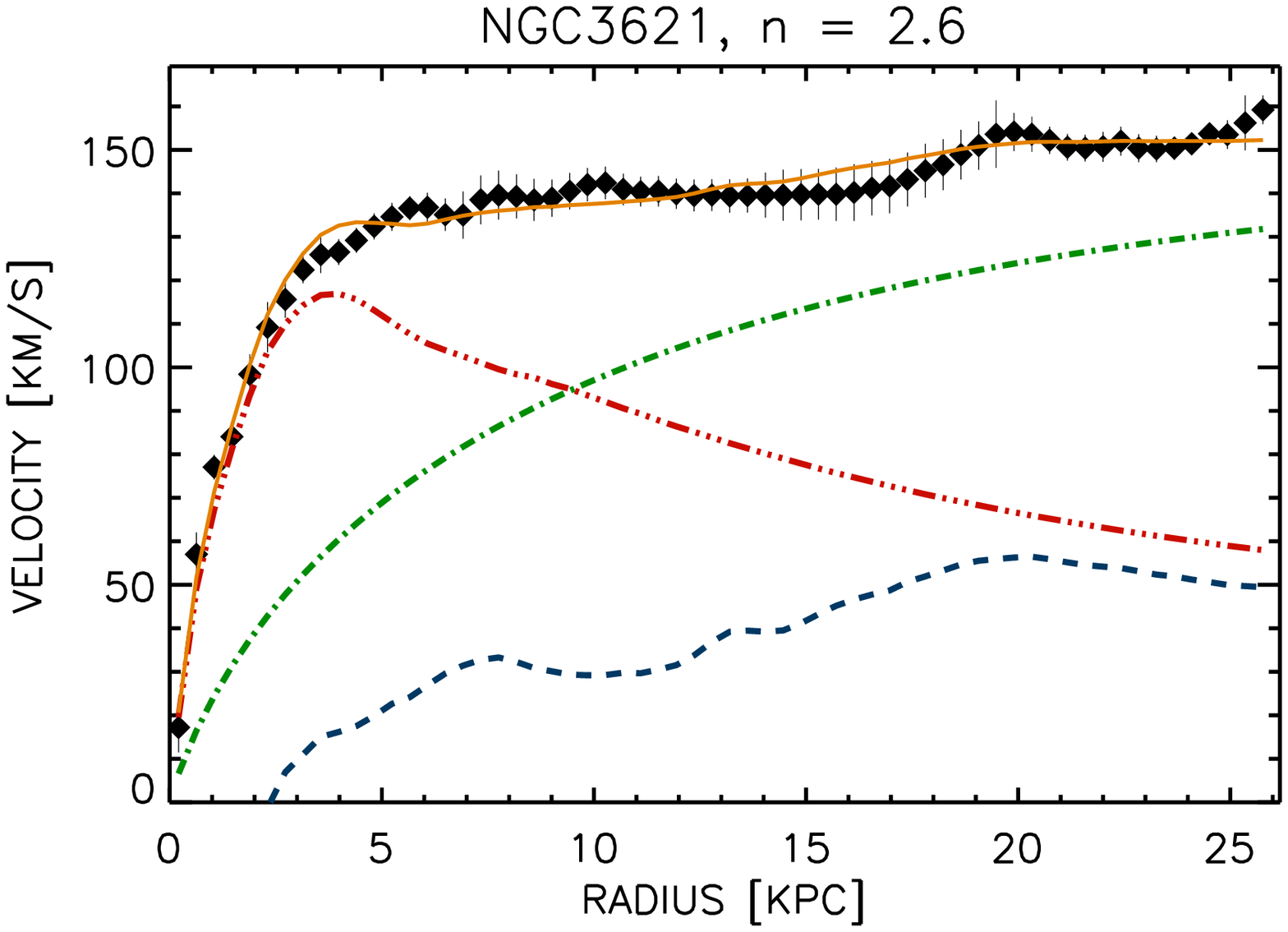}\includegraphics[width=0.25\textwidth]{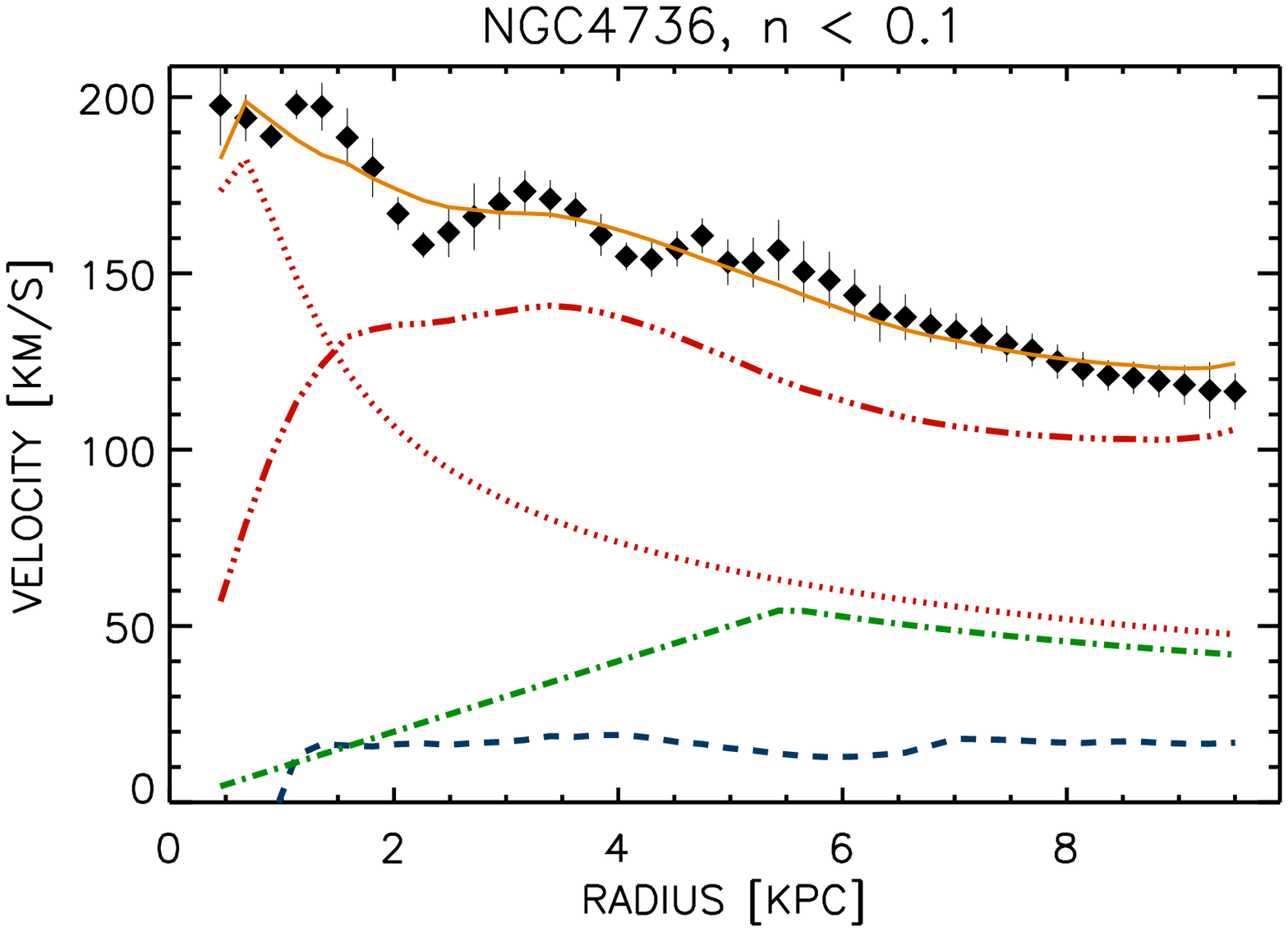}\includegraphics[width=0.25\textwidth]{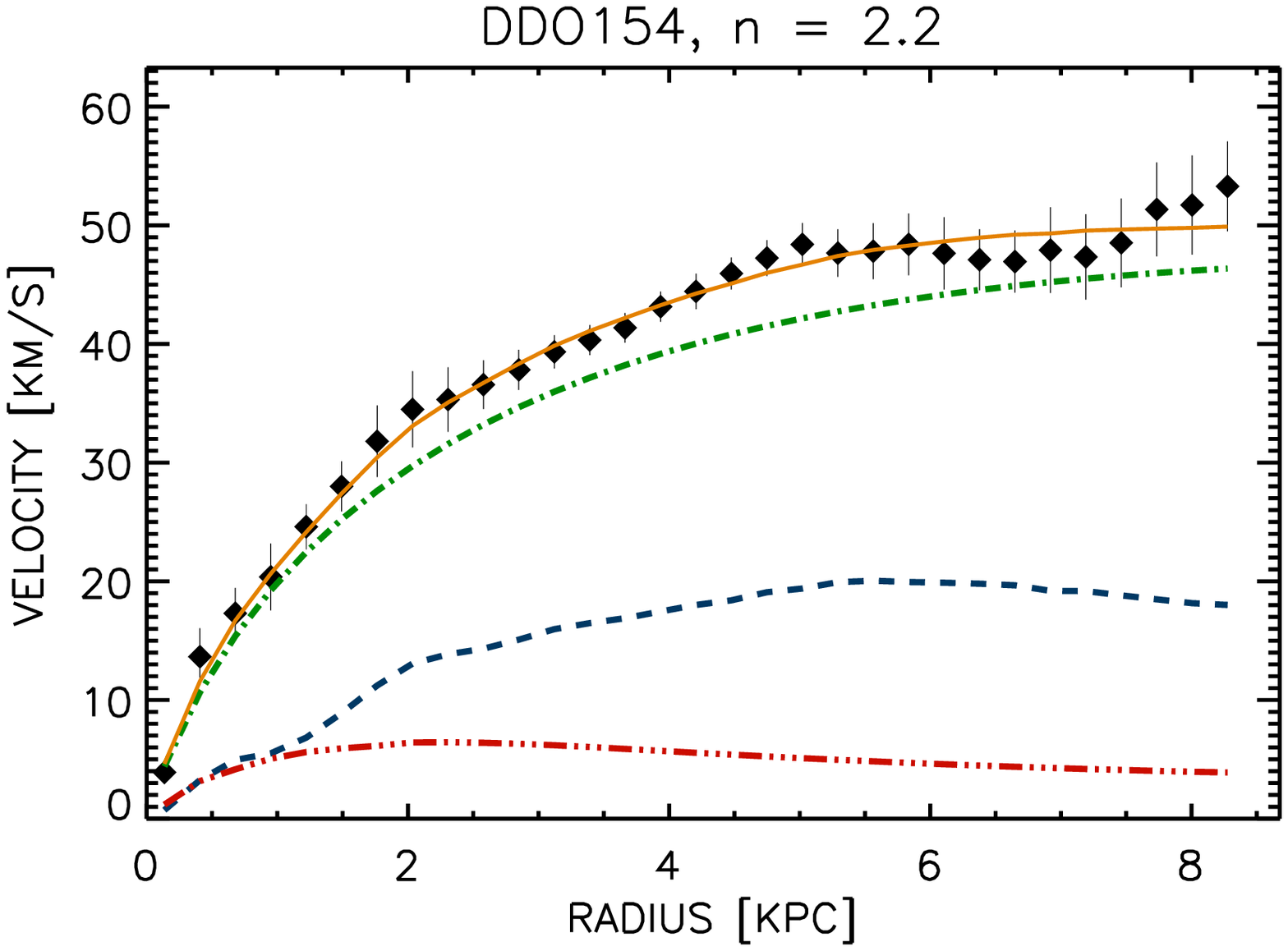}\includegraphics[width=0.25\textwidth]{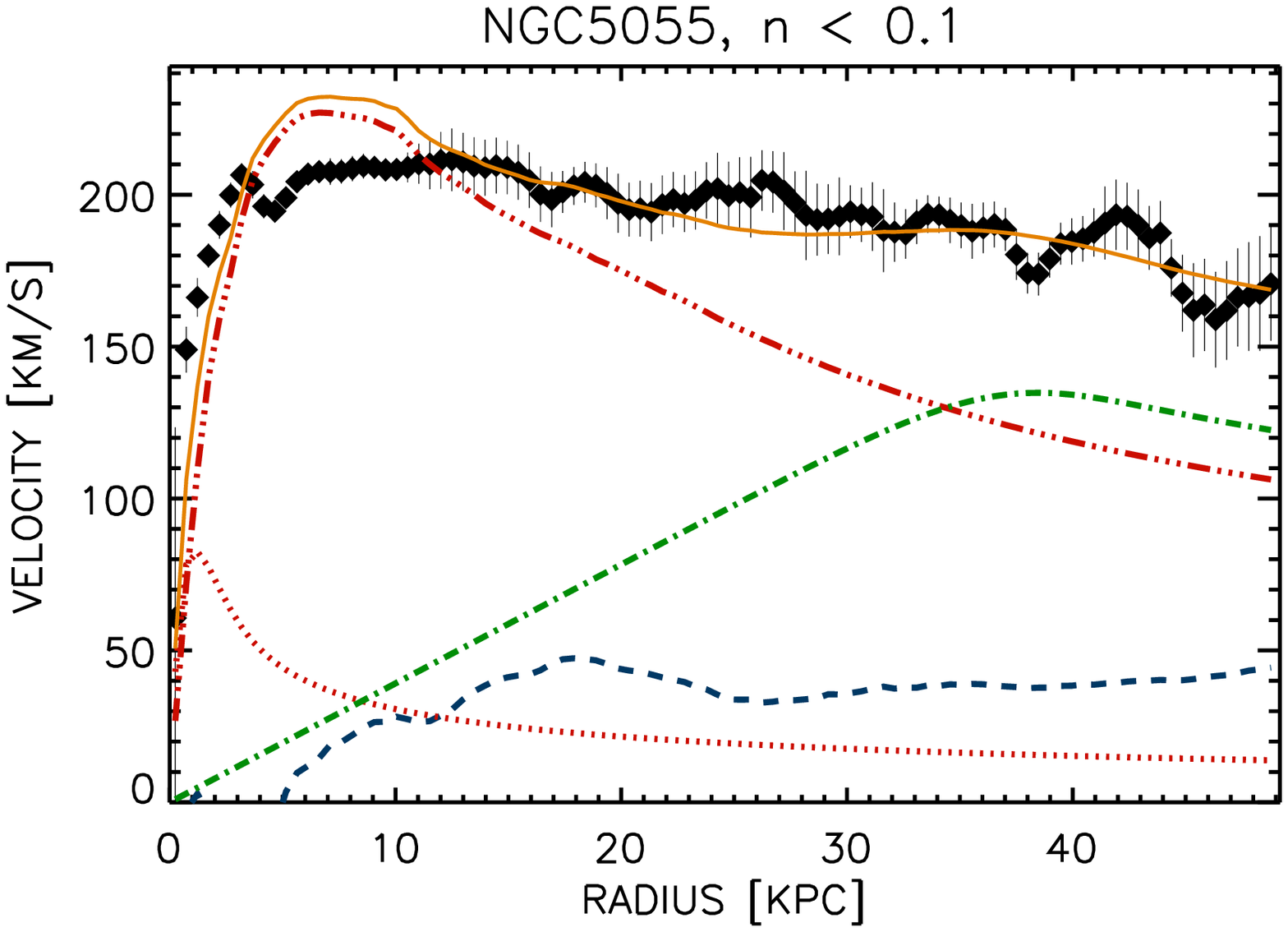}
\includegraphics[width=0.25\textwidth]{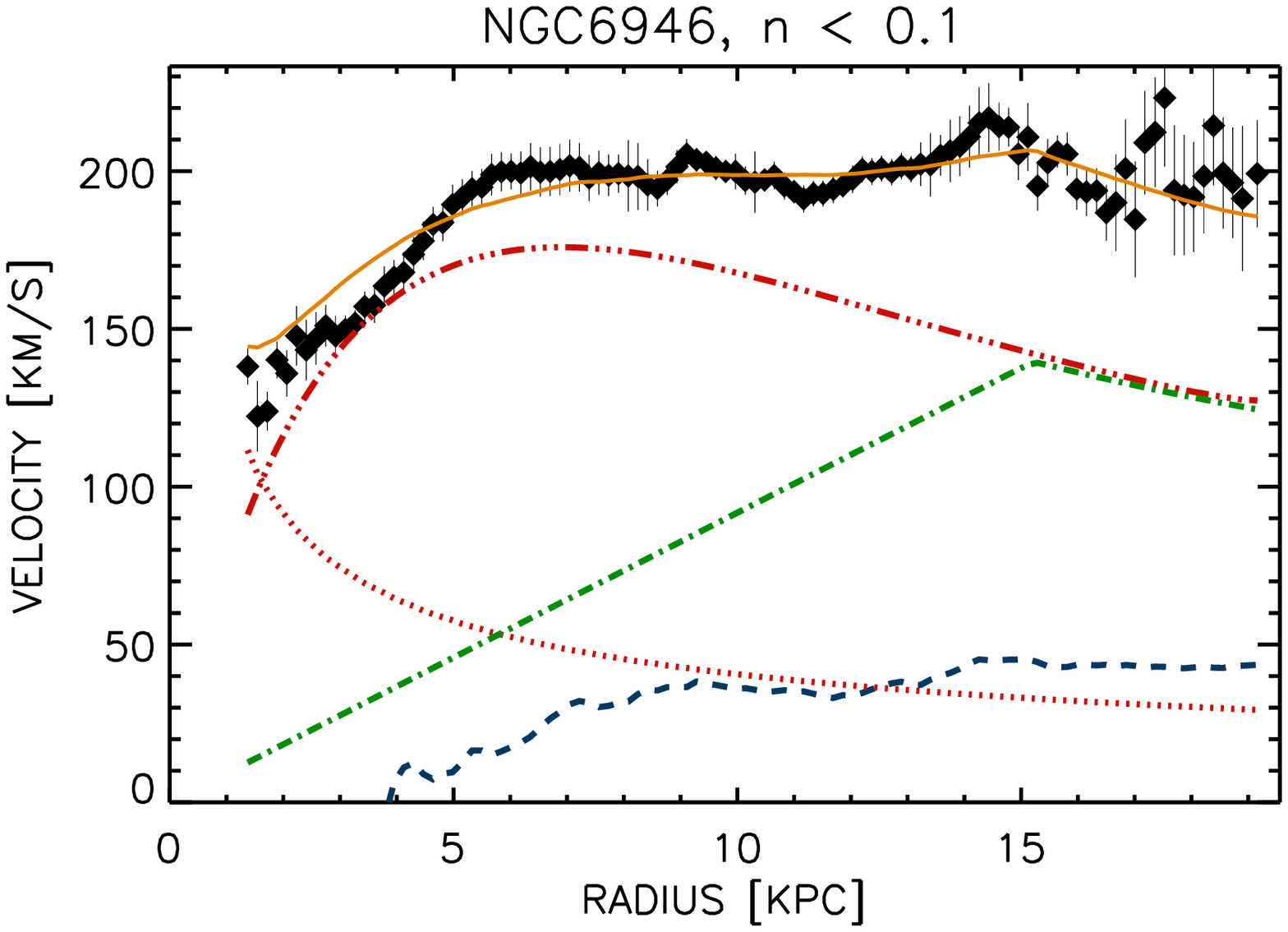}\includegraphics[width=0.25\textwidth]{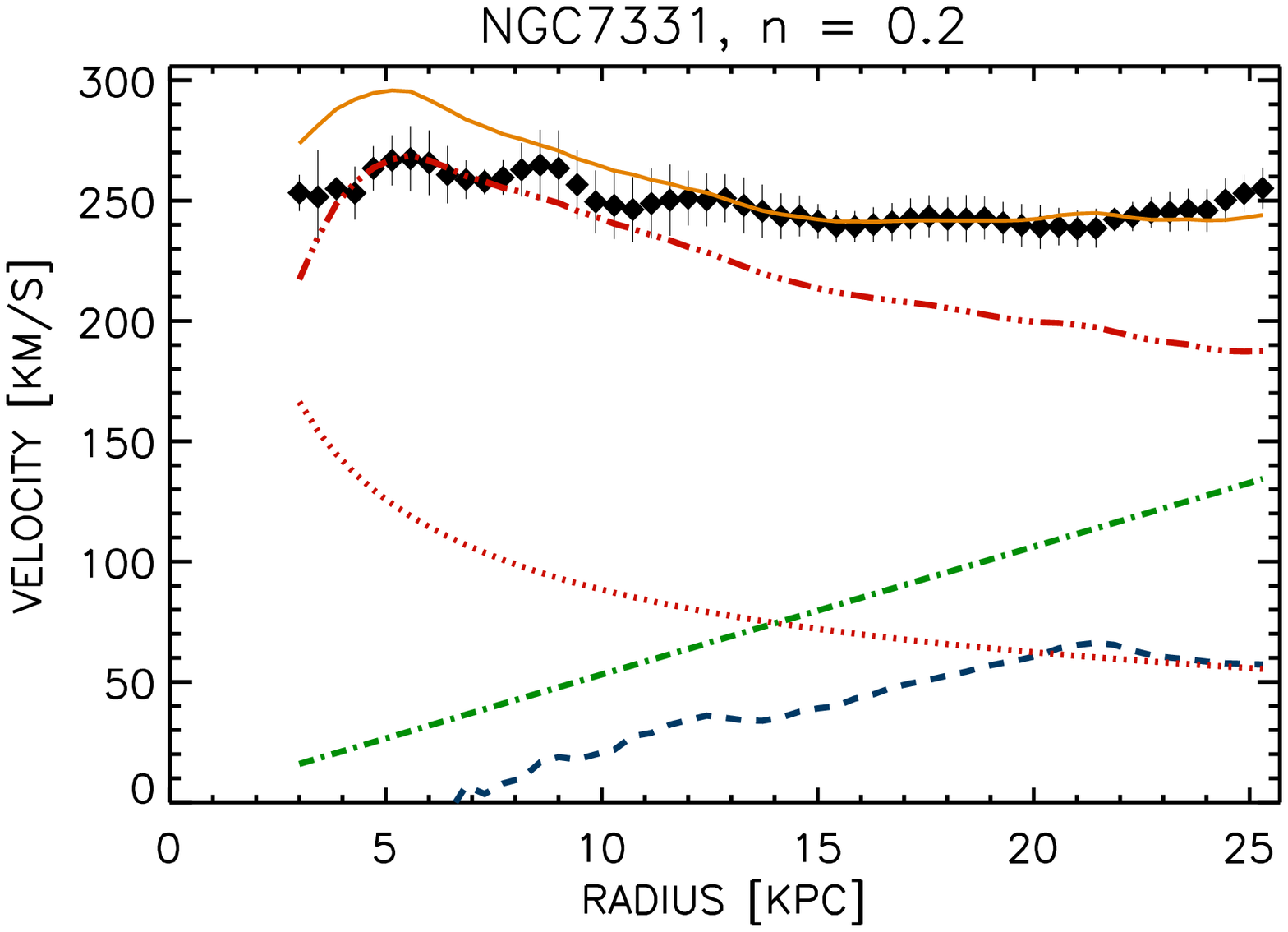}\includegraphics[width=0.25\textwidth]{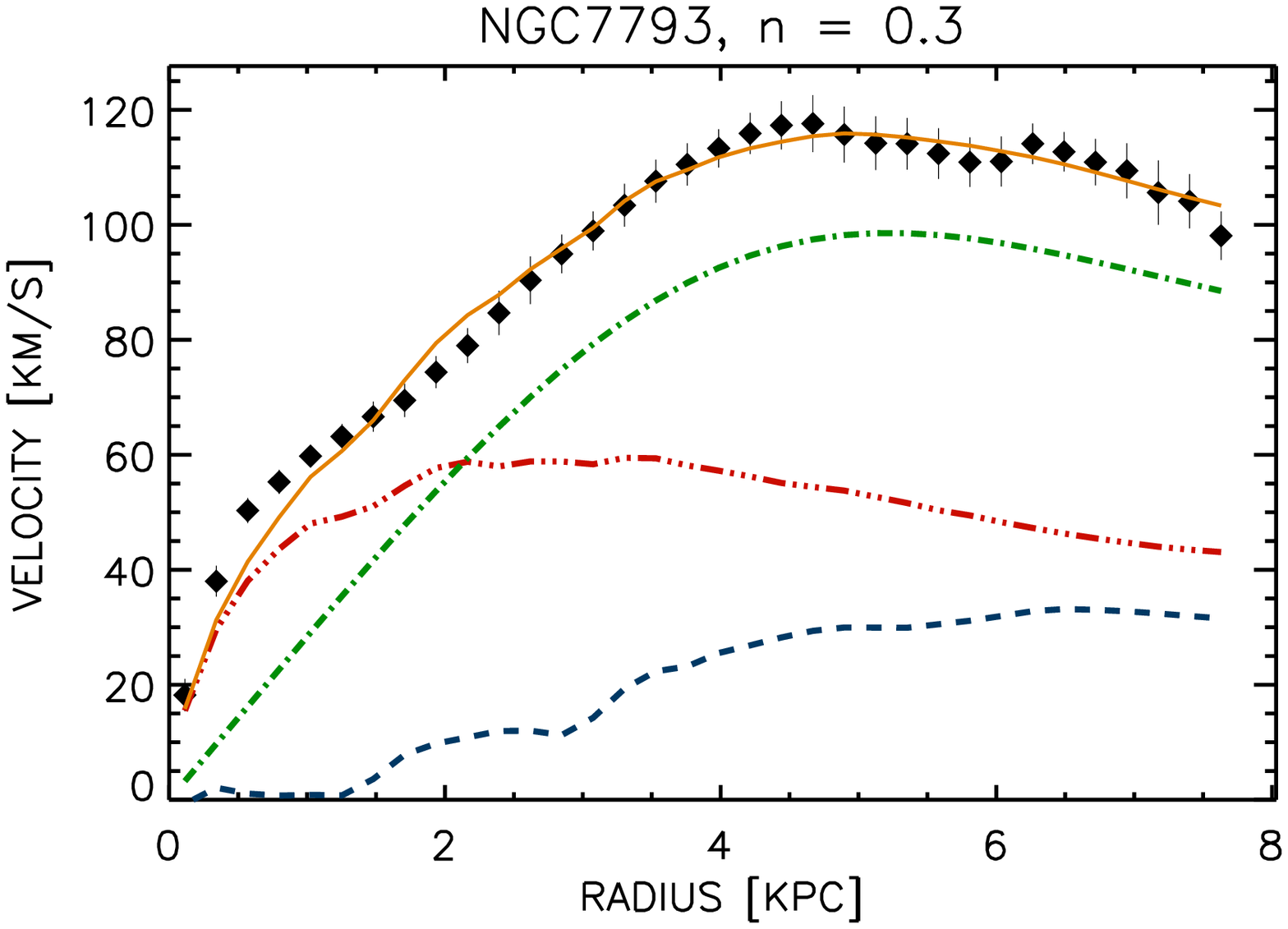}\includegraphics[width=0.25\textwidth]{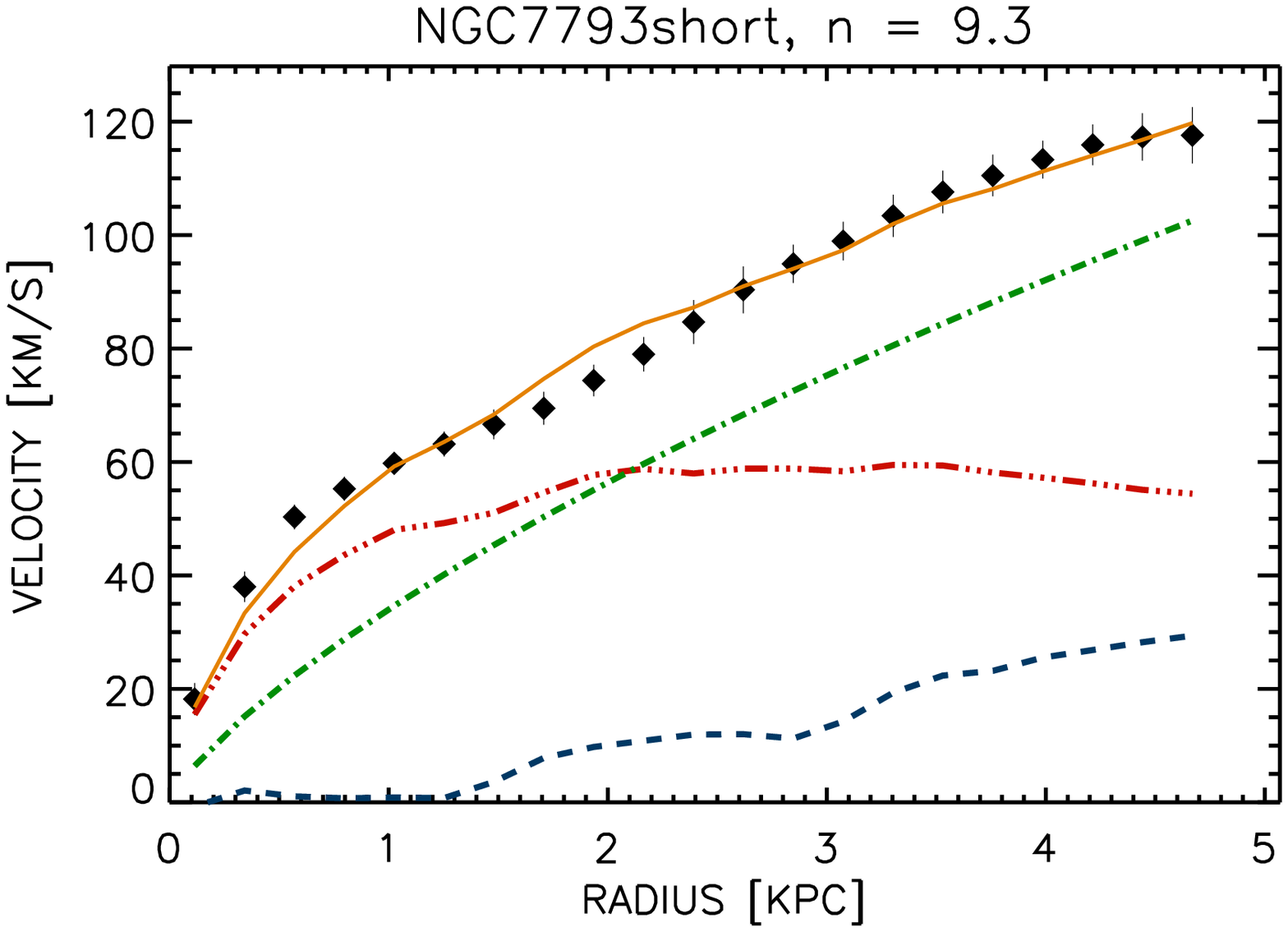}
 \caption{Mass models for the Einasto halo derived from the \hi\ RC of
   THINGS galaxies. Displayed results are for fixed stellar
   mass-to-light ratios derived from stellar population synthesis with
   a diet-Salpeter initial mass function and whose values are given in
   D08.  A red dashed-dotted (dotted) line is for the stellar disk
   (bulge, respectively) components, a blue dashed line for the atomic
   gas, a green dashed-dotted line for the dark halo and an orange
   thick line for the overall model.}
 \label{fig:rc-salpeter}
 \end{center}
 \end{figure*}

\begin{figure*}
 \begin{center}
\includegraphics[width=0.25\textwidth]{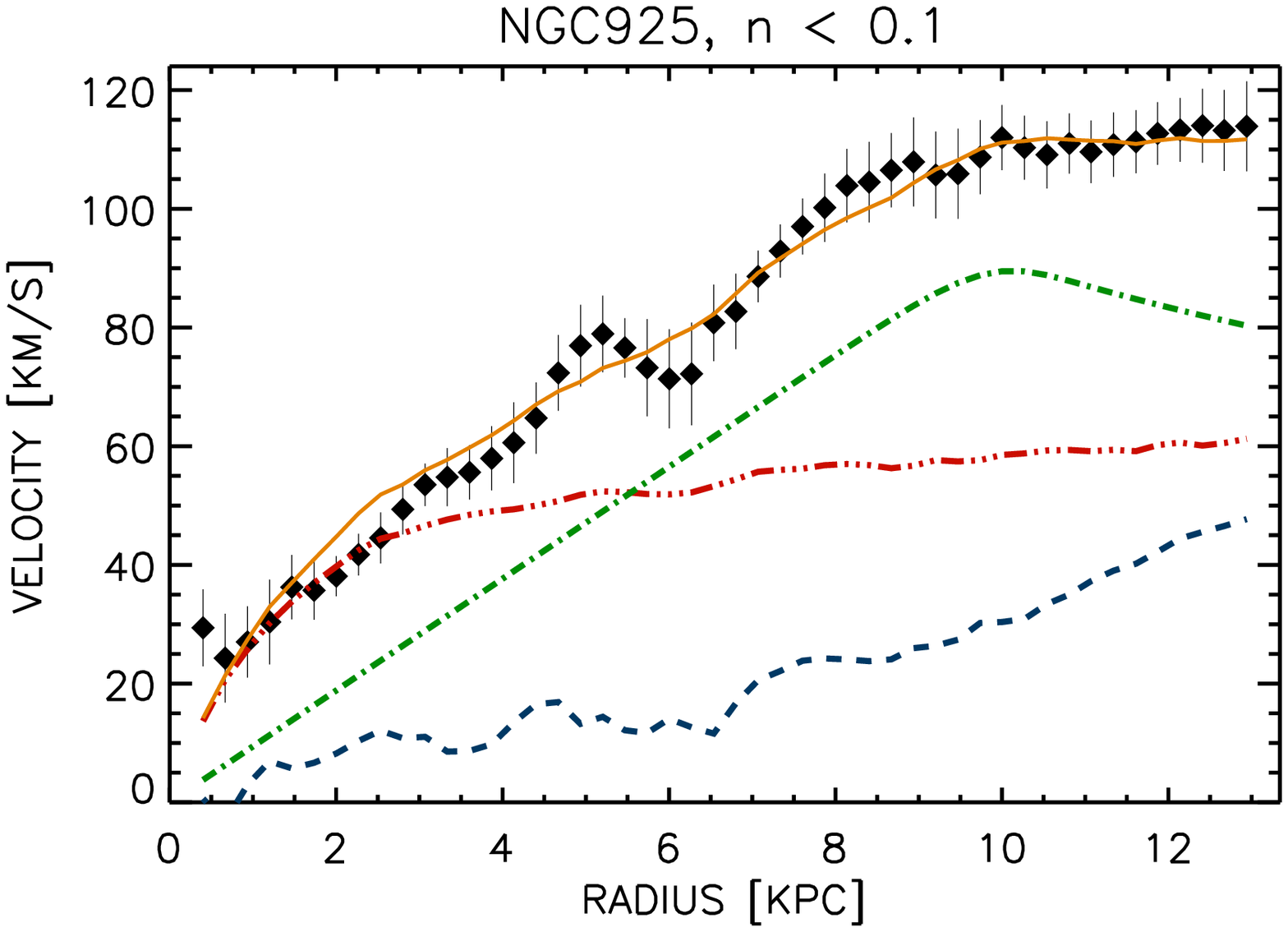}\includegraphics[width=0.25\textwidth]{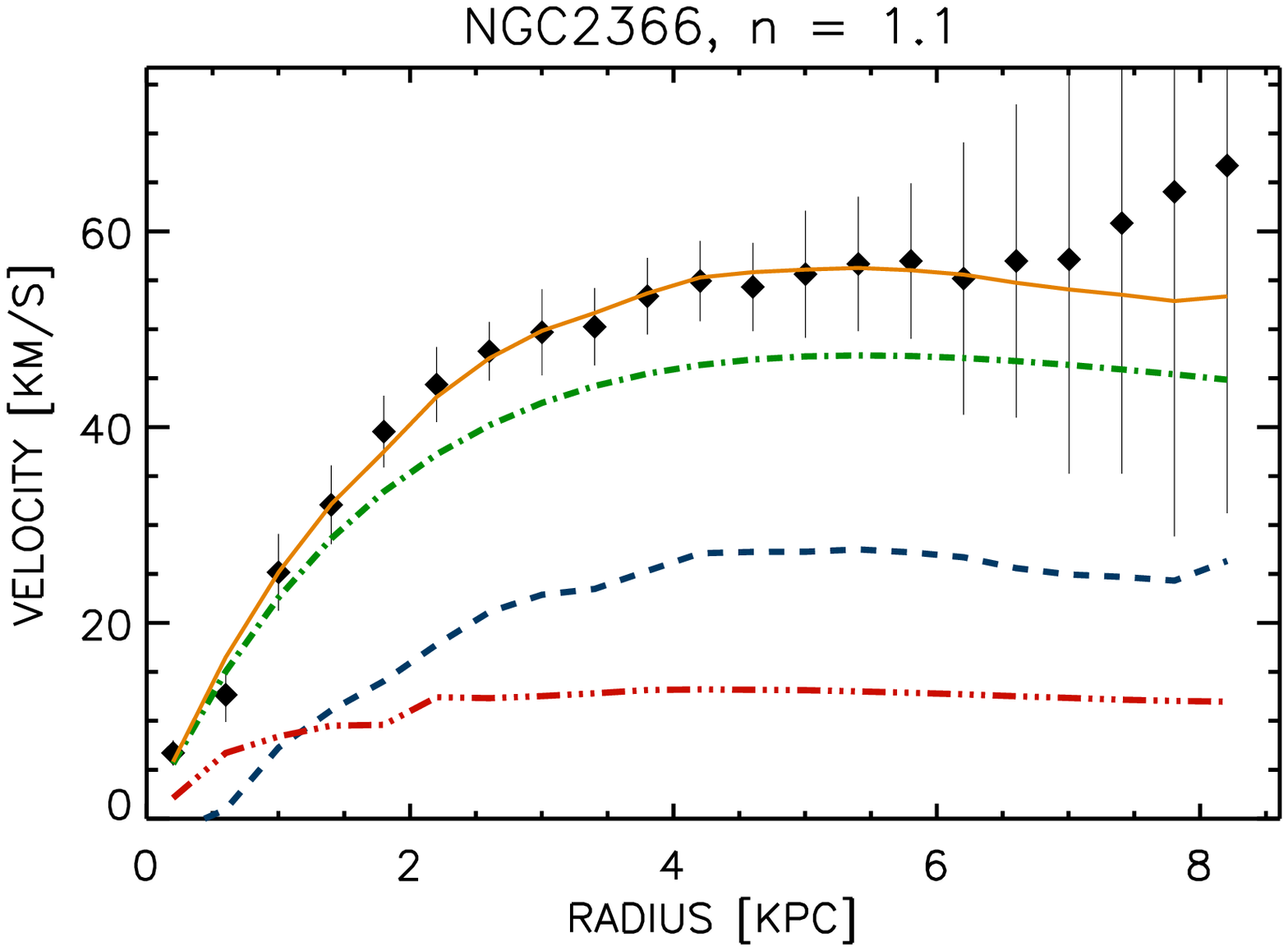}\includegraphics[width=0.25\textwidth]{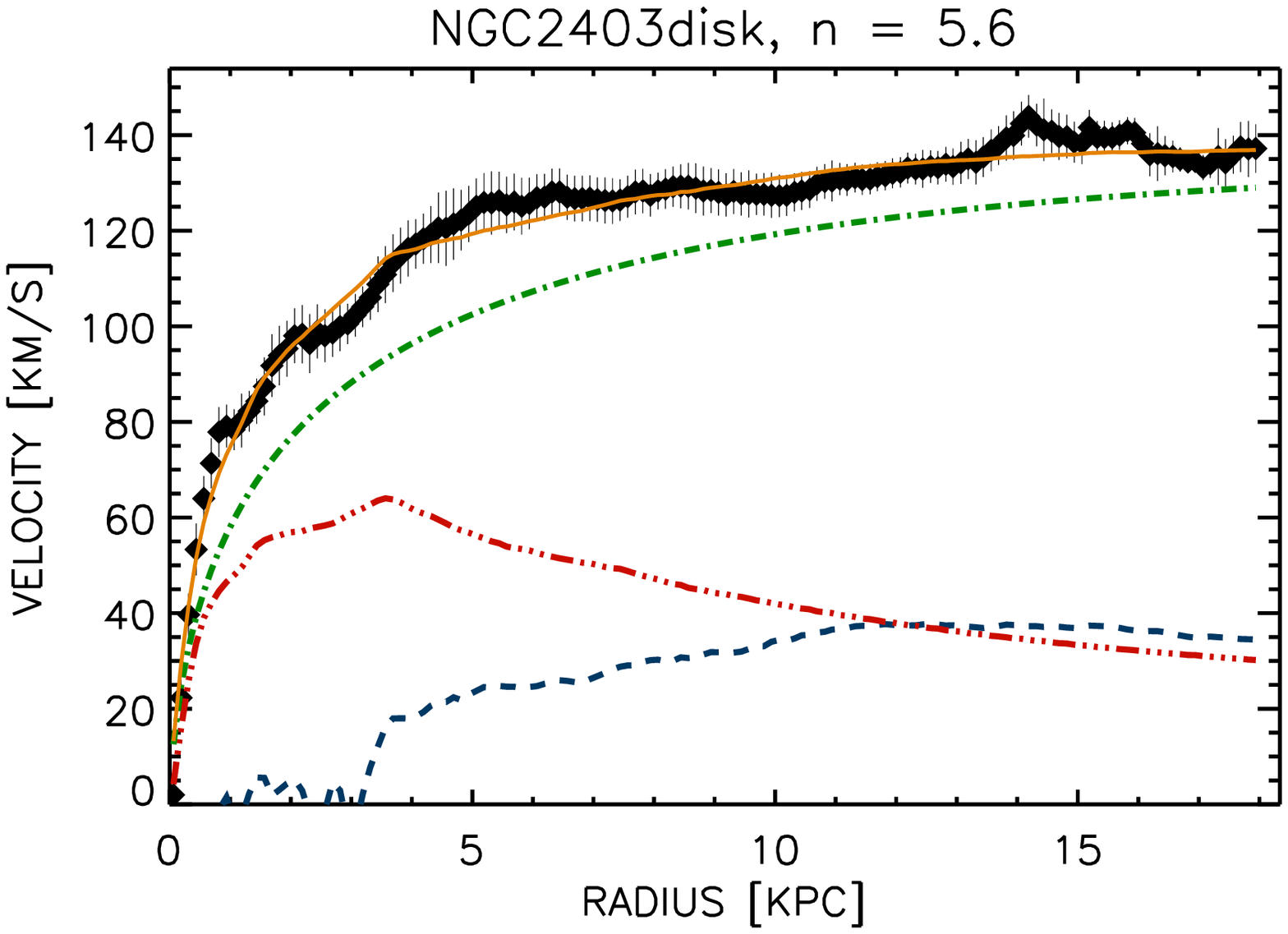}\includegraphics[width=0.25\textwidth]{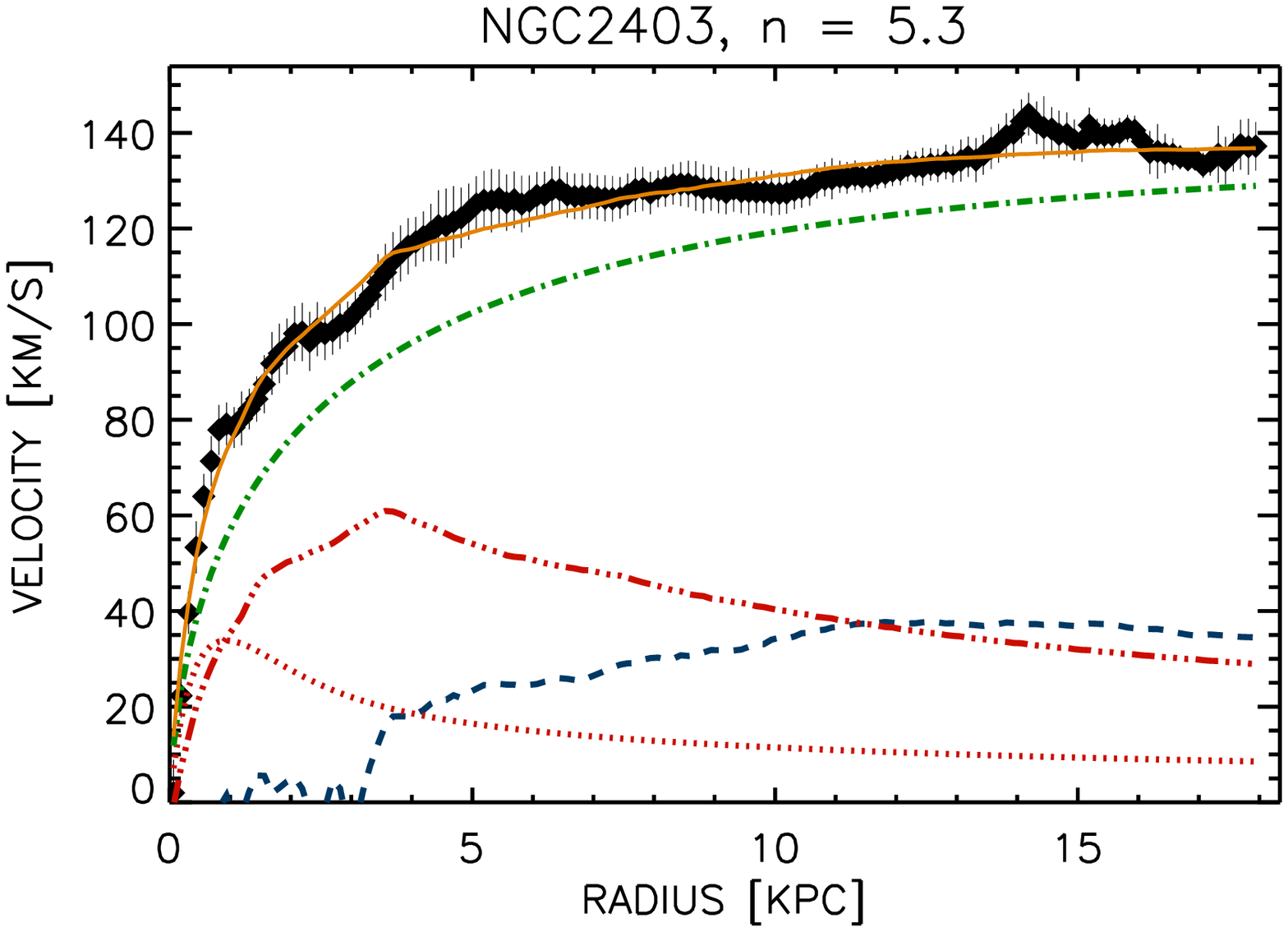}
\includegraphics[width=0.25\textwidth]{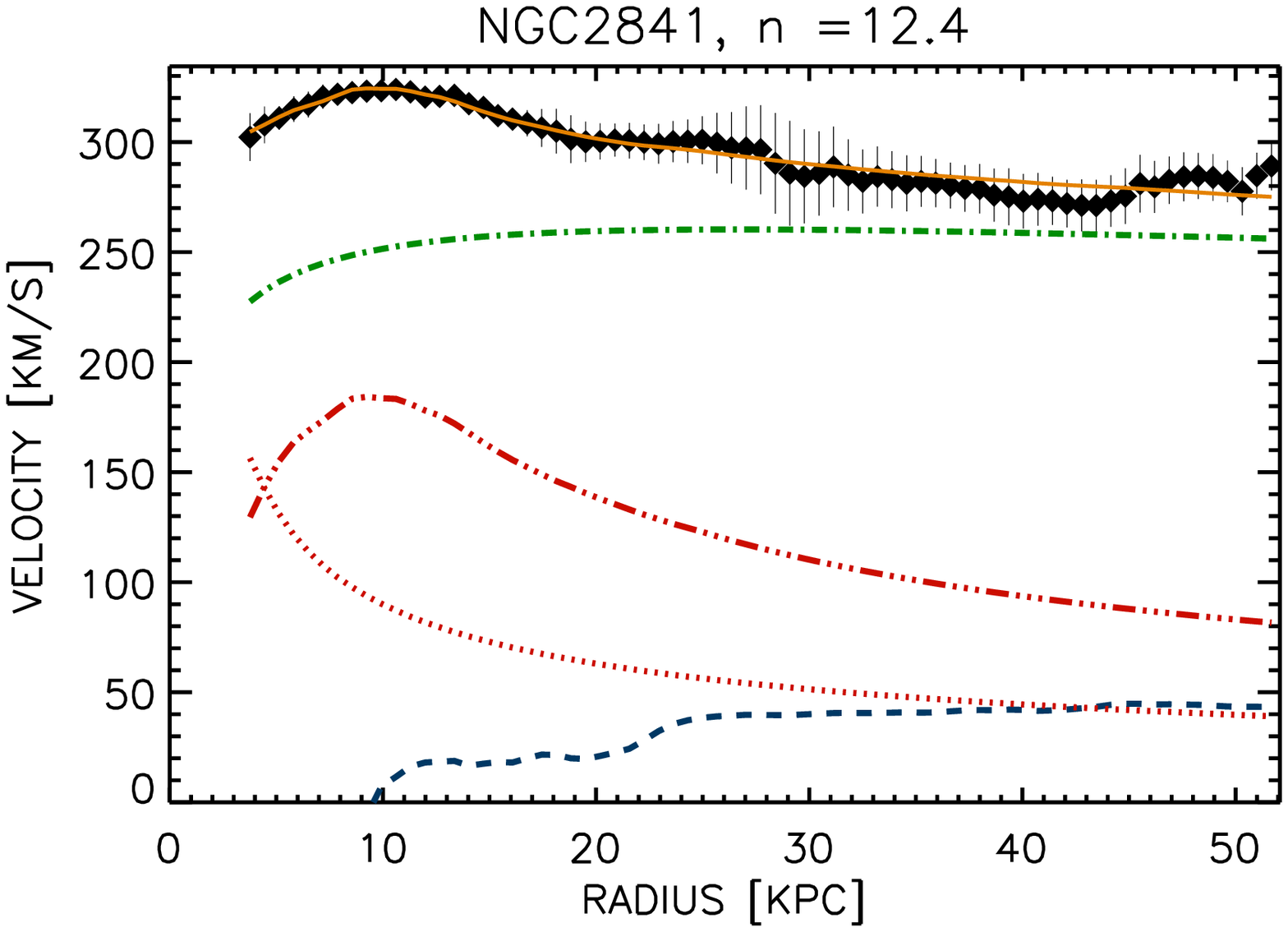}\includegraphics[width=0.25\textwidth]{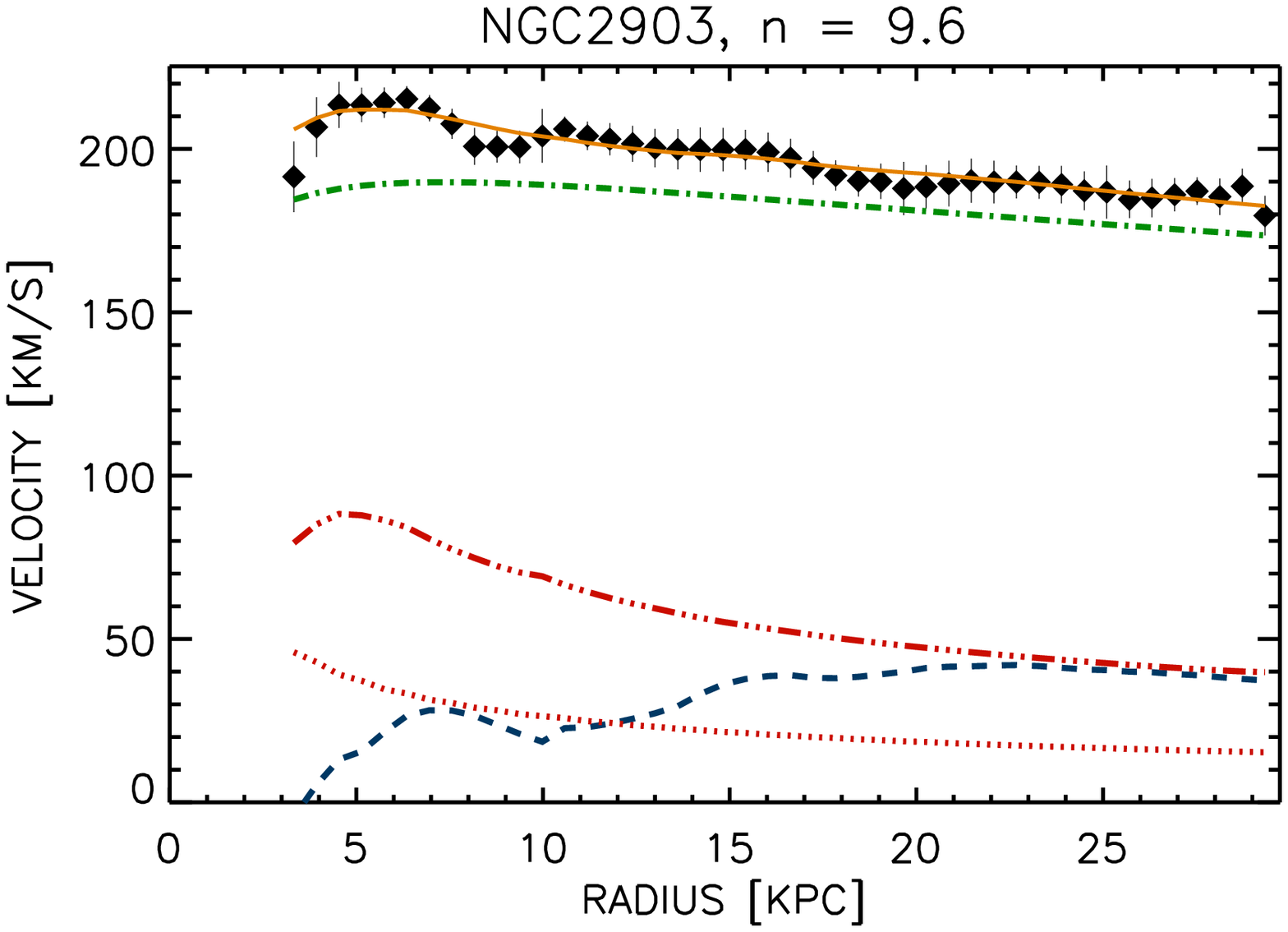}\includegraphics[width=0.25\textwidth]{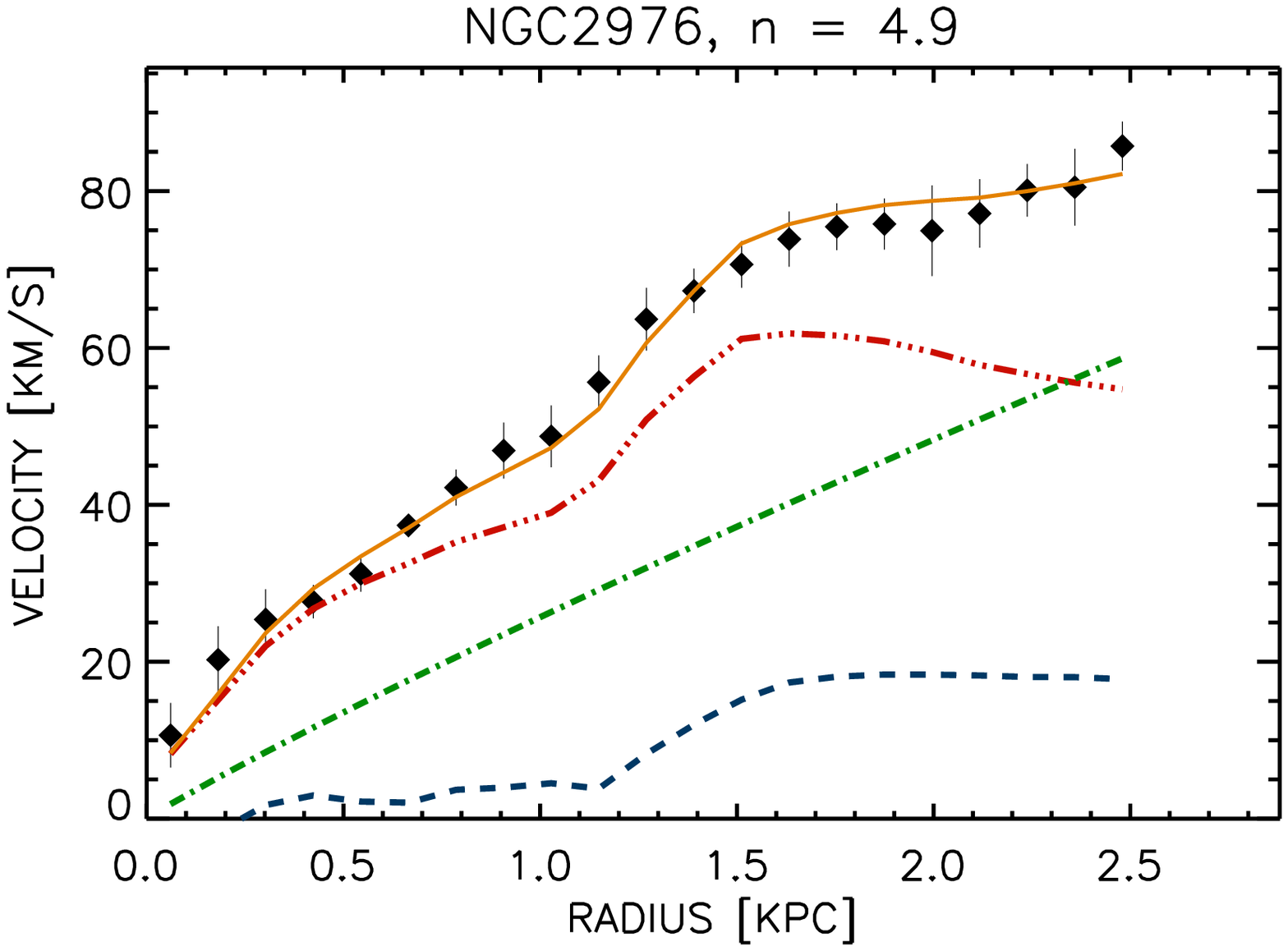}\includegraphics[width=0.25\textwidth]{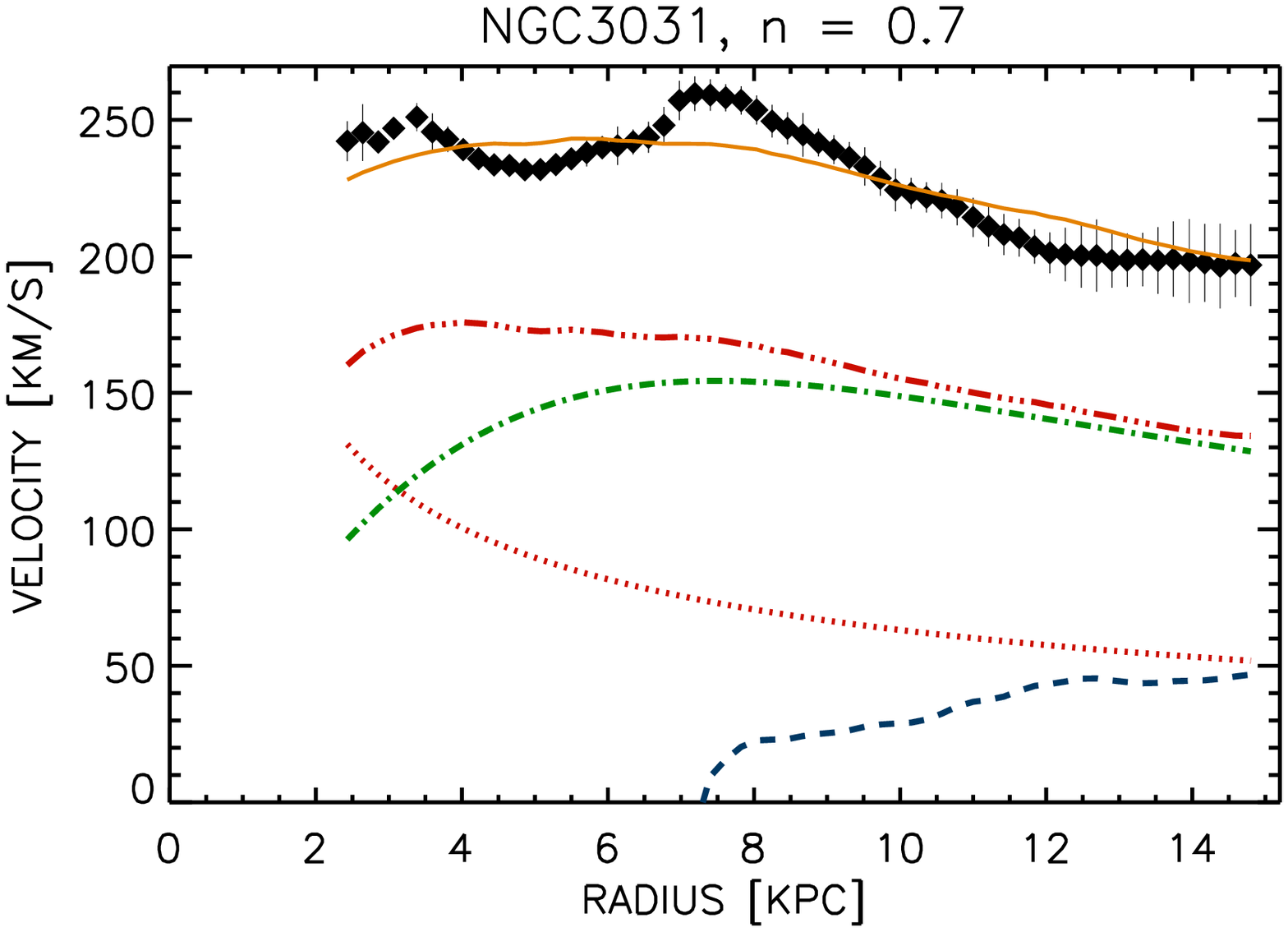}
\includegraphics[width=0.25\textwidth]{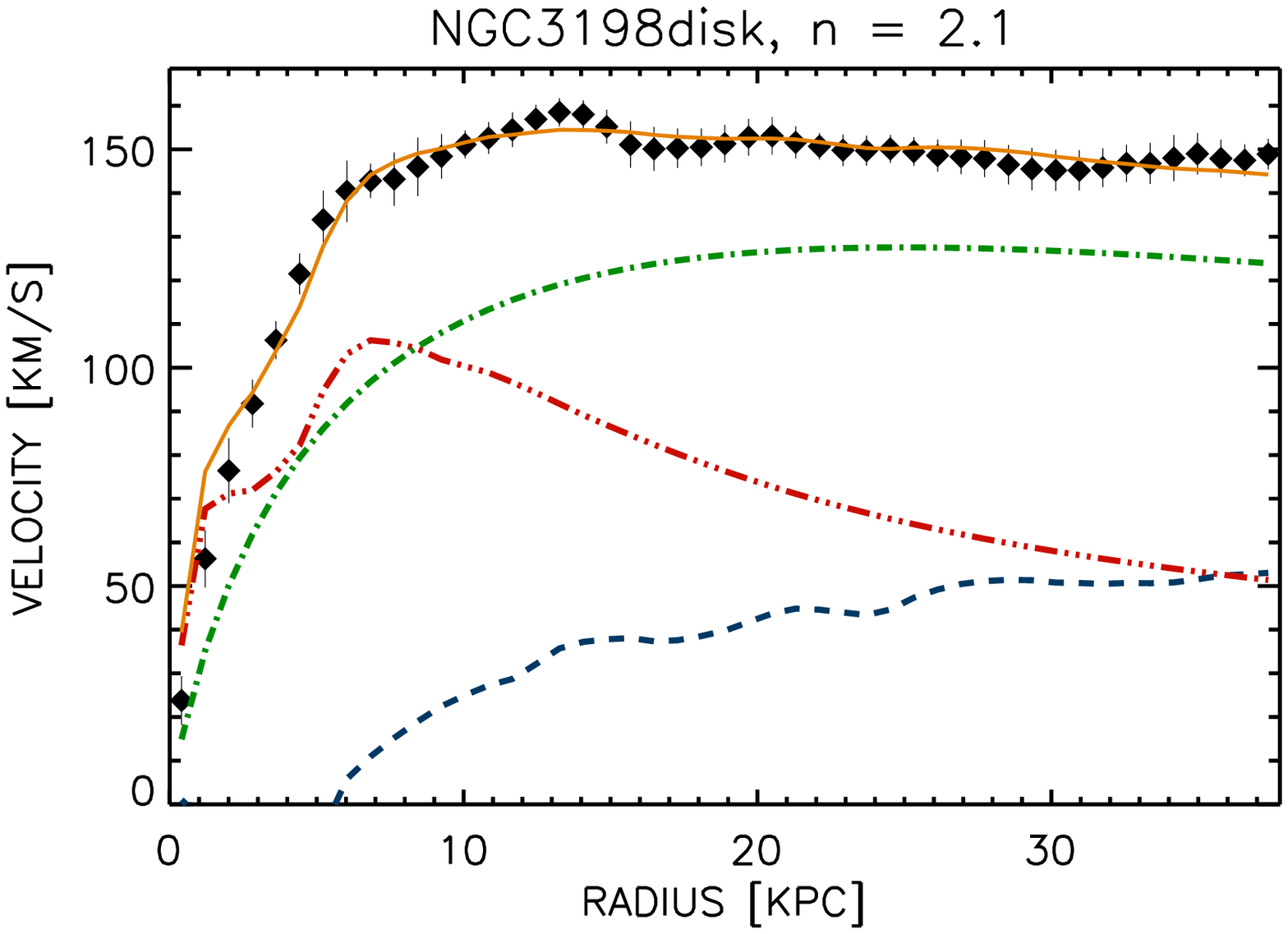}\includegraphics[width=0.25\textwidth]{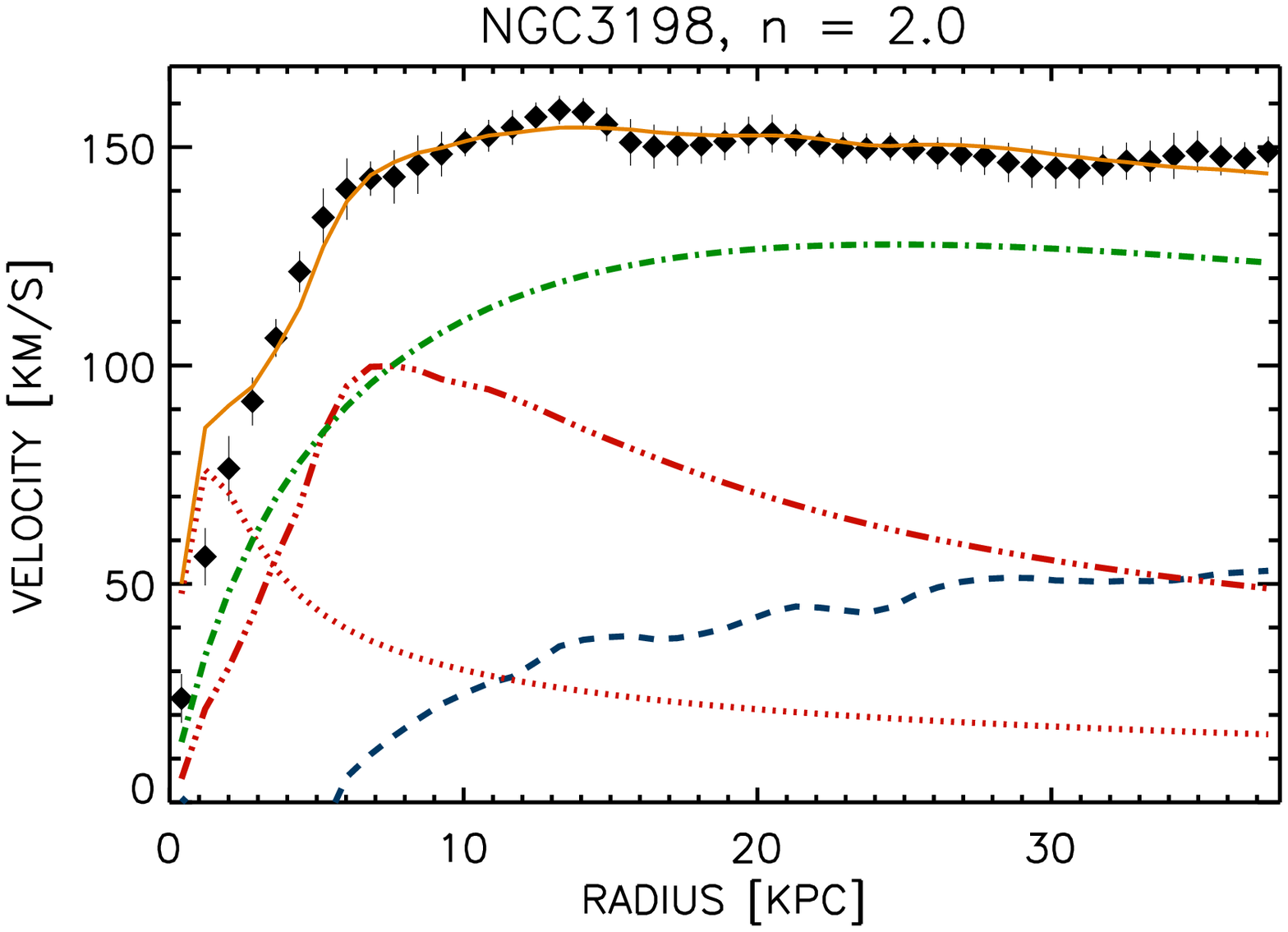}\includegraphics[width=0.25\textwidth]{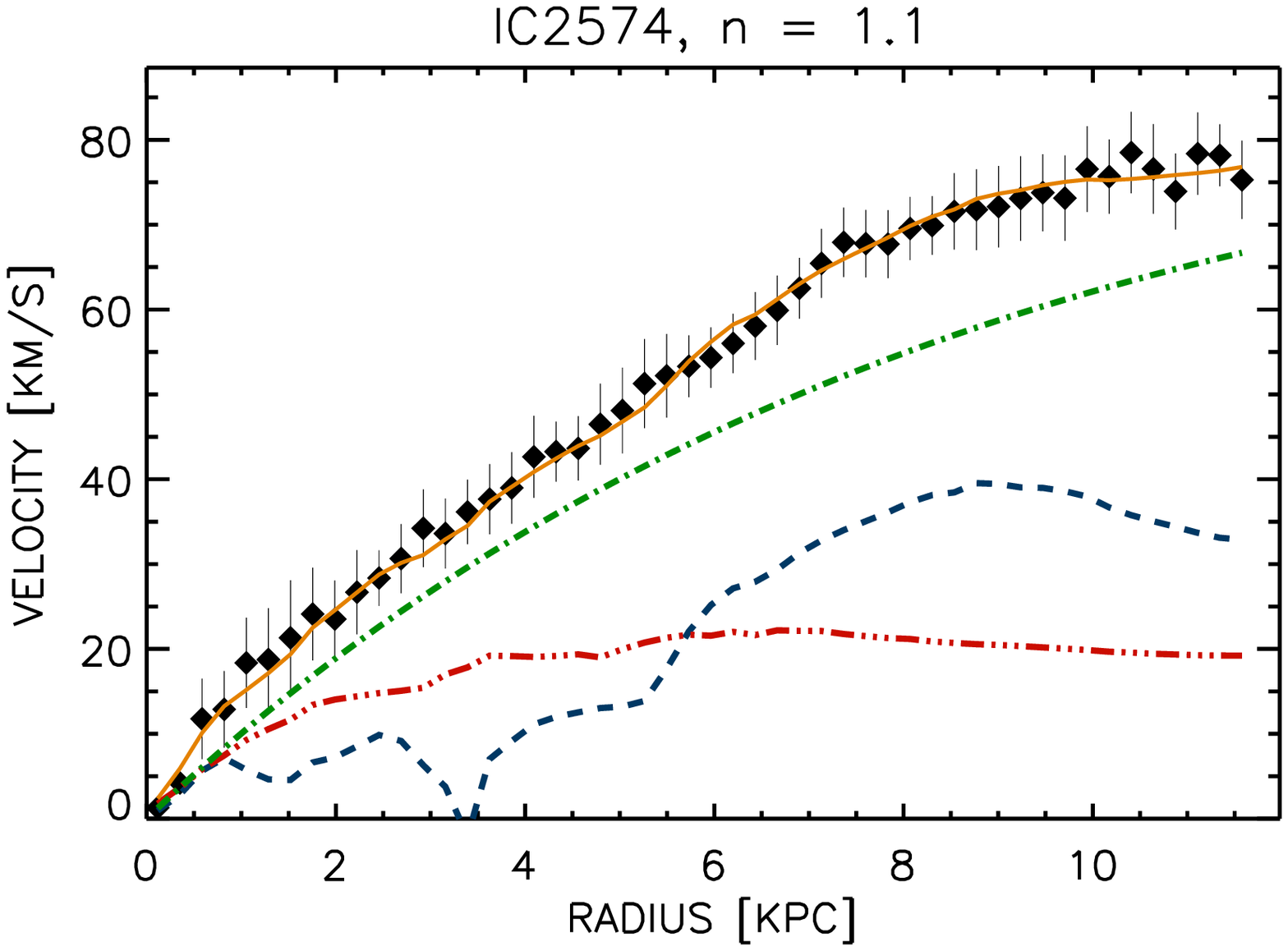}\includegraphics[width=0.25\textwidth]{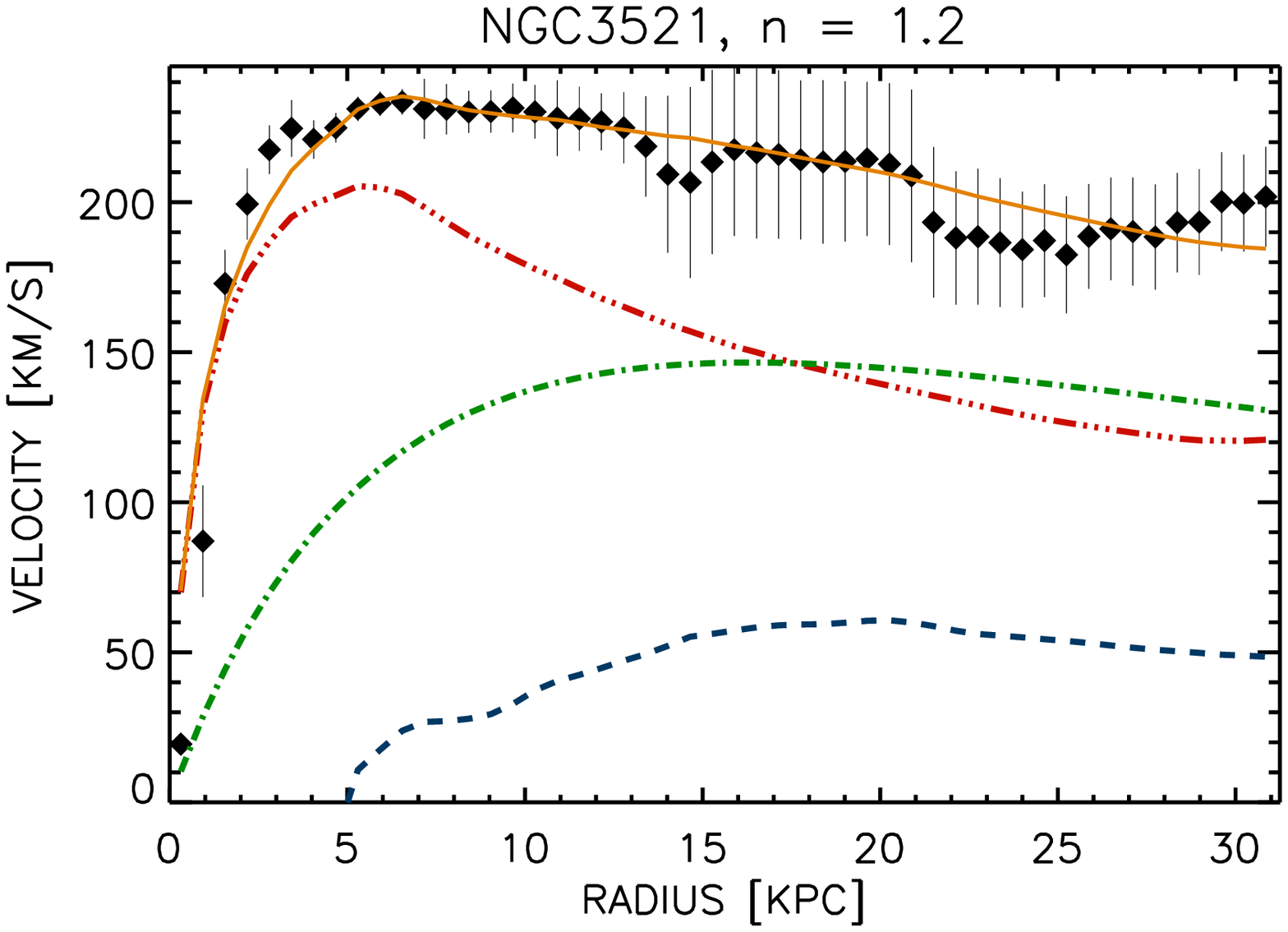}
\includegraphics[width=0.25\textwidth]{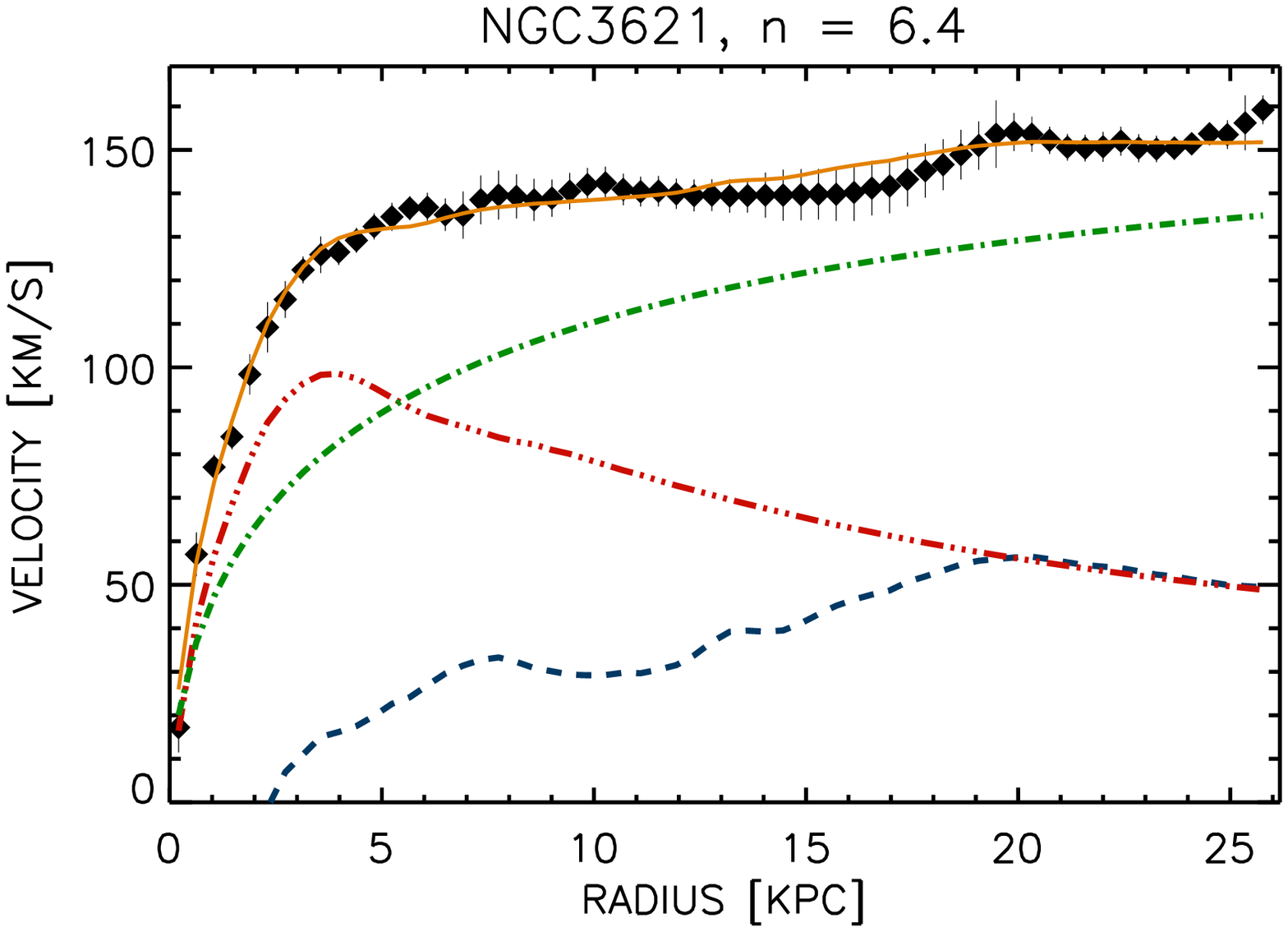}\includegraphics[width=0.25\textwidth]{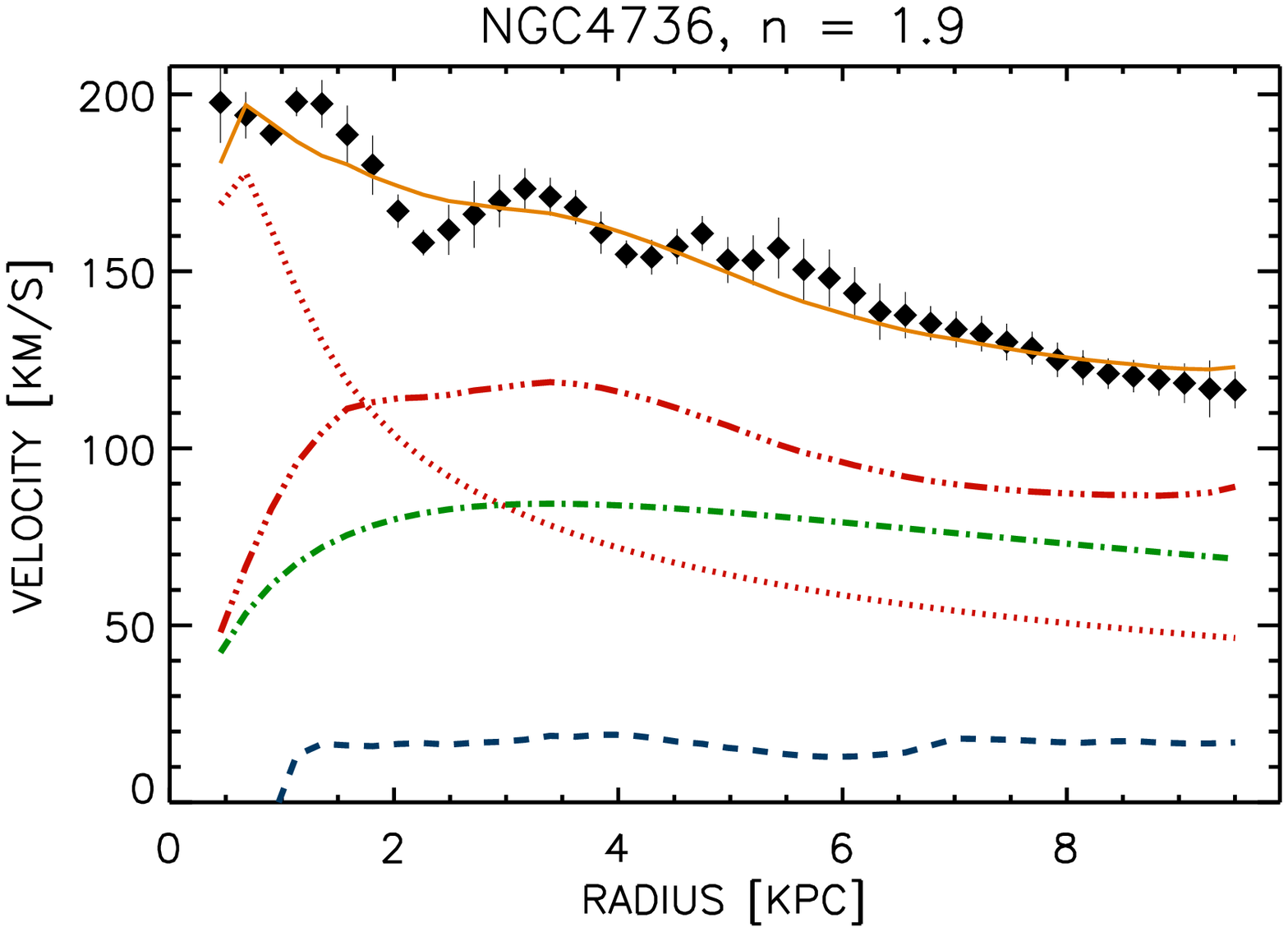}\includegraphics[width=0.25\textwidth]{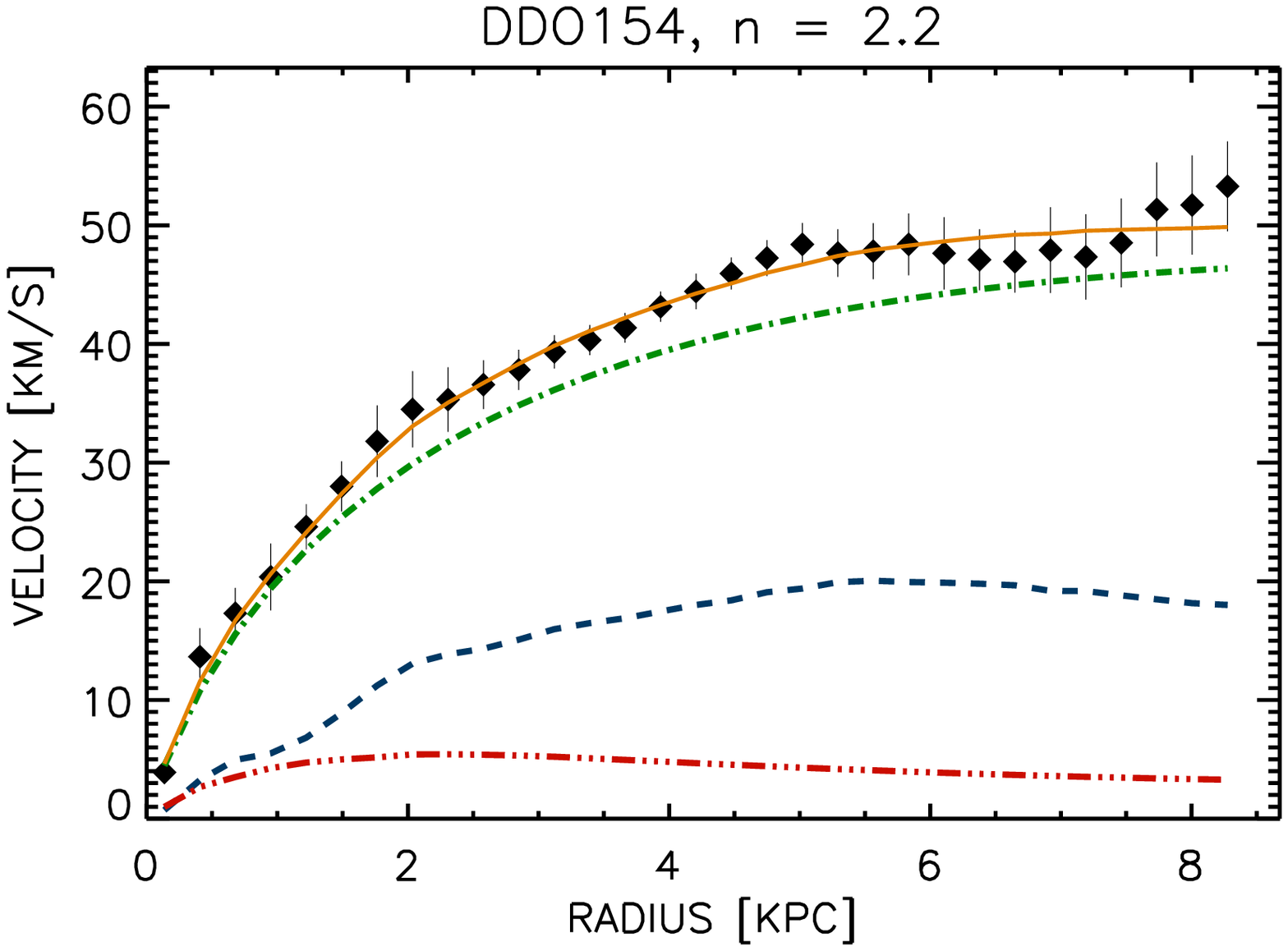}\includegraphics[width=0.25\textwidth]{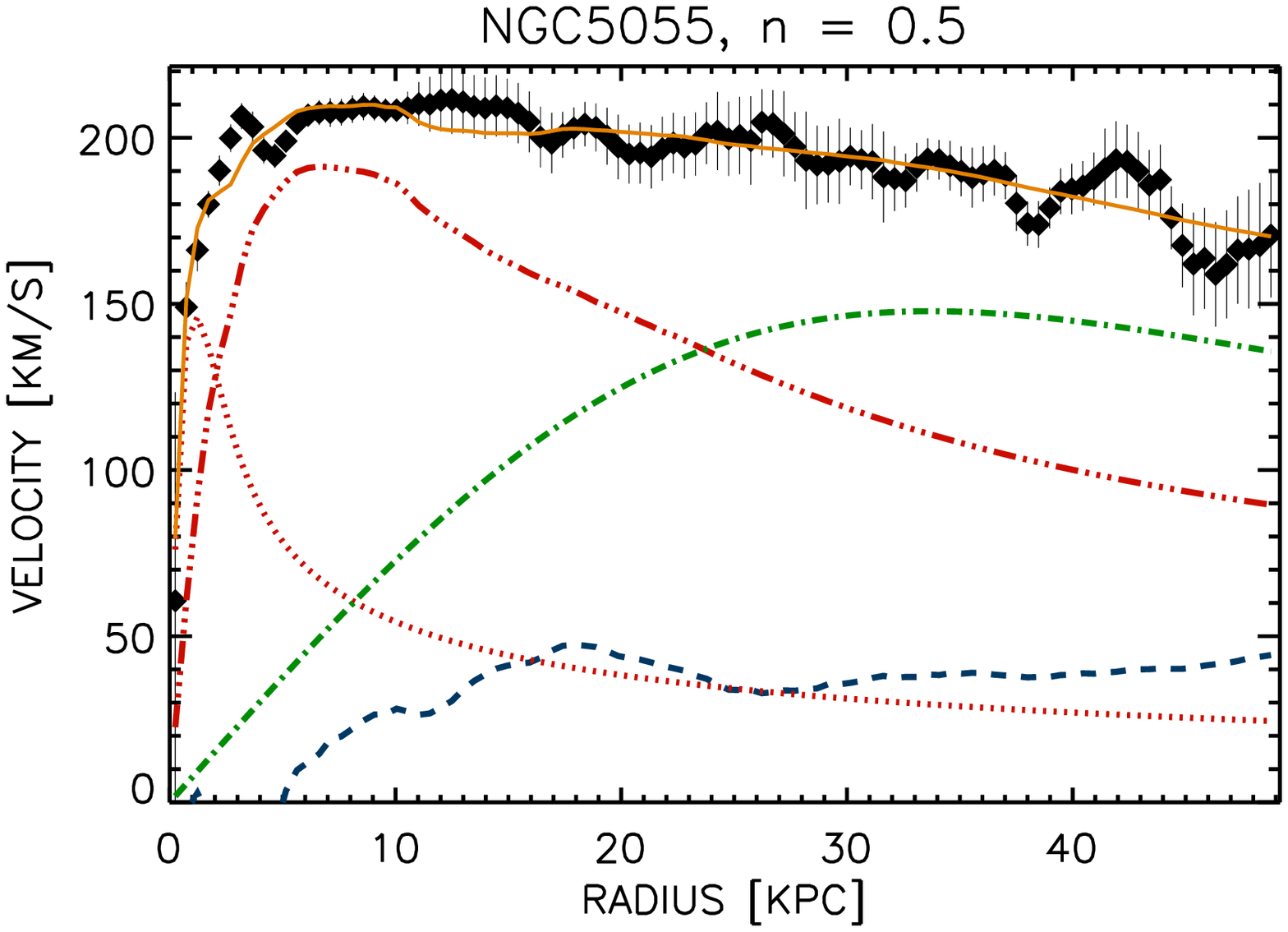}
\includegraphics[width=0.25\textwidth]{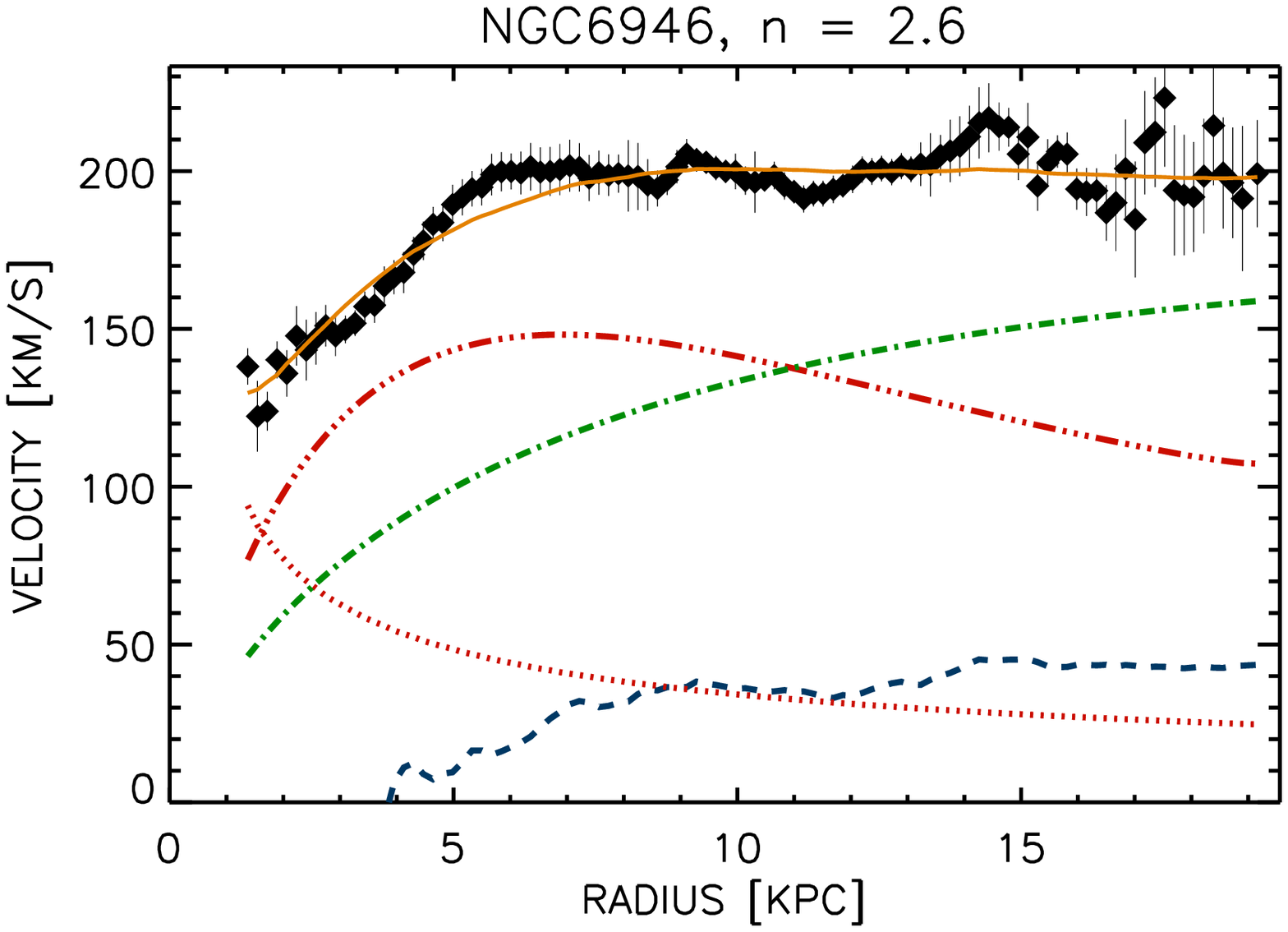}\includegraphics[width=0.25\textwidth]{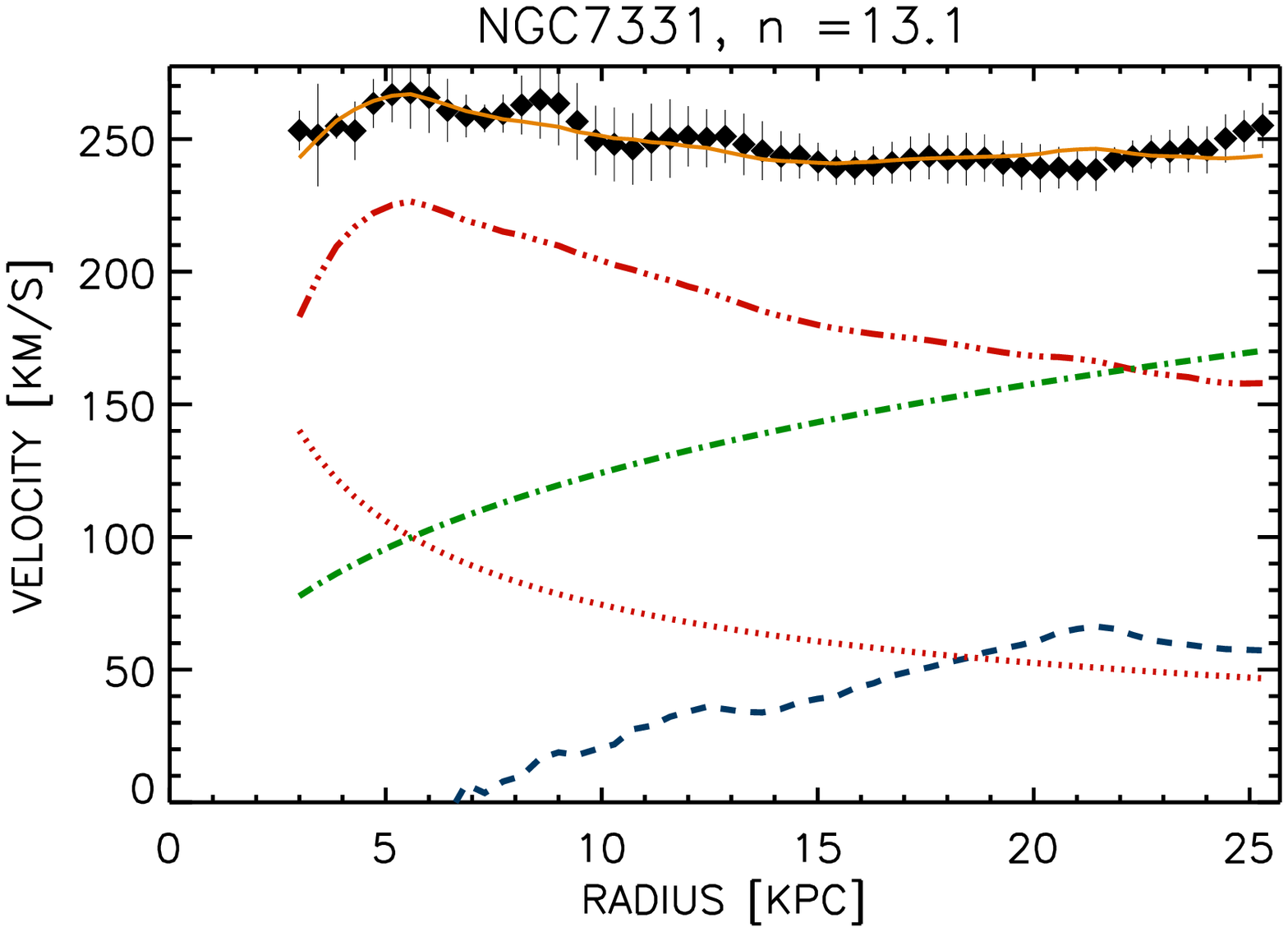}\includegraphics[width=0.25\textwidth]{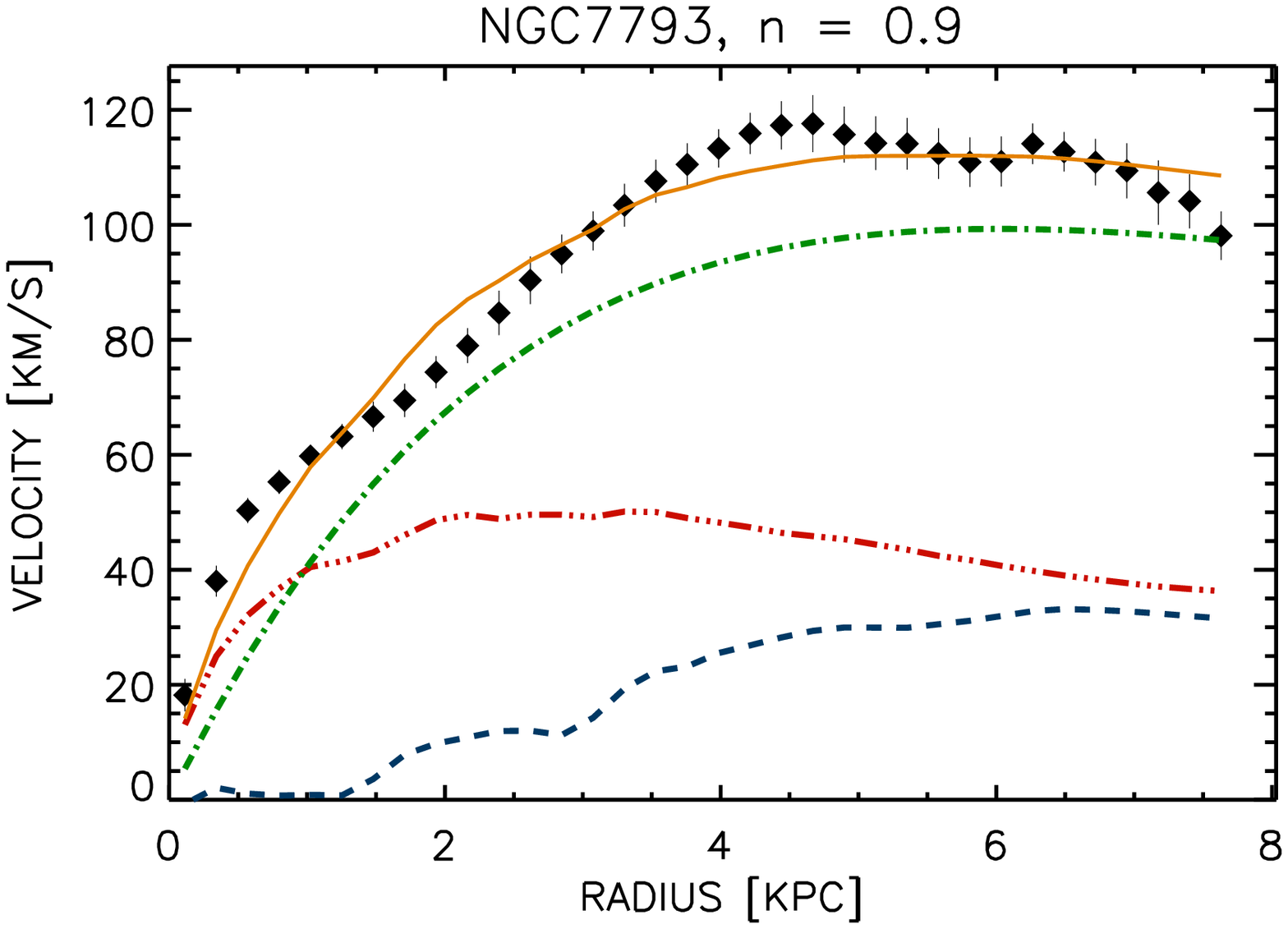}\includegraphics[width=0.25\textwidth]{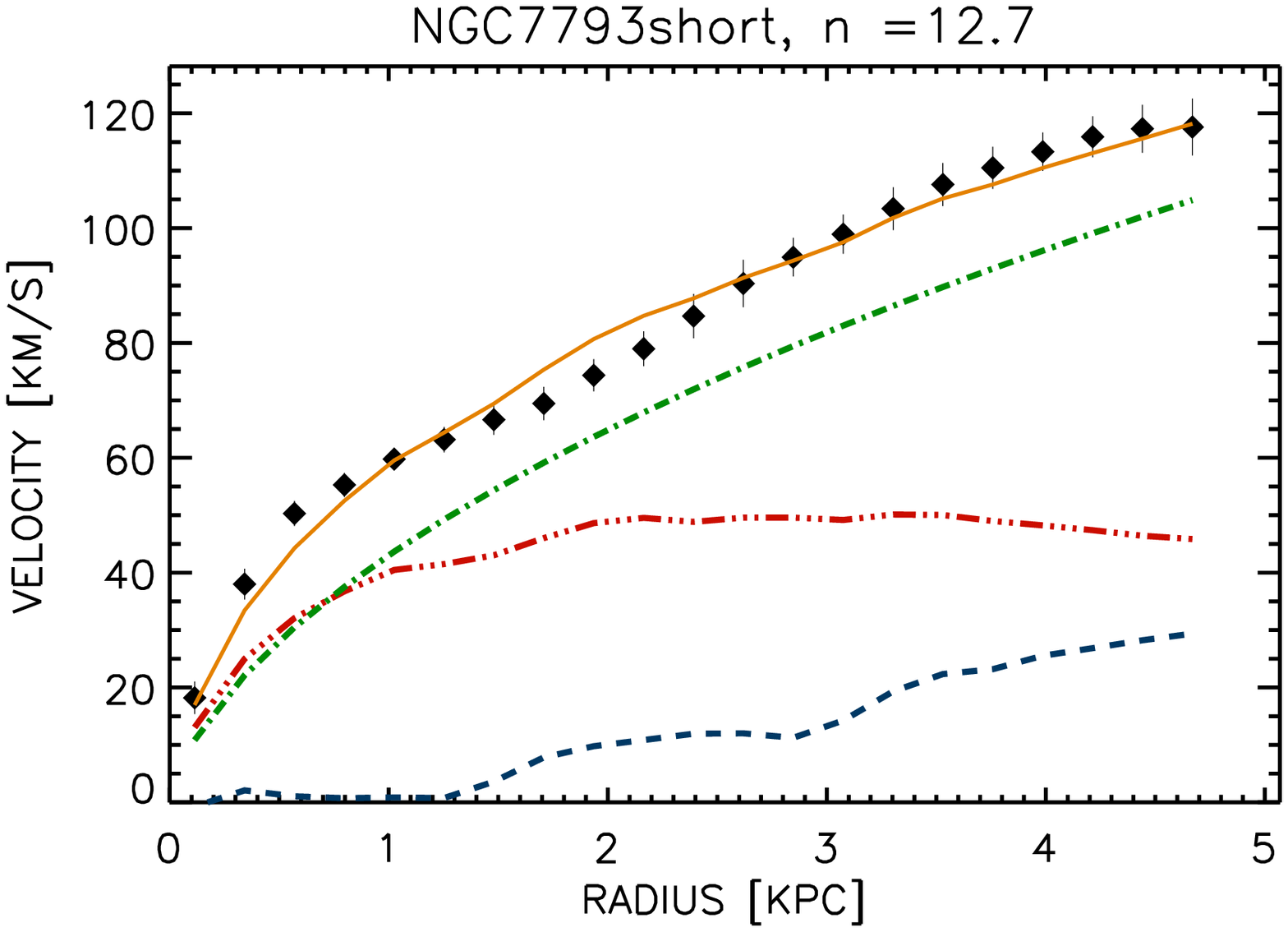}
 \caption{Same as Fig.~\ref{fig:rc-salpeter} but with a Kroupa initial mass
   function.}
 \label{fig:rc-kroupa}
 \end{center}
 \end{figure*}

\begin{figure}[ht]
 \begin{center}
\includegraphics[width=0.5\columnwidth]{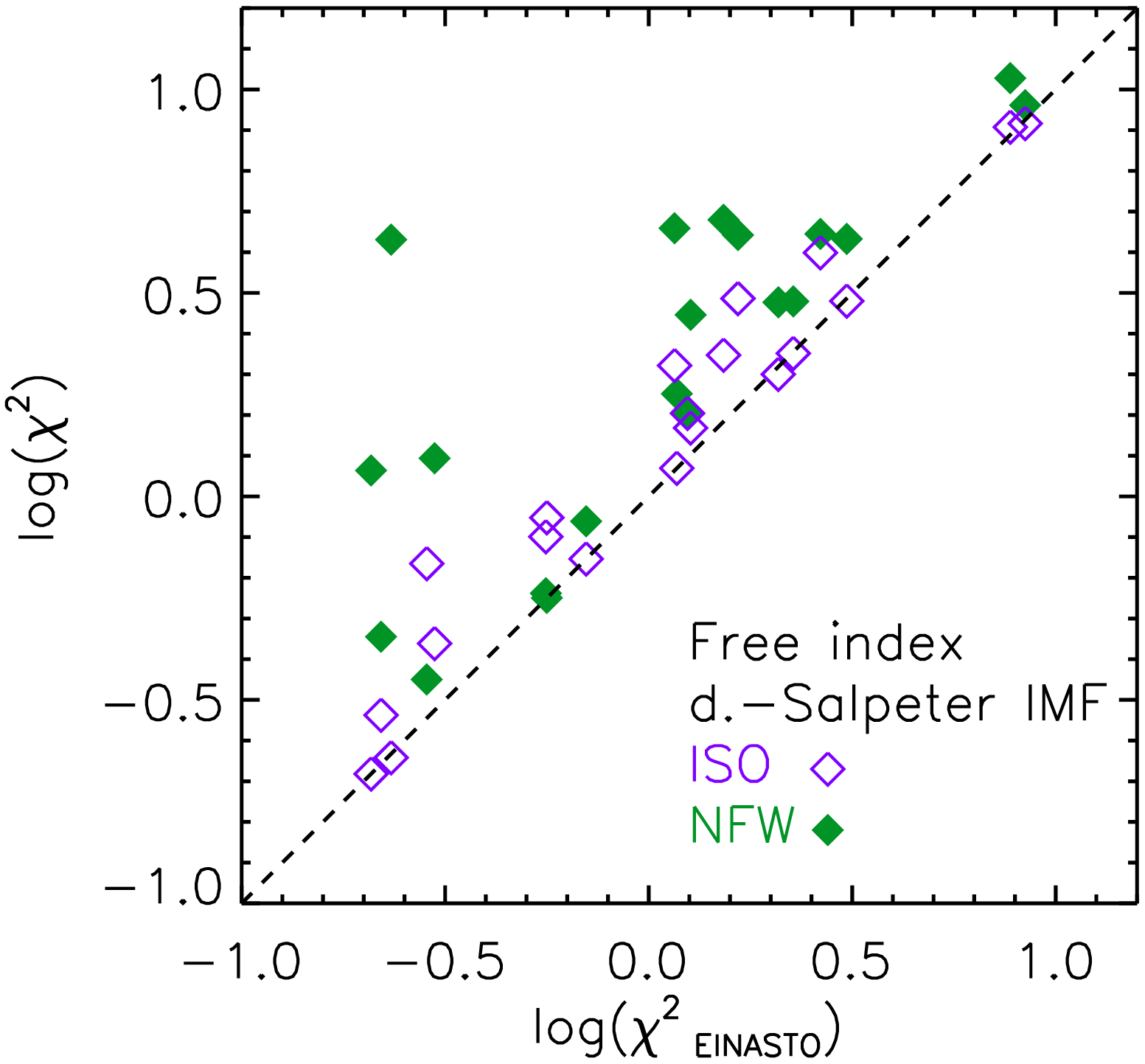}\includegraphics[width=0.5\columnwidth]{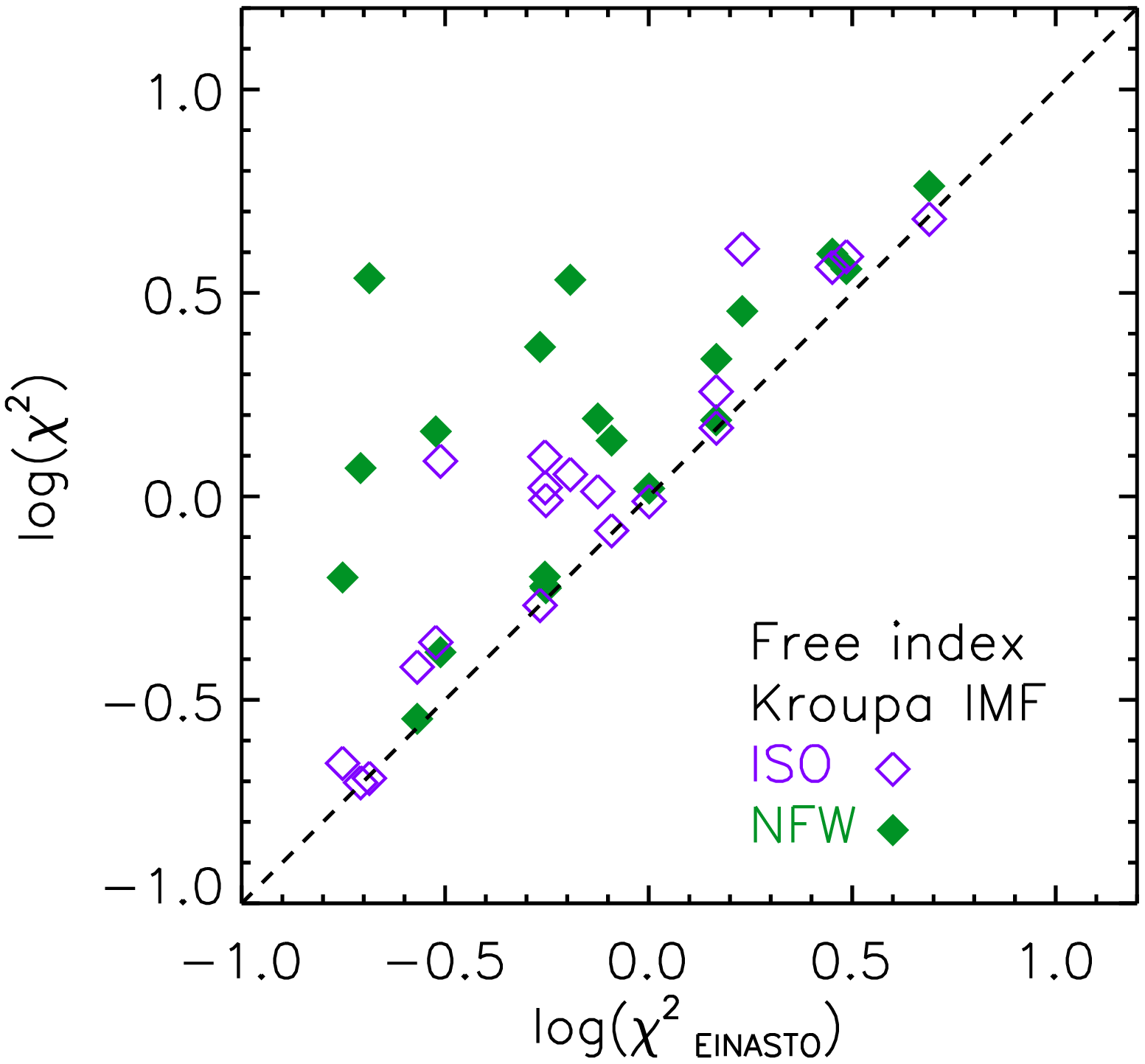}
 \caption{Comparison of reduced $\chi^2$ for the Einasto halo with the NFW
   and Iso halos (respectively filled and open symbols), using the 
   diet-Salpeter (\emph{left}) and Kroupa (\emph{right}) IMFs.}
 \label{fig:compredchi2}
 \end{center}
 \end{figure}

Figures~\ref{fig:rc-salpeter} and~\ref{fig:rc-kroupa}
 show the decomposition of the RCs  into    baryonic and dark matter
components  for the Einasto model.  The fit parameters and their associated $1\sigma$
error, as well as the reduced chi-squared values,  $\chi^2_r$, are  given  in
Tabs~\ref{tab:resfit1} and~\ref{tab:resfit2} for the Einasto, NFW and Iso halos.
Figure~\ref{fig:compredchi2} displays the values of Einasto $\chi^2_r$
and compares them with NFW and Iso $\chi^2_r$  values.

 %%%%%%%%%%%%%%% table EINASTO - NFW - Iso - FREE INDEX - diet salpeter IMF %%%%%%%%%%%%%%%%%%%%
\begin{deluxetable*}{l||clll|cll|cll}
%\rotate
\tablecaption{Fit parameters of the Einasto halo to  THINGS galaxies for a free index and fixed mass-to-light ratios derived using a diet-Salpeter IMF.}
\tablehead{
   &    &    &      EINASTO &   &     &  NFW &  &   & Iso & \\ 
 Galaxy  & 
 $\chi^2_r$ &  $ \rho_{-2}$ &      $r_{-2}$ & $n$ & 
 $\chi^2_r$  &  $v_{200}$   & $c$   &
 $\chi^2_r$  &   $\rho_0$ &     $r_c$  \\
 &    & \densunit& (kpc)&  &  & (\kms)&  &   &  \densunit &  (kpc) \\
  &  (1) & (2)& (3)& (4)& (5)& (6)& (7)&  (8)& (9)&  (10) }
\startdata
    NGC925 &  1.2 &  3.4 $\pm$ 171.6 &   10.0 $\pm$  1.1$\times 10^3$ &  $10^{-3}$ $\pm$ 25.0 &  4.6 & 172.6 $\pm$  17.6 &  1.3 $\pm$ 0.2 &  2.1 &    3.4 $\pm$   0.8 &   16.8 $\pm$  10.5 \\ 
   NGC2366 &  0.2 &  5.5 $\pm$   1.2 &    3.2 $\pm$    0.4 &  1.1 $\pm$  0.2 &  1.2 &  71.2 $\pm$  30.9 &  4.3 $\pm$ 2.3 &  0.2 &   37.6 $\pm$   4.3 &    1.3 $\pm$   0.1 \\ 
  NGC2403d &  0.6 &  1.3 $\pm$   0.3 &   17.1 $\pm$    2.4 &  4.6 $\pm$  0.5 &  0.6 & 109.7 $\pm$   1.0 & 10.0 $\pm$ 0.2 &  0.9 &   81.8 $\pm$   5.2 &    2.1 $\pm$   0.1 \\ 
   NGC2403 &  0.6 &  1.6 $\pm$   0.3 &   15.3 $\pm$    1.8 &  4.1 $\pm$  0.4 &  0.6 & 110.9 $\pm$   1.5 &  9.8 $\pm$ 0.3 &  0.8 &   78.4 $\pm$   4.7 &    2.1 $\pm$   0.1 \\ 
   NGC2841 &  0.2 &  1.6 $\pm$   0.5 &   28.5 $\pm$    5.0 & 10.4 $\pm$  1.6 &  0.5 & 182.8 $\pm$   1.7 & 16.4 $\pm$ 0.3 &  0.3 &  298.0 $\pm$  21.5 &    2.0 $\pm$   0.1 \\ 
   NGC2903 &  0.3 & 36.5 $\pm$   5.5 &    4.3 $\pm$    0.4 &  8.2 $\pm$  1.6 &  0.4 & 112.2 $\pm$   0.8 & 31.5 $\pm$ 0.8 &  0.7 &  $> 1000$ &    0.0 $\pm$   0.1 \\ 
   NGC2976 &  2.1 & 10.1 $\pm$ 3.3$\times 10^{6}$ &    8.0 $\pm$  6.6$\times 10^{7}$ &  0.1 $\pm$  2$\times 10^5$ &  3.0 &  62.1 $\pm$  12.7 &  1.8 $\pm$ 0.4 &  2.0 &   11.6 $\pm$   4.6 &   10.6 $\pm$  54.9 \\ 
   NGC3031 &  2.6 &  7.0 $\pm$   1.7 &    7.7 $\pm$    0.9 &  0.1 $\pm$  0.2 &  4.4 & 184.0 $\pm$ 161.0 &  3.2 $\pm$ 3.2 &  4.0 &   14.7 $\pm$   6.0 &    5.3 $\pm$   2.0 \\ 
  NGC3198d &  1.2 &  1.2 $\pm$   0.2 &   17.2 $\pm$    1.2 &  1.5 $\pm$  0.3 &  1.8 & 122.8 $\pm$   5.6 &  5.2 $\pm$ 0.5 &  1.2 &   15.1 $\pm$   2.2 &    4.9 $\pm$   0.4 \\ 
   NGC3198 &  2.3 &  1.2 $\pm$   0.3 &   17.0 $\pm$    1.5 &  1.4 $\pm$  0.4 &  3.0 & 122.1 $\pm$   7.1 &  5.2 $\pm$ 0.7 &  2.2 &   14.4 $\pm$   2.8 &    5.0 $\pm$   0.6 \\ 
    IC2574 &  0.2 &  0.7 $\pm$   0.4 &   15.8 $\pm$    6.9 &  1.0 $\pm$  0.4 &  4.3 &  69.3 $\pm$   5.8 &  3.4 $\pm$ 0.1 &  0.2 &    4.0 $\pm$   0.2 &    7.4 $\pm$   0.5 \\ 
   NGC3521 &  8.4 &  0.7 $\pm$   3.8 &   28.6 $\pm$  114.0 &  0.3 $\pm$  3.4 &  9.1 &  89.5 $\pm$ 155.9 &  2.1 $\pm$ 7.2 &  8.2 &    1.3 $\pm$   1.8 &   36.3 $\pm$ 107.4 \\ 
   NGC3621 &  0.7 &  0.6 $\pm$   0.3 &   27.2 $\pm$    6.7 &  2.6 $\pm$  0.6 &  0.9 & 168.8 $\pm$   7.3 &  3.7 $\pm$ 0.2 &  0.7 &   14.4 $\pm$   0.9 &    5.6 $\pm$   0.2 \\ 
   NGC4736 &  1.2 &  5.5 $\pm$   7.4 &    5.4 $\pm$   16.8 &  $10^{-3}$ $\pm$  0.7 &  1.6 &  39.5 $\pm$  16.5 &  7.9 $\pm$ 7.9 &  1.6 &   17.0 $\pm$  28.4 &    1.8 $\pm$   1.9 \\ 
    DDO154 &  0.3 &  1.4 $\pm$   0.3 &    6.0 $\pm$    0.8 &  2.2 $\pm$  0.3 &  1.2 &  63.8 $\pm$   2.6 &  4.1 $\pm$ 0.1 &  0.4 &   27.6 $\pm$   2.4 &    1.3 $\pm$   0.1 \\ 
   NGC5055 &  7.7 &  0.7 $\pm$   0.2 &   32.2 $\pm$    4.3 &  0.1 $\pm$  0.2 & 10.7 & 234.6 $\pm$  26.1 &  0.9 $\pm$ 0.1 &  8.1 &    0.9 $\pm$   0.3 &   44.7 $\pm$  30.2 \\ 
   NGC6946 &  1.3 &  4.6 $\pm$   2.9 &   14.9 $\pm$    8.3 &  2$\times 10^{-3}$ $\pm$  0.3 &  2.8 & 307.4 $\pm$  37.5 &  1.9 $\pm$ 0.3 &  1.5 &    5.3 $\pm$   0.5 &   21.4 $\pm$   5.8 \\ 
   NGC7331 &  3.1 &  1.0 $\pm$ 4.8$\times 10^5$ &  $>1000$ &  0.2 $\pm$  3$\times 10^5$ &  4.3 & 275.0 $\pm$ 131.7 &  1.1 $\pm$ 0.9 &  3.0 &    1.6 $\pm$   0.4 &  113.3 $\pm$ 519.4 \\ 
   NGC7793 &  1.7 & 22.7 $\pm$   1.9 &    3.7 $\pm$    0.1 &  0.3 $\pm$  0.1 &  4.4 & 141.4 $\pm$   8.1 &  6.6 $\pm$ 0.2 &  3.1 &   78.2 $\pm$  11.2 &    1.9 $\pm$   0.2 \\ 
  NGC7793s &  1.5 &  $< 10^{-3}$ &  $>1000$ &  9.3 $\pm$ 42.3 & 4.8 & 181.0 $\pm$   20.8 &  5.6 $\pm$ 0.4 &  2.2 &   51.8 $\pm$   6.8 &    3.5 $\pm$   0.7   
 \enddata
\tablecomments{Columns (1)-(5)-(8) Reduced $\chi^2$. Columns (2)-(9)  Characteristic density. Columns (3)-(10) 
Characteristic radius. Column (4) Einasto index. Columns (6)-(7) NFW
halo characteristic velocity and concentration.}
 \label{tab:resfit1}
 \end{deluxetable*}
 %%%%%%%%%%%%%%% end table EINASTO - NFW - Iso - FREE INDEX - diet salpeter IMF %%%%%%%%%%%%%%%%%%%%

 %%%%%%%%%%%%%%%   table EINASTO - NFW - Iso - FREE INDEX - Kroupa IMF %%%%%%%%%%%%%%%%%%%%
\begin{deluxetable*}{l||clll|cll|cll}
%\rotate
\tablecaption{Same as in Tab.~\ref{tab:resfit1} but with mass-to-light ratios derived using a Kroupa IMF.}
\tablehead{
   &    &    &      EINASTO &   &     &  NFW &  &   & Iso & \\ 
 Galaxy  &  $\chi^2_r$ &  $ \rho_{-2}$ &      $r_{-2}$ & $n$ &  $\chi^2_r$  &  $v_{200}$ & $c$ &  $\chi^2_r$  & $\rho_0$ & $r_c$  \\
 &    & \densunit& (kpc)&  &  & (\kms)&  &   &  \densunit &  (kpc) }
\startdata
    NGC925 &  0.6 &  4.5 $\pm$   0.5 &    8.9 $\pm$    0.6 &   0.1 $\pm$   0.1 &  3.4 & 171.4 $\pm$  15.2 &  2.2 $\pm$  0.1 &  1.1 &	6.0 $\pm$    0.8 &  9.6 $\pm$ 1.8 \\ 
   NGC2366 &  0.2 &  5.7 $\pm$   1.1 &    3.2 $\pm$    0.3 &   1.1 $\pm$   0.2 &  1.2 &  75.4 $\pm$  16.3 &  4.2 $\pm$  0.8 &  0.2 &   40.1 $\pm$    4.2 &  1.3 $\pm$ 0.1 \\ 
  NGC2403d &  0.6 &  1.5 $\pm$   0.4 &   15.6 $\pm$    2.0 &   5.6 $\pm$   0.5 &  0.6 & 101.2 $\pm$   1.0 & 12.6 $\pm$  0.3 &  1.1 &  154.0 $\pm$   10.8 &  1.5 $\pm$ 0.1 \\ 
   NGC2403 &  0.6 &  1.7 $\pm$   0.4 &   14.6 $\pm$    1.8 &   5.3 $\pm$   0.5 &  0.6 & 102.0 $\pm$   0.9 & 12.4 $\pm$  0.2 &  1.0 &  147.0 $\pm$    9.9 &  1.5 $\pm$ 0.1 \\ 
   NGC2841 &  0.2 & 10.7 $\pm$   1.2 &   11.2 $\pm$    0.6 &  12.4 $\pm$   1.5 &  0.6 & 172.0 $\pm$   1.4 & 25.2 $\pm$  0.6 &  0.2 &  $> 1000$ &  0.6 $\pm$ 0.1 \\ 
   NGC2903 &  0.3 & 72.4 $\pm$  19.4 &    3.2 $\pm$    0.4 &   9.6 $\pm$   2.1 &  0.4 & 111.2 $\pm$   0.8 & 36.2 $\pm$  1.0 &  1.2 &  $> 1000$ &  0.0 $\pm$ 0.1 \\ 
   NGC2976 &  0.5 &  0.0 $\pm$   0.6 &             $>1000$ &   4.9 $\pm$  77.8 &  2.3 & 108.6 $\pm$   5.9 &  4.5 $\pm$  0.2 &  0.5 &   36.6 $\pm$    4.7 &  4.5 $\pm$ 2.8 \\ 
   NGC3031 &  3.1 & 30.0 $\pm$   3.0 &    4.7 $\pm$    0.2 &   0.7 $\pm$   0.2 &  3.6 &  94.3 $\pm$   5.5 & 26.8 $\pm$  3.6 &  3.9 &  740.0 $\pm$  434.0 &  0.8 $\pm$ 0.3 \\ 
  NGC3198d &  0.8 &  2.2 $\pm$   0.2 &   13.1 $\pm$    0.6 &   2.1 $\pm$   0.3 &  1.4 & 109.3 $\pm$   2.5 &  8.9 $\pm$  0.6 &  0.8 &   47.1 $\pm$    5.9 &  2.7 $\pm$ 0.2 \\ 
   NGC3198 &  1.5 &  2.2 $\pm$   0.3 &   13.0 $\pm$    0.7 &   2.0 $\pm$   0.3 &  2.2 & 109.9 $\pm$   3.2 &  8.7 $\pm$  0.7 &  1.5 &   44.3 $\pm$    7.4 &  2.8 $\pm$ 0.3 \\ 
    IC2574 &  0.2 &  0.7 $\pm$   0.3 &   15.1 $\pm$    5.0 &   1.1 $\pm$   0.3 &  3.4 &  94.6 $\pm$   4.3 &  2.2 $\pm$  0.2 &  0.2 &	4.9 $\pm$    0.2 &  6.3 $\pm$ 0.3 \\ 
   NGC3521 &  4.9 &  6.1 $\pm$   2.5 &    9.4 $\pm$    1.6 &   1.2 $\pm$   0.9 &  5.8 & 121.7 $\pm$  22.1 &  9.7 $\pm$  3.1 &  4.8 &   75.6 $\pm$   45.8 &  2.5 $\pm$ 0.9 \\ 
   NGC3621 &  0.6 &  0.3 $\pm$   0.2 &   39.9 $\pm$   12.3 &   6.4 $\pm$   1.0 &  0.6 & 119.8 $\pm$   1.9 &  7.9 $\pm$  0.3 &  1.3 &   49.3 $\pm$    4.2 &  2.8 $\pm$ 0.1 \\ 
   NGC4736 &  1.5 & 50.8 $\pm$  15.5 &    1.8 $\pm$    0.3 &   1.9 $\pm$   0.8 &  1.5 &  44.1 $\pm$   2.3 & 43.8 $\pm$  9.0 &  1.8 &  $> 1000$ &  0.1 $\pm$ 0.1 \\ 
    DDO154 &  0.3 &  1.5 $\pm$   0.3 &    5.9 $\pm$    0.8 &   2.2 $\pm$   0.3 &  1.4 &  91.3 $\pm$   4.0 &  2.7 $\pm$  0.1 &  0.4 &   28.5 $\pm$    2.5 &  1.3 $\pm$ 0.1 \\ 
   NGC5055 &  0.8 &  1.3 $\pm$   0.1 &   22.6 $\pm$    0.6 &   0.5 $\pm$   0.1 &  1.6 & 234.4 $\pm$  22.5 &  1.9 $\pm$  0.3 &  1.0 &	4.8 $\pm$    0.5 & 11.7 $\pm$ 0.9 \\ 
   NGC6946 &  1.0 &  1.8 $\pm$   1.2 &   18.7 $\pm$    7.4 &   2.6 $\pm$   0.9 &  1.0 & 187.9 $\pm$  17.0 &  6.1 $\pm$  0.7 &  1.0 &   44.2 $\pm$    4.1 &  3.7 $\pm$ 0.2 \\ 
   NGC7331 &  0.3 &  0.0 $\pm$   0.0 & $>1000$ &  13.1 $\pm$  19.7 &  0.3 & 195.6 $\pm$  15.3 &  5.2 $\pm$  0.6 &  0.4 &   26.1 $\pm$    3.6 &  5.2 $\pm$ 0.5 \\ 
   NGC7793 &  2.8 & 19.8 $\pm$   3.7 &    3.6 $\pm$    0.3 &   0.9 $\pm$   0.2 &  3.9 & 114.4 $\pm$  10.2 &  9.2 $\pm$  0.8 &  3.7 &  128.0 $\pm$   17.6 &  1.5 $\pm$ 0.1 \\ 
  NGC7793s &  1.7 &   $< 10^{-3}$    &             $>1000$ &  12.7 $\pm$  32.9 &  2.9 & 191.9 $\pm$   9.7 &  6.1 $\pm$  0.4 &  4.1 &   96.7 $\pm$   15.1 &  2.0 $\pm$ 0.3 
 \enddata
 \label{tab:resfit2}
 \end{deluxetable*}
 %%%%%%%%%%%%%%% end table EINASTO - NFW - Iso - FREE INDEX - Kroupa IMF %%%%%%%%%%%%%%%%%%%%

% GRAPHS OF RC fits  

%%%%%%%%%%%%%%%%%%%%% diet salpeter imf - free index %%%%%%%%%%%%%%%%%%%%%%%
%%%%%%%%%%%%%%%%%%%%%%%%%%%%%%%%%%%%%%%%%%%%%%%%%%%%%%%%%%%

%%%%%%%%%%%%%%%%%%%%% kroupa imf - free index %%%%%%%%%%%%%%%%%%%%%%%
%%%%%%%%%%%%%%%%%%%%%%%%%%%%%%%%%%%%%%%%%%%%%%%%%%%%%%%%%%%

  The most important results from the analysis of the fits are: 
 \begin{itemize}
 \item The quality of the fits is very good for the majority of the
   galaxies.  $80\%$ (16  out of  20) of the fits done with a Kroupa
   IMF  have  $\chi^2_r < 1.5$.  This fraction falls to $60\%$ (12
    out of  20) when using a diet-Salpeter IMF.  The Einasto model is thus
   highly constrained (low $\chi^2$) owing to the large numbers of
   degrees of freedom of each fit.
 \item Most of the fits ($70\%$)   with a Kroupa IMF are better
   (i.e.,  have lower $\chi^2_r$) than those   with a
   diet-Salpeter IMF.  That trend is in good agreement with the
   results obtained for the NFW and Iso halos
   (see D08 and our current study in Tabs.~\ref{tab:resfit1}
   and~\ref{tab:resfit2}).
    Both our analysis and that of D08 show  how some mass models done under the prescriptions of \citet{bel01} with 
   the assumption of a diet-Salpeter IMF can be  unphysical due to 
      modelled velocities exceeding   observed ones in the inner disk regions (for, e.g.,  
      NGC 3521, NGC 5055, NGC 6946 or NGC 7331). It explains why the Einasto 
      halo properties deduced from the diet-Salpter IMF often present  scattered distributions 
      and/or  large errors, as will be shown in following sections.
    It is thus tempting to reject that IMF hypothesis. It is the reason
   why we focus more on the results obtained with the Kroupa IMF hypothesis hereafter, except where mentioned. 
  \item The Einasto models give better results than the Iso and NFW
    models, irrespective of the IMF: $60\%$ ($80\%$) of the Einasto
    halos  fit  better than Iso halos for the diet-Salpeter IMF
    (Kroupa IMF, respectively);  all of them fit better than the
    NFW cusp  model, irrespective of the IMF. This result is  expected 
    owing to the effect of the third modeling parameter (the Einasto index)  
    in addition to the usual characteristic scale halo density and size. 
    We analyse the statistical significance of that modeling improvement 
    in \S\ref{sec:comparisoneinastoisonfw}. Furthermore it is  likely 
    that the  exponential-like  decrease of the dark matter volumic density 
    described by Eq.~\ref{eq:rhoeinasto}  contributes to the fit improvement as well, as 
    will be shown in \S\ref{sec:twoparam}. 
   \end{itemize}

  %%%%%%%%%%%%%%%%%%%%% comparison of reduced chi2 - Einasto-NFW-Iso%%%%%%%%%%%%%%%%%%%%%%%
%%%%%%%%%%%%%%%%%%%%%%%%%%%%%%%%%%%%%%%%%%%%%%%%%%%%%%%%%%%

\section{The Einasto model as the preferred model}
\label{sec:comparisoneinastoisonfw}

In this section, we investigate the significance of the improvement of
the RC fits with the Einasto halo over those with the 
Iso and NFW models.

We first analyze the number of fits that  are  statistically
significant with the Einasto model and that were not significant with
any of the two-parameter Iso and NFW models.  By ``statistically
significant'' we mean that a model and the observations follow the
same distribution, i.e., that there is less than 
5\% probability of obtaining as good a fit by chance according to chi-squared statistics.
For
the Kroupa IMF, the fraction of the fits that  were  not significant
with the Iso model,  but are  with the Einasto
halo  consists of  30\% of the sample (6  out of  20 fits).  This
fraction becomes 25\%  when going  from NFW  to  Einasto. For the
diet-Salpeter IMF, only two galaxies (10\%) have  gone  from a not
significant result with the Iso model to significant with the Einasto
halo, while it is $20\%$ from NFW to Einasto (4 galaxies).

 %%%%%%%%%%%%%%%%%%%%% Einasto Parameter Space diet-Salpeter IMF%%%%%%%%%%%%%%%%%%%%%%%
 \begin{figure*}[ht]
 \begin{center}
\includegraphics[width=0.3\textwidth]{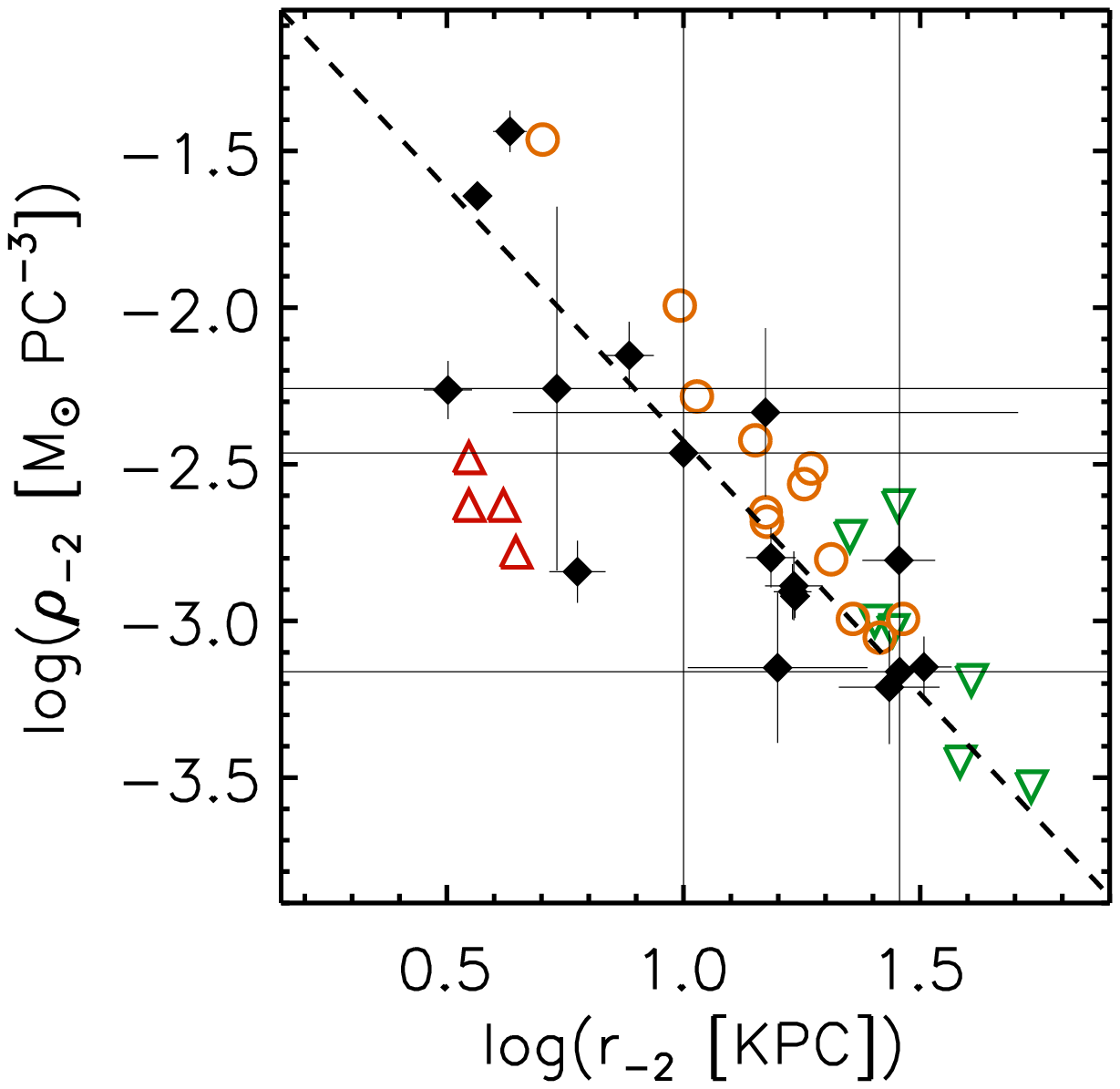}\includegraphics[width=0.3\textwidth]{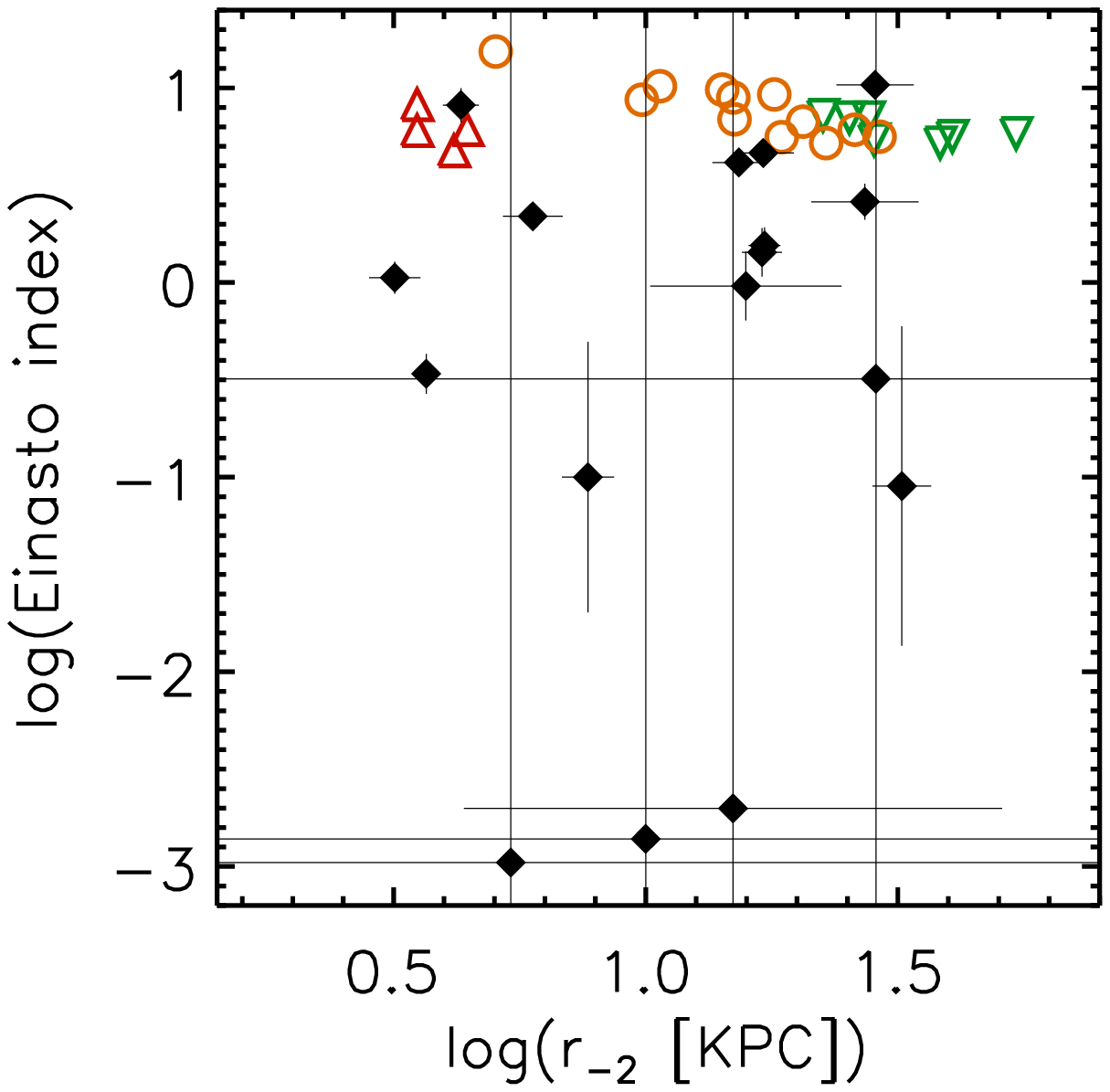}\includegraphics[width=0.3\textwidth]{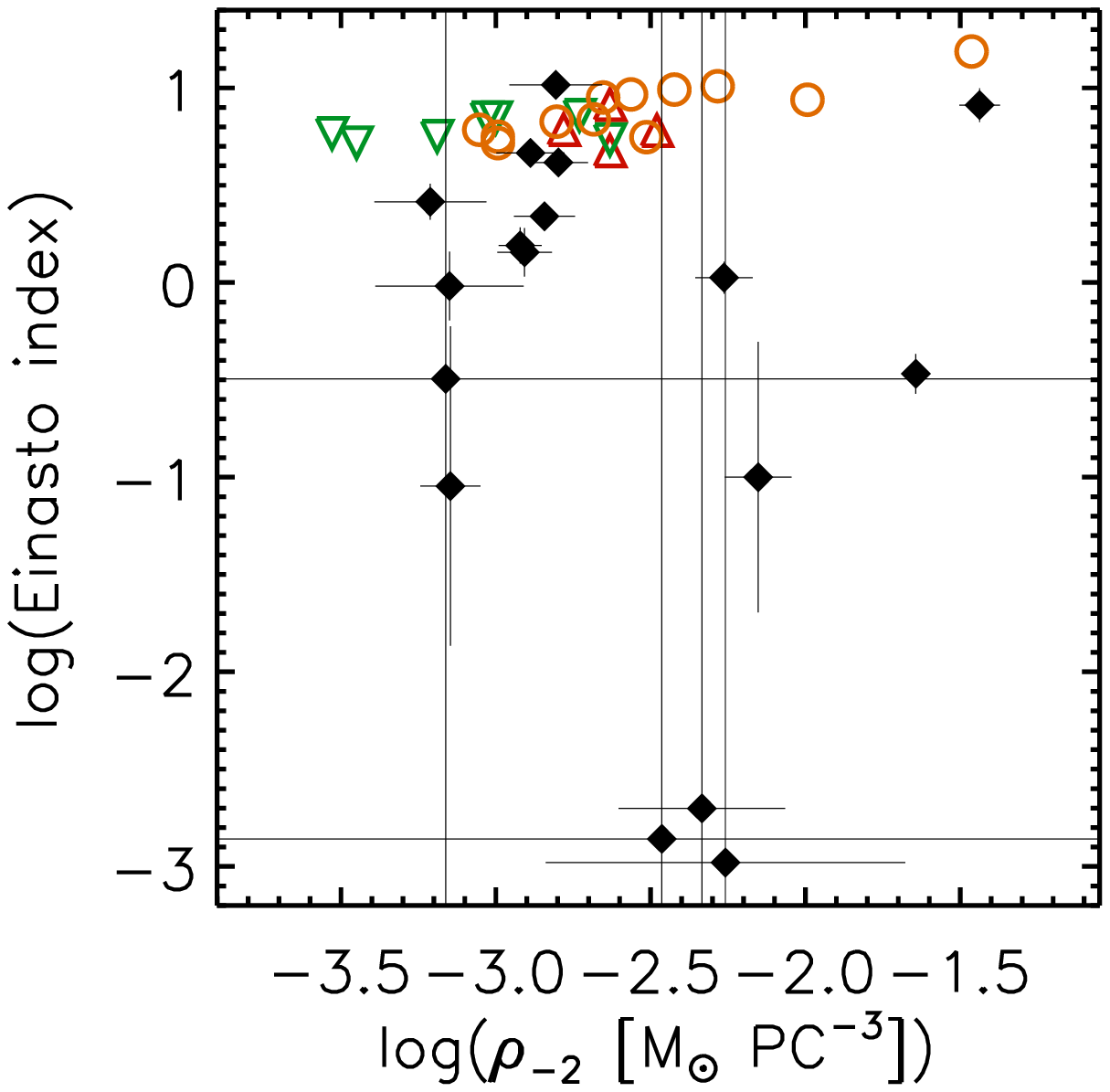}
 \caption{Parameter  space [$\rho_{-2},r_{-2},n$] of
   Einasto halos for the THINGS sample (filled diamonds).  Fixed
   stellar mass-to-light ratios are derived from stellar population
   synthesis using a diet-Salpeter initial mass function.  Colored
   symbols show the parameters of dwarf- (upward red triangles) and
   galaxy- (downward green triangles)  halos modeled in $\Lambda$CDM simulations
   \citep{nav04}. Galaxy-sized halos from \citet{tis10} are displayed
   with orange circles.  A dashed line is a power law fit to the relationship
   linking $\rho_{-2}$ and $r_{-2}$: $\log(\rho_{-2}) = (-1.61 \pm
   0.07) \log(r_{-2})-(0.81 \pm 0.06)$.
  The degenerate fits with $\log r_{-2} > 3$ and $\log \rho_{-2} > -5$ are not drawn for clarity.}
 \label{fig:haloparam1}
 \end{center}
 \end{figure*}
%%%%%%%%%%%%%%%%%%%%%%%%%%%%%%%%%%%%%%%%%%%%%%%%%%%%%%%%%%%

Though these numbers do not represent the majority of the sample, they
are not negligible.  As a comparison, $15\%$ ($20\%$) of the sample
acquires significant fits when passing from  NFW to Iso
for the Kroupa (respectively diet-Salpeter) IMF.  Therefore, the
improvement of fitting the \hi\ RCs by the Einasto model
instead of the Iso or NFW models is  equal to or  better than the
improvement implied when switching from NFW to Iso.  Conversely, the
Einasto  model does not diminish the significance
of the fits :  no fit has turned from being significant with Iso/NFW
to not significant with the Einasto halo, regardless of the IMF.

However, it is not surprising that the three-parameter Einasto model is
better able to reproduce the RCs than the two-parameter Iso and NFW models.
Because neither Iso nor NFW models are nested\footnote{A model with $k$
  parameters is considered nested in a more 
complex model with  e.g.,  $k+1$ parameters when fixing one of the
$k+1$ parameters allows  one  to reproduce the simplest $k$
parameter model.} into the Einasto family of
models, we cannot simply perform an $F$-test to check for the significance of
the improvement of the fits with the three-parameter Einasto model.
Instead,
we   verify the significance of this improvement by performing
Akaike information criterion tests
\citep[hereafter, AIC,][]{aka74}, which are designed to compare non-nested models of different
numbers of parameters.   The goal of an AIC test is not to statistically reject one model
with respect to another,  but rather to find the model
that is   more likely to be correct.  The criterion is expressed
 as  AIC $= \chi^2 +2N$, where $N$ is the number of parameters of
the model.   An  Einasto model is considered more likely to be
correct than the Iso or NFW model when $\rm AIC_{Einasto} <
AIC_{Iso/NFW}$.  We find  that 65\% (60\%) of the fits with the
Einasto model are more likely to be correct than  those  with the
ISO model assuming a Kroupa (diet-Salpeter) IMF.
These percentages increase to 95\% when comparing with
the NFW model,  independent of  the assumed IMF.

In summary, the Einasto model provides more fits that are
statistically significant and is more likely to be correct than the
two other models.  For these reasons, we conclude that fitting RCs 
of the current THINGS sample with an Einasto halo is a
significant  improvement  with respect to the NFW and Iso models.

\section{Dynamical properties of Einasto halos}   
\label{sec:halofamily}

\subsection{Non-universality of the halo}
\label{sec:nonuniversality}

\begin{figure*}[ht]
 \begin{center}
\includegraphics[width=0.3\textwidth]{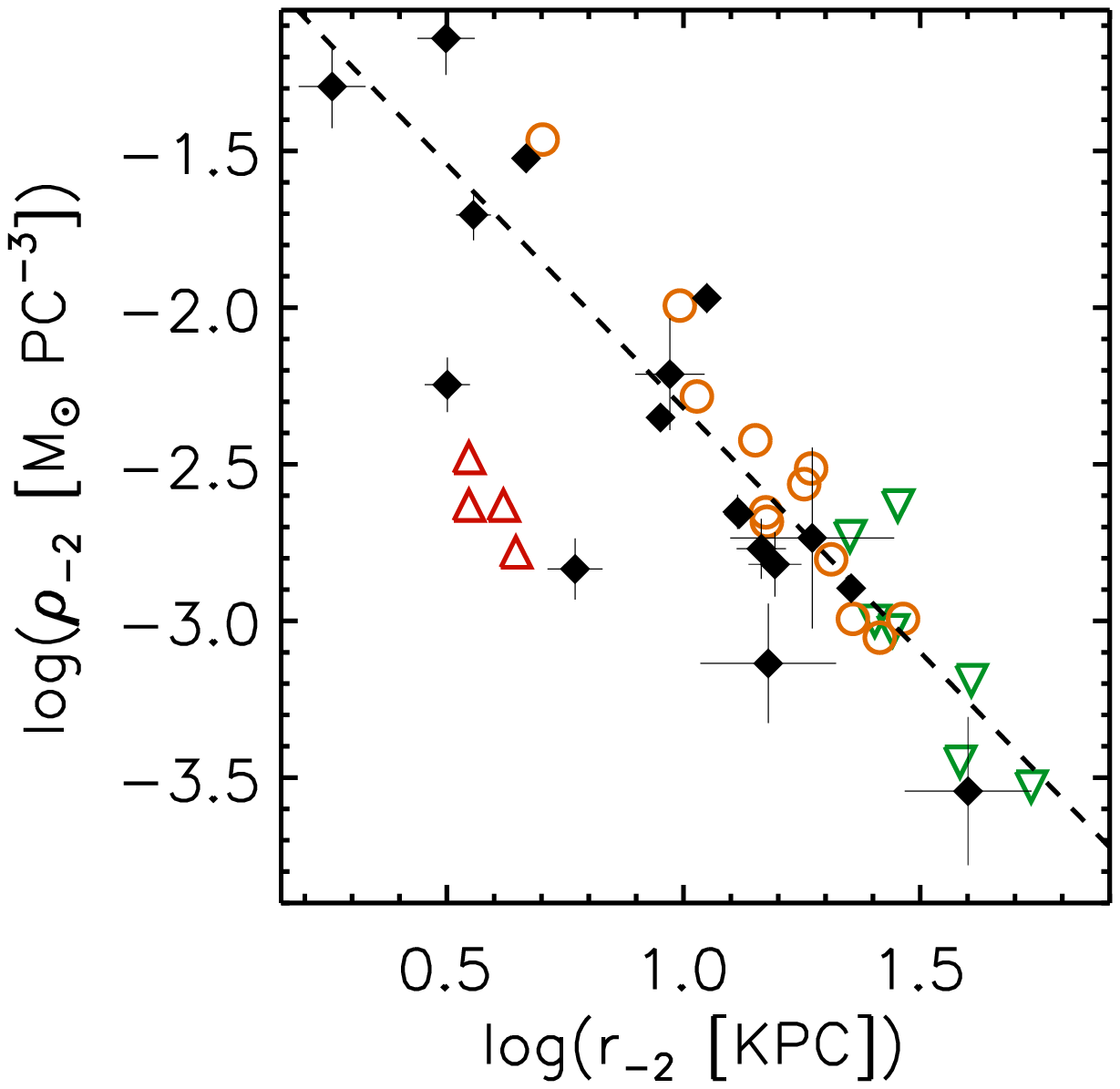}\includegraphics[width=0.3\textwidth]{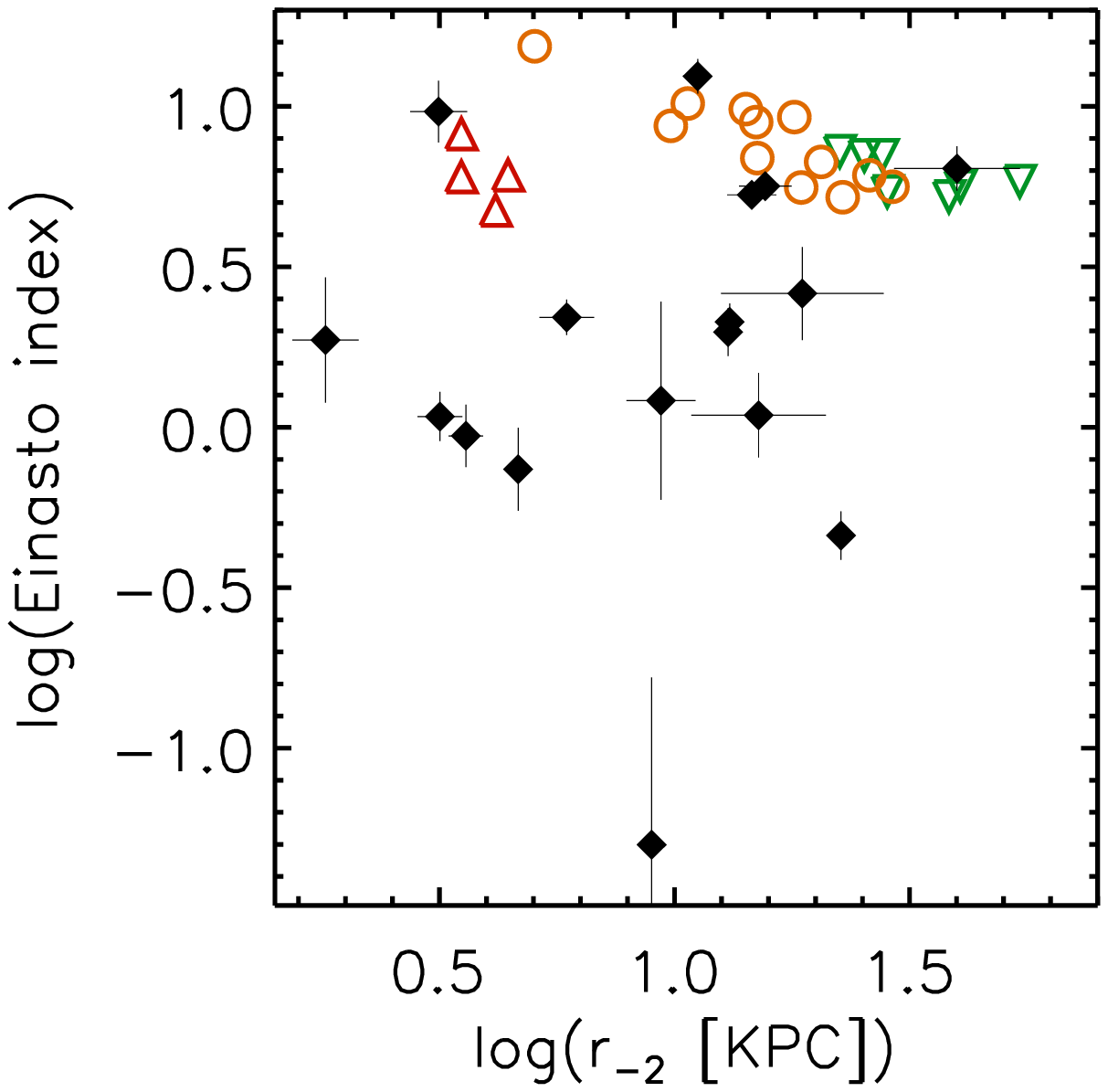}\includegraphics[width=0.3\textwidth]{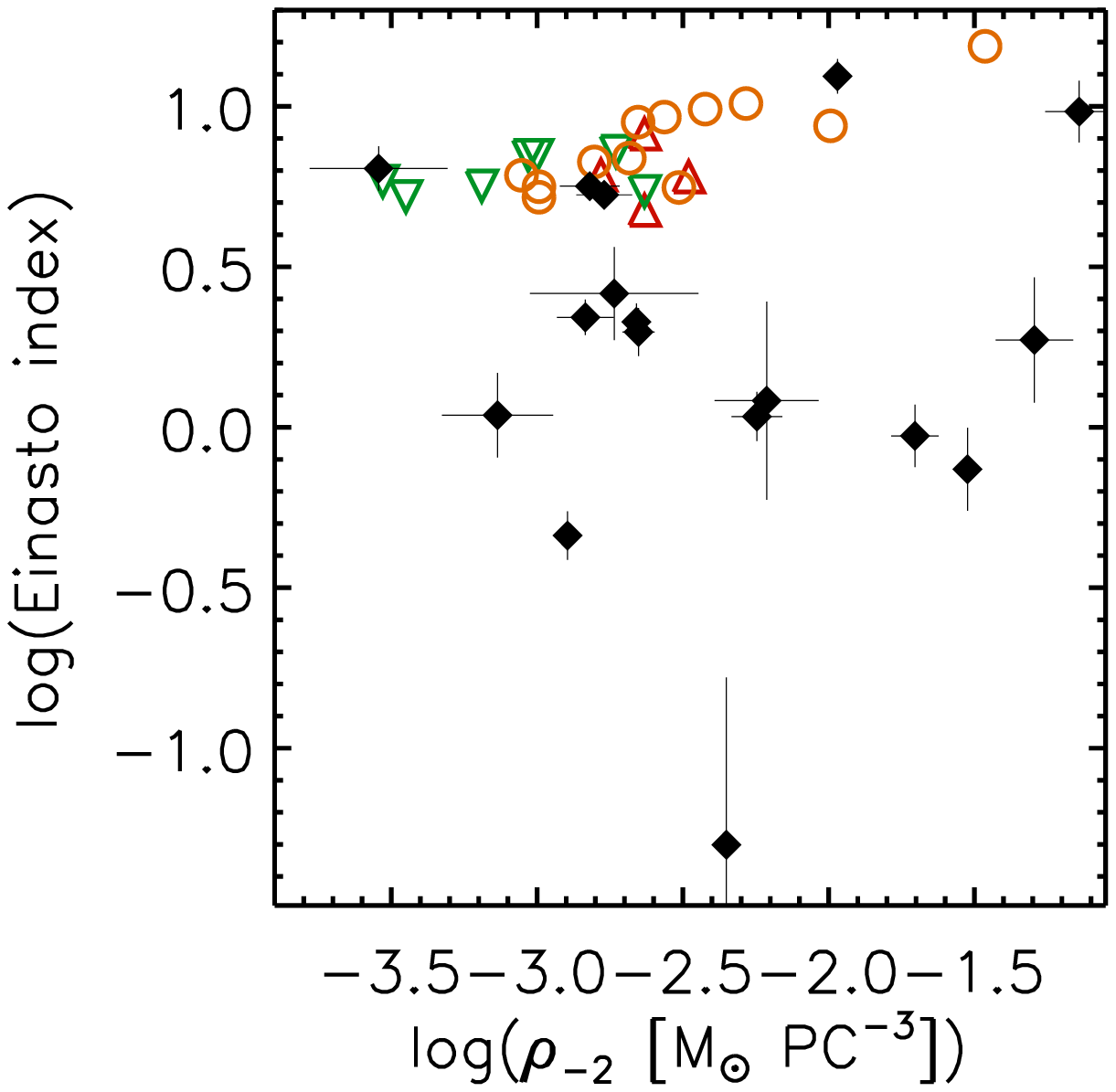}
 \caption{Same as Figure~\ref{fig:haloparam1} but for a Kroupa initial mass function.  
  The  dashed line is a power law fit   $\log(\rho_{-2}) = (-1.56 \pm 0.06)\log(r_{-2})-(0.76 \pm 0.06)$.}
 \label{fig:haloparam2}
 \end{center}
 \end{figure*}

\begin{figure}[ht]
 \centering
\includegraphics[width=0.7\columnwidth]{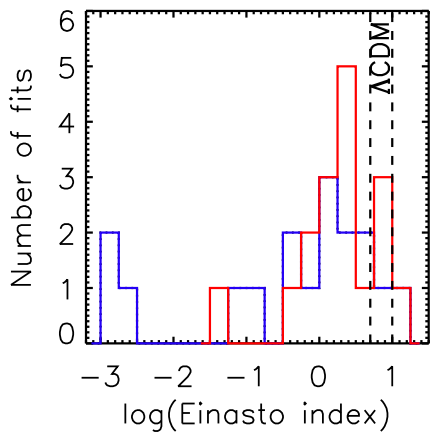}
 \caption{Distribution of Einasto indices. Results for the diet-Salpeter and Kroupa 
 IMFs are shown by blue and red histograms, respectively. 
 The range of Einasto indices from
 $\Lambda$CDM simulations with and without the physics of baryons ($5
 \lesssim n \lesssim 10$) is displayed by dashed lines.}
 \label{fig:histoindex}
 \end{figure}

Figures~\ref{fig:haloparam1} and~\ref{fig:haloparam2}  show  the
fit parameters   for the diet-Salpeter and Kroupa IMFs,
respectively.   Figure~\ref{fig:histoindex}  shows  the
distribution of Einasto indices.   Based on these distributions,  we
can distinguish several families of halos.     A
first family of halos has well constrained parameters with large indices ($n > 4$). 
 We refer to this family as cuspy Einasto halos hereafter.   
Indeed  it corresponds to halos whose indices are comparable with  the  index of a typical
galaxy-sized halo from numerical simulations \citep[$5 \leq n_{\rm simulated} \leq 7$,][]{nav04,mer06,gra06,nav10}. 
Their innermost density slopes are steeper than $-0.8$, hence consistent 
with cosmological slopes \citep[$\alpha = -0.9\pm 0.1$,][]{nav10}.  
  A second family of galaxies  has  well-constrained parameters with low  index
values  ($0.1 < n \leq 4$). It contains the
majority of the galaxies.   A third halo family has very low indices ($n \leq 0.1$). 
Most of those fits occur with the diet-Salpeter IMF. 
Finally a fourth family of
halos has unrealistic, extremely large scale radii, with large associated 
uncertainties (the rising part of the RC of NGC~7793, NGC~7331, NGC~2976).  
Those fits are extremely degenerate, which is very likely caused by an almost complete dominance of the 
baryonic material over the dark component, at least for NGC~2976 and NGC~7331. 
   
Figures~\ref{fig:haloparam1} and~\ref{fig:haloparam2} 
show a clear relation linking the characteristic density to the
radius. Small halos are denser than large ones.  A power-law fit to
the observed relationship  gives  $\rho_{-2} \propto r_{-2}^{-1.6
  \pm 0.1}$ for both IMFs.  This implies that the
surface density of dark matter $\Sigma_{-2} = \rho_{-2} r_{-2}$
derived at $r_{-2}$ scales as $\Sigma_{-2} \propto r_{-2}^{-0.6}
\propto \rho_{-2}^{3/8}$. Those relations have been determined without 
taking account the fourth halo family fits and by averaging the fit parameters 
of galaxies that have duplicated RCs but a different stellar decomposition  (NGC~2403 and NGC~3198).
 It is worthwhile to note that the correlation we find between the scale density and radius of the Einasto model 
differs from the relations found between the scale density and radius of two-parameter NFW and Iso
models.  For instance  \citet{kor04} and \citet{spa08} have  derived  a
relation $\rho_0 \propto 1/r_0$
\citep[but see][]{bar04},
where the scale density $\rho_0$ and and $r_0$ are the characteristic
density and radius of two-parameter NFW and Iso models.  As a
consequence the dark matter surface density $\Sigma_0 = \rho_0\, r_0$
remains roughly constant at the core radius of the halo \citep{kor04,
  spa08, don09}.  This  trend is not observed with the Einasto halo.
  In this context, it is interesting to note that   we have also
fit   the two-parameter Iso model to the current sample (Iso
column in Tabs.~\ref{tab:resfit1} and~\ref{tab:resfit2}).  The
results  do not imply a similar  relation  as in \citet{kor04},
\citet{spa08} or \citet{don09}, but agree more with \citet{bar04}.  We
indeed find inner surface densities for the core halo  that
depend  on the scale parameters.  A  linear fit  yields 
$\rho_{0} \propto r_{0}^{-1.4 \pm 0.03}$ for the diet-Salpeter IMF and
$\rho_{0} \propto r_{0}^{-1.5 \pm 0.04}$ for the Kroupa IMF. 
It may seem   odd that these results disagree  with \citet{don09}
because part of their sample  consists  of the current THINGS
galaxies (though using the D08 RCs).   A possible
explanation of the difference could be the choice of stellar
mass-to-light ratios   (\citeauthor{don09} use best fit mass-to-light
ratios),  the use of other kinematical tracers (mostly the ionized gas) as well as
 the choice of  the outer slope of the density profile of dark matter
($\rho \propto r^{-3}$ in \citeauthor{spa08} and \citeauthor{don09}, while our
Iso model implies $\rho \propto r^{-2}$).  A
 detailed  investigation of both analysis methods is   beyond the
scope of this article.   Clearly though, the implied (non-)constancy
of the dark matter surface density warrants further investigation.

 %%%%%%%%%%%%%%%%%%%%% Einasto Parameter Space diet-Salpeter IMF%%%%%%%%%%%%%%%%%%%%%%%
%%%%%%%%%%%%%%%%%%%%%%%%%%%%%%%%%%%%%%%%%%%%%%%%%%%%%%%%%%%

As for the Einasto index, we  note  the more scattered
distribution and the presence of more halos having a small index for
the diet-Salpeter IMF than for the Kroupa IMF. This is caused by a larger contribution of
stellar baryons to the RCs.  The
effect of a (very) low index  (illustrated in Fig.~\ref{fig:rceinasto}
as declining velocities in the outer parts of the dark matter halo) 
 are clearly visible  for NGC~4736,   NGC~6946 (Fig.~\ref{fig:rc-salpeter},
diet-Salpeter IMF only)  and  NGC~925 (Figs.~\ref{fig:rc-salpeter}
and~\ref{fig:rc-kroupa}).
         
No obvious correlation is found between the Einasto index  and 
scale density or radius.  Using halos with $n \leq 4$ (second and third 
  halo families) we derive an average index of $\bar{n} = 0.8 \pm
0.3$ with the diet-Salpeter IMF and $\bar{n} = 1.3 \pm 0.2$ with the
Kroupa IMF (using averaged indices for NGC~2403/NGC~2403d and NGC~3198/NGC~3198d). 
 Note that the   amplitude and  scatter become more important if the first halo
family is  also  taken into account, with $\bar{n} = 2.1 \pm 0.8$
(diet-Salpeter) and $\bar{n} = 3.2 \pm 0.9$ (Kroupa). 

   The cuspiest Einasto halos are typically found in
galaxies where the dark matter dominates the visible matter at almost
all radii (as seen in Figs.~\ref{fig:rc-salpeter} and~\ref{fig:rc-kroupa}),  coupled with a relatively extended rotation
curve and mass distribution ($r > 15$ kpc)  with  rotation
velocities larger than 120 \kms\ at large galactocentric radius. 
Note also  they correspond to galaxies whose RCs start at $r \geq 3$ kpc. 
 The missing information at $r < 3$ kpc might affect our conclusions for those galaxies.

Perhaps the most important result from the analysis of the indices 
is that no  single  value or well-defined  relation  can
reproduce all types of halos. The index is also a complex function of the 
 halo  mass (see \S\ref{sec:haloconcentration}). 
 We can thus conclude  that no obvious ``universal'' Einasto index, hence Einasto halo, 
 can be deduced from the  current galaxy sample.

 %%%%%%%%%%%%%%%%%%%%% histogram of einasto indices %%%%%%%%%%%%%%%%%%%%%%%
%%%%%%%%%%%%%%%%%%%%%%%%%%%%%%%%%%%%%%%%%%%%%%%%%%%%%%%%%%%

\subsection{Intermediate class halo density and density slope profiles} 
\label{sec:densityprofile}
\begin{figure}[h!]
 \begin{center}
\includegraphics[width=\columnwidth]{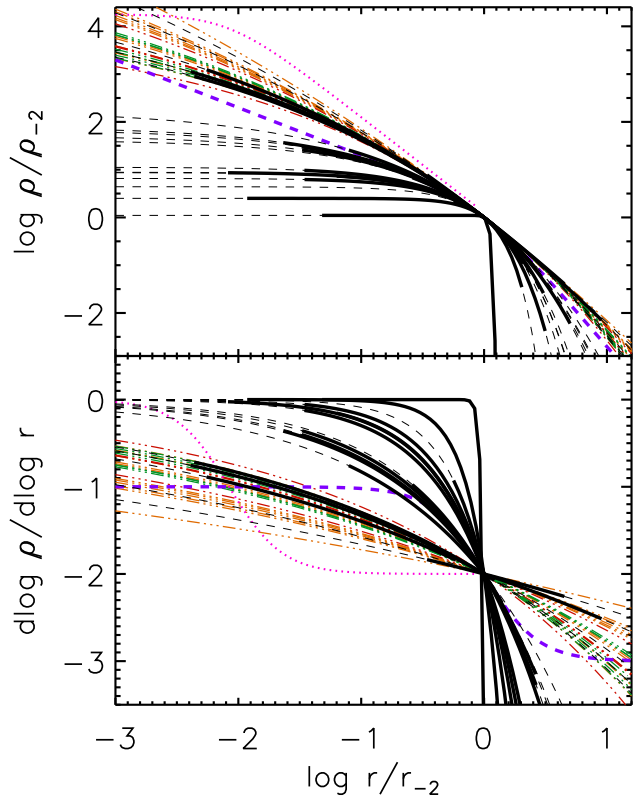}
 \caption{Dark matter density profiles (\emph{top}) and
   density slope profiles (\emph{bottom}) for the fit Einasto
   halos (\emph{black, solid lines}). Only results for the Kroupa IMF
   are shown for clarity. For each halo the radial range 
   of the observations is highlighted by a thick line.
   Colored dashed-dotted lines represent Einasto halos from $\Lambda$CDM
     simulations whose parameters are shown in Fig.~\ref{fig:haloparam1}.
      A purple dashed line represents a generic galaxy-sized NFW halo, a magenta dotted line a generic galaxy-sized Iso halo.} 
 \label{fig:compdens}
 \end{center}
 \end{figure}

Figure~\ref{fig:compdens} displays the radial profiles of the   mass
volume  density and the radial profiles of the logarithmic density
slope of the THINGS sample.  For clarity, only results for the
Kroupa IMF are represented, excluding the fourth halo family. 
 Our conclusions are similar for the diet-Salpeter IMF.   Also shown are the
radial profiles of  generic galaxy-sized Iso and NFW  halos 
with comparable densities,  giving a halo RC with a velocity amplitude 
of $\sim 200$ \kms\ at a typical radius of 25 kpc.  Other dashed colored curves
correspond to Einasto halos  with parameters derived  from
cosmological simulations (see \S\ref{sec:compsimu} for details).
 
 Figure~\ref{fig:compdens} (top panel) shows that the density profiles  typically 
decrease very smoothly in the inner regions.    Further out,  the profiles become very
steep and the densities very  low   past  the scale radius
$r_{-2}$.  It  can also be  seen that most of the halos derived from the
observations  are less dense in their inner parts than  any of the 
simulated galaxy-sized halos.  All of this is explained by the fact
that THINGS galaxies generally  have  small Einasto indices whereas
simulated halos have   large index  values  due to their cuspy
nature (see \S\ref{sec:compsimu} for a complete discussion).

%%%%%%%%%%%%%%%%% Table logarithmic slope%%%%%%%%%%%%%%%%%%%%%%
\begin{deluxetable}{lrr}
\tablecaption{Logarithmic slope $\alpha$ of the dark matter density profile
  derived at $\log(r/r_{-2}) = -1.5$.}
\tablehead{ Galaxy  &   $\alpha$ (diet-Salpeter IMF) &  $\alpha$ (Kroupa IMF)}
\startdata
    NGC925  &  $\lesssim 0  \pm   0.01  $  &  $\lesssim 0  \pm  0.01$	    \\
   NGC2366  &  $-0.08	    \pm   0.01  $  &  $-0.08	   \pm  0.01$	    \\
   NGC2403  &  $-0.91	    \pm   0.03  $  &  $-1.06	   \pm  0.03$	    \\
   NGC2841  &  $-1.43	    \pm   0.03  $  &  $-1.51	   \pm  0.01$	    \\
   NGC2903  &  $-1.31	    \pm   0.01  $  &  $-1.40	   \pm  0.02$	     \\
   NGC3031  &  $\lesssim 0  \pm   0.01  $  &  $-0.02	   \pm  0.01$	 \\
   NGC3198  &  $-0.20	    \pm   0.01  $  &  $-0.37	   \pm  0.01$	\\
    IC2574  &  $-0.05	    \pm   0.03  $  &  $-0.08	   \pm  0.03$	 \\
   NGC3521  &  $\lesssim 0  \pm   0.01  $  &  $-0.12	   \pm  0.02$	 \\
   NGC3621  &  $-0.53	    \pm   0.05  $  &  $-1.17	   \pm  0.07$	 \\
   NGC4736  &  $\lesssim 0  \pm   0.01  $  &  $-0.32	   \pm  0.03$	 \\
    DDO154  &  $-0.41	    \pm   0.03  $  &  $-0.42	   \pm  0.03$	      \\
   NGC5055  &  $\lesssim 0  \pm   0.01  $  &  $\lesssim 0  \pm  0.01$	      \\
   NGC6946  &  $\lesssim 0  \pm   0.01  $  &  $-0.53	   \pm  0.09$	    \\
   NGC7793  &  $\lesssim 0  \pm   0.01  $  &  $-0.05	   \pm  0.01$
\enddata
\label{innerslope}
\end{deluxetable}
%%%%%%%%%%%%%%%%% end Table logarithmic slope%%%%%%%%%%%%%%%%%%%%%%
  
The bottom panel of Fig.~\ref{fig:compdens} shows that 
on the one hand,  the logarithmic slopes in the innermost regions 
 are more reminiscent of the slope of a Iso halo than that
of  cosmological Einasto or NFW cusps,  for the majority of the
sample.   On the other hand,  because of the 
progressively steepening nature implied by the Einasto model (Eq.~\ref{eq:rhoeinasto}), the observed slope profiles become  
similar to a NFW cusp at larger radii. The slopes are even larger than those of  any 
cosmological halos beyond $r_{-2}$, as a result of the small amplitude of the observed Einasto indices. A  density slope
profile  of a typical  observed  Einasto halo ($\log(\rho_{-2}$ \msol pc$^{-3}) \sim -2$, $r_{-2} \sim 10$ kpc, 
 $n \sim 1.3$, as fit with the Kroupa IMF) thus represents  an
intermediate  case  between a pseudo-isothermal core halo and  cosmological cusps inside $r_{-2}$. 
 As a consequence of these results, we estimate that half of the total mass of such  halos 
  is contained within about 15 kpc, and the mass profile has rapidly converged 
 towards, e.g. 95\% its total mass at $r \sim 40$ kpc only 
 (or four times the average scale radius $r_{-2}$). 

We report the inner slopes $\alpha = \rm d\log (\rho)/d\log (r)$ of the dark matter density profiles in
Tab.~\ref{innerslope}. 
For sake of uniformity with cosmological simulations whose inner  
slopes are derived at a fraction of the Einasto radius $r_{-2}$
 (basically $-2 < \log(r/r_{-2}) < -1$, Navarro et al. 2010),  
we have derived the inner slopes of the THINGS sample at $\log(r/r_{-2}) = -1.5$ (or $r = 0.03 r_{-2}$), discarding 
all fits of the fourth halo family.
  The mean logarithmic slope of the
density profiles is $\bar{\alpha} = -0.3 \pm 0.1$ for the
diet-Salpeter IMF and $\bar{\alpha} = -0.5 \pm 0.1$ for the Kroupa
IMF. Those numbers become $\bar{\alpha} = -0.1 \pm 0.1$ and $\bar{\alpha} = -0.2 \pm 0.1$ (respectively)
when  the  cuspiest halos of the sample ($n > 4$) are removed from the distribution.  
 Note  that the mean inner density slope  is  $-1.3 \pm 0.2$ for those cuspiest
 halos only.  
  For comparison, the innermost slope in the Aquarius pure-$\Lambda$CDM 
 simulation (at $r_{-2}/100$) is $-0.9\pm 0.1$, 
 which is shallower than the cuspiest of our halos, but steeper than most of them. 
 
%%%%%%%%%%%%%%%%%%%%% figure radial profiles of Einasto densities and density slopes %%%%%%%%%%%%%%%%%%%%%%%
%%%%%%%%%%%%%%%%%%%%%%%%%%%%%%%%%%%%%%%%%%%%%%%%%%%%%%%%%%%  

As a reminder, an almost constant dark matter density has been found
for a large sample of dark matter dominated galaxies
\citep[e.g.][]{deb02}. These authors have found a mean inner density
slope $\alpha \sim -0.2$, as measured ``directly'' from optical spectroscopy by inverting the
RCs into  volume  densities and extracting the slope at the innermost 
data point  of the observations.  We do not know yet the
average slope that would be fit with the Einasto profile for
similar low surface brightness objects.     The current
THINGS sample consists mostly of galaxies more massive than those in
\citet{deb02} (stellar and total masses), despite the presence of a few low surface density
objects \citep[among which NGC~2366 is common with][]{deb02}.  Note,
however, that both samples show  comparable  average inner density
slopes of dark matter. 
 At first it may seem surprising because the quantity of baryons 
strongly differs in those samples. That particularity will be 
investigated in a future paper of our series.

\subsection{Halo concentration versus Einasto index and   mass}
 \label{sec:haloconcentration}
\begin{figure*}[t!]
 \begin{center}
\includegraphics[width=0.28\textwidth]{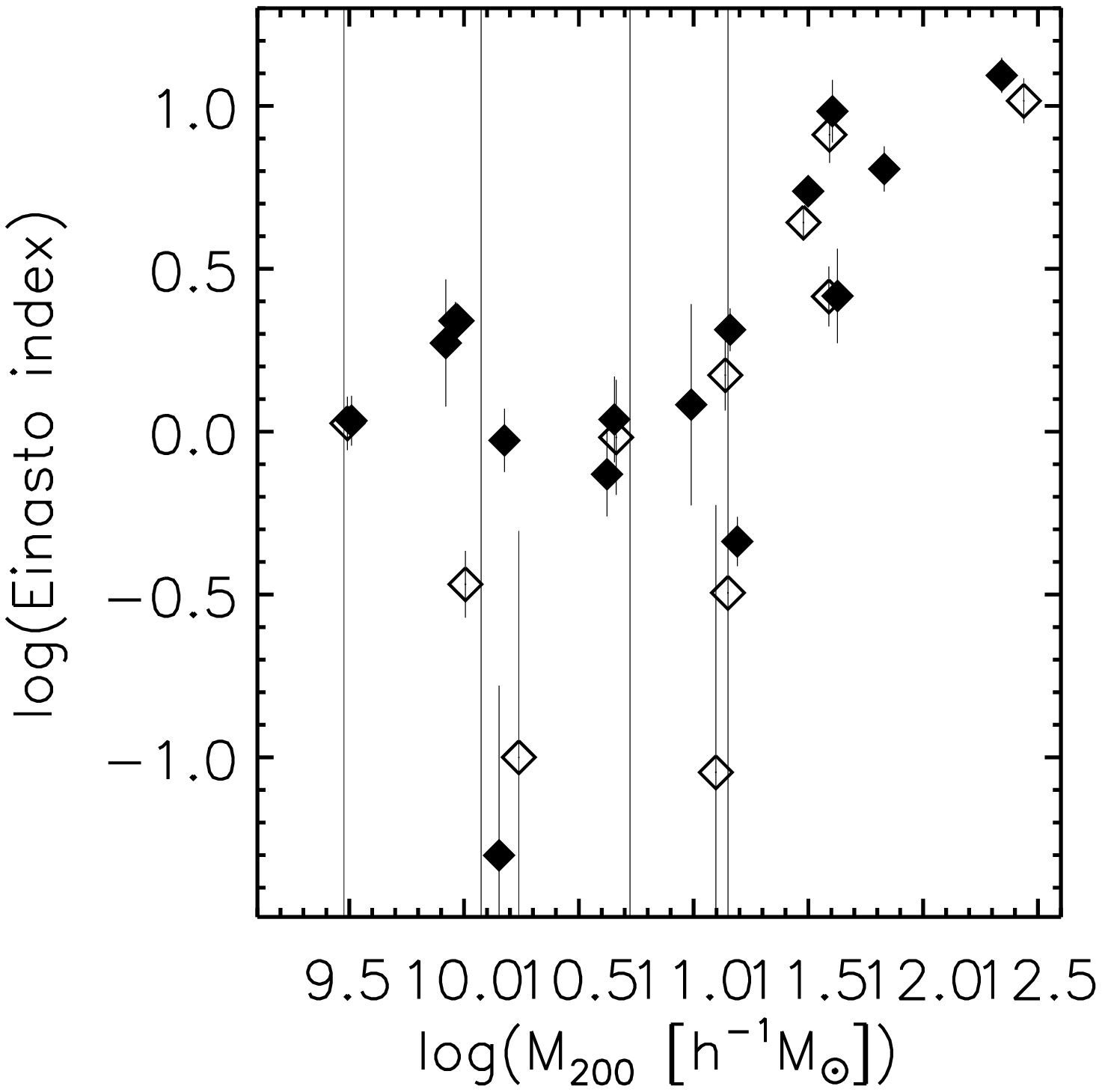}\includegraphics[width=0.28\textwidth]{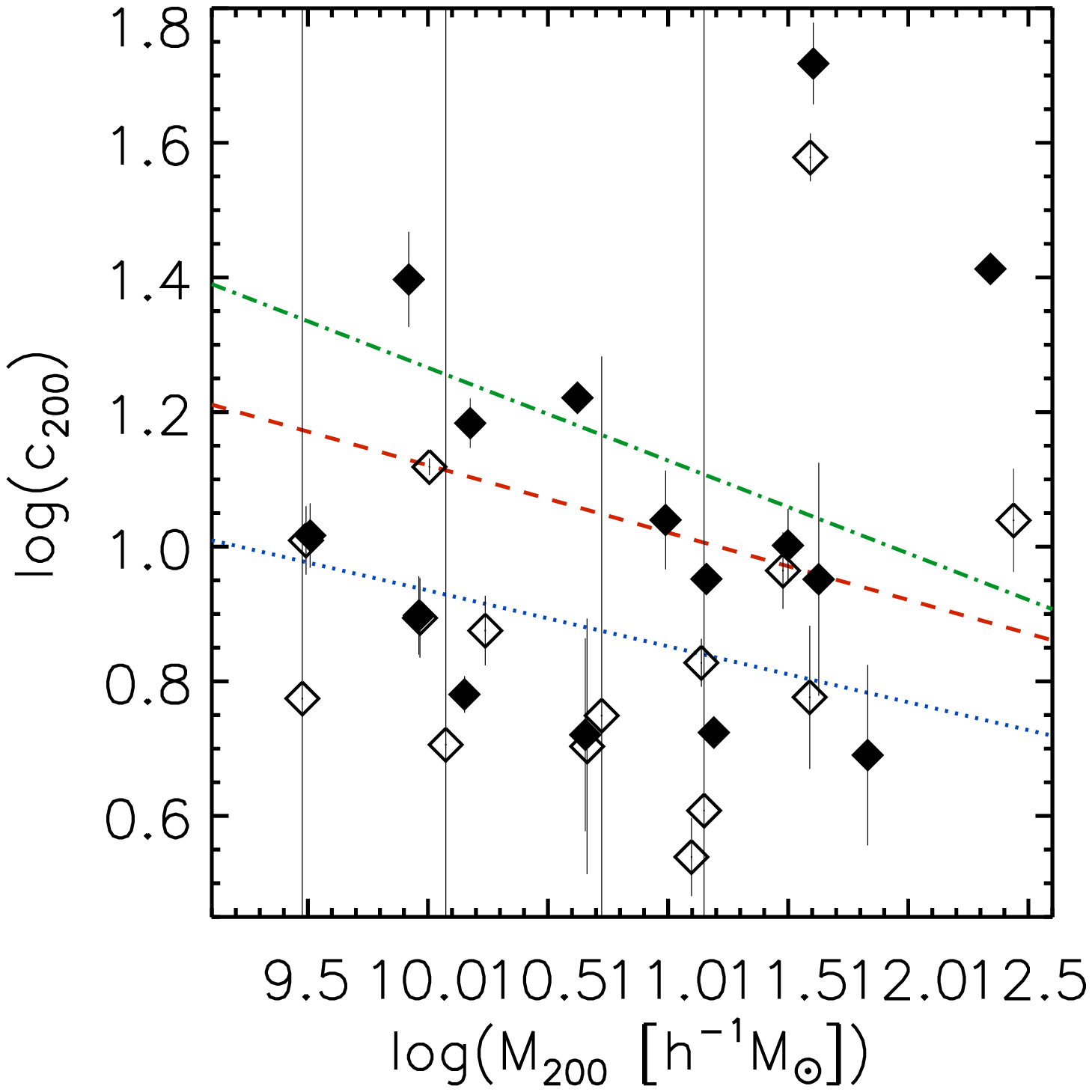}\includegraphics[width=0.28\textwidth]{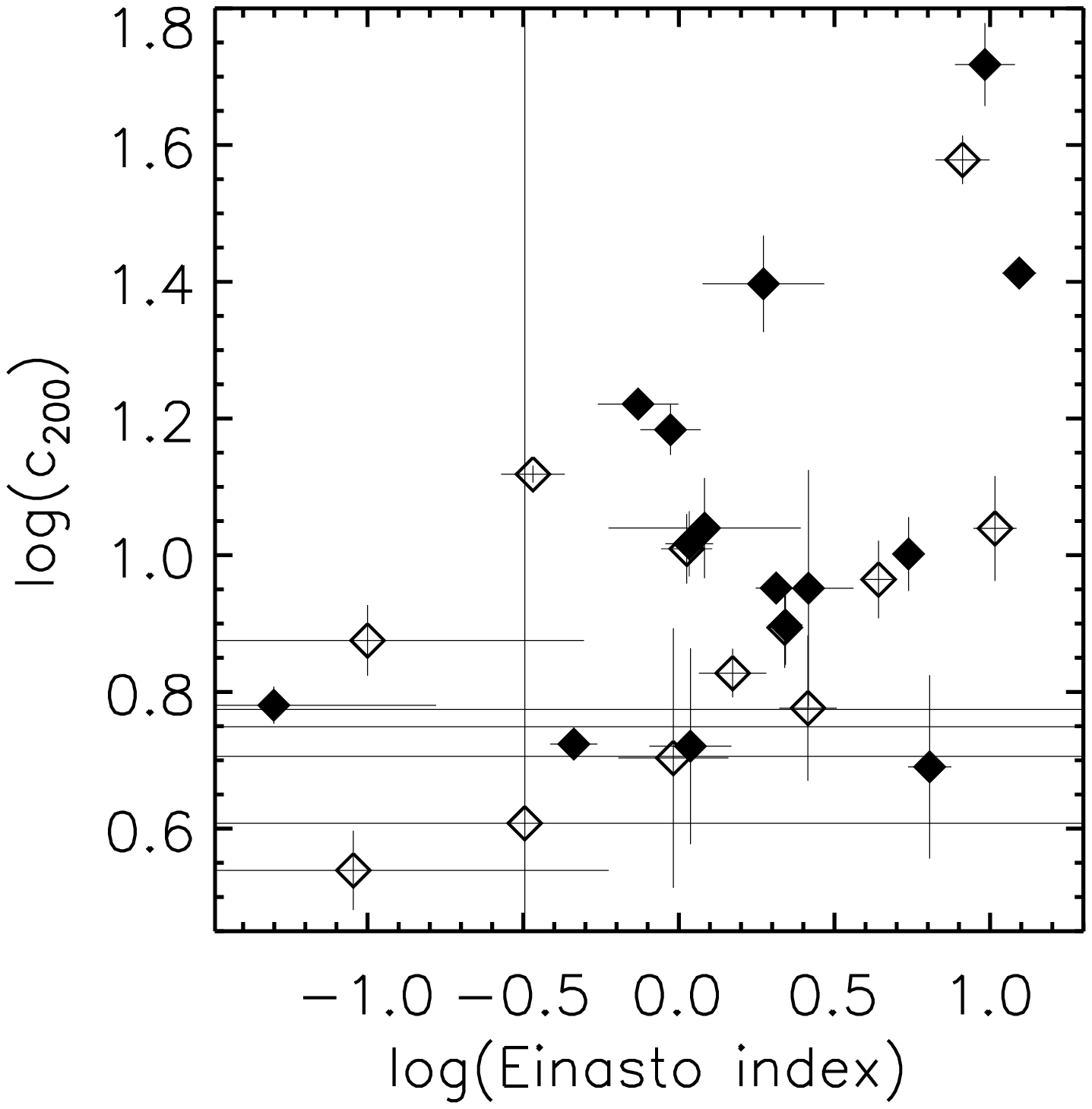}
 \caption{Virial masses $M_{200}$, concentrations $c_{200}$ and indices of Einasto halos.  
 Open (filled) symbols are for the diet-Salpeter (Kroupa, respectively) IMF. 
 In the middle panel  a green dashed-dotted line represents  the dependency of the Einasto halo concentrations on the mass 
 $\log(c_{200})=2.646-0.138\log(M_{200})$ from numerical simulations \citep[][concentrations derived with fixed Einasto indices]{gao08}, 
 a blue dotted the dependency of the NFW halo concentrations  $\log(c_{200})=1.765-0.083\log(M_{200})$ \citep{mac08} and
 a red dashed line the dependency of the NFW halo concentrations $\log(c_{200})=2.121-0.1\log(M_{200})$ \citep{net07}. 
 Halos with $\log(n) \sim -3$ are not displayed for clarity reasons.}
 \label{fig:haloprop1}
 \end{center}
 \end{figure*}

We have derived the virial radius $r_{200}$ and halo concentration $c_{200} = r_{200}/r_{-2}$
   as the radius of a sphere of mean density $\rho_{200}$ which equals
two hundred times the critical density for closure of the Universe, $\rho_{\rm crit}=3H^2_0/8\pi G$.
We derive the (halo) mass $M_{200}$ at $r_{200}$ using
Eq.~\ref{eq:Meinasto}.  Figure~\ref{fig:haloprop1} compares the
virial masses, halo concentrations and indices. Concentrations and
virial masses are reported in Tab.~\ref{tab:concvirmass}.  For
clarity, the fourth halo family has been omitted in
Fig.~\ref{fig:haloprop1} and Tab.~\ref{tab:concvirmass}.
 
Adopting the Kroupa stellar mass scaling, it is observed that 
the Einasto index increases with the halo mass for halos more massive than  
$M_{200} \sim  2 \times 10^{11}\ h^{-1}$ \msol\ while its distribution
 is more scattered for less massive halos, showing 
an almost flat part. 
That curved trend is less evident with the diet-Salpeter IMF results.
Though the mass range spanned by the observations is relatively small in comparison to what can be probed 
by numerical models, the complex relation 
between Einasto index and mass illustrates that no ``universal" galactic dark matter halo can be identified from these
  observations: the Einasto index cannot be simply deduced from the
halo mass,  nor does a simple scaling law allows one to scale the
Einasto index/mass profile  to those of a halo with a different
mass. 
 Our observational result thus confirms the non-universality of
Einasto halos seen in numerical simulations \citep{nav10}.

The concentration is not  correlated with the Einasto index (right-hand panel), but tends to decrease
with the mass (middle panel). The scatter of the trend is however significant. 
Halo fits with the diet-Salpeter IMF are less concentrated than halo fits with the Kroupa IMF, 
as caused by the more important contribution of the stellar component.    
 As a comparison, the expected dependencies of the Einasto and NFW 
concentrations on the halo virial mass from cosmological simulations
\citep{net07,mac08,gao08} are displayed as well.  These authors have found
concentrations basically scaling with the total mass as $c_{200}
\propto M_{200}^{-0.1}$. If one omits the two outliers which are the most concentrated and cuspiest halos,  
the relations of \citet{net07} and \citet{gao08} are
 in good agreement with the seven halos at the upper end of the $c_{200}-M_{200}$ trend 
for the Kroupa IMF, and thus above most of concentrations derived with the diet-Salpeter IMF. The NFW relation 
of \citet{mac08} is more consistent with the lower end of the remaining halos at the lower end of the concentrations.
Notice however that  the comparison is of  limited  value as  the halo masses can be measured directly from the
numerical simulations whereas we can only provide   estimates for the observations.

%%%%%%%%%%%%%%%%%%%% Figure Einasto concentrations, masses, indices %%%%%%%%%%%%%%%%%%%%%%%
%%%%%%%%%%%%%%%%%%%%%%%%%%%%%%%%%%%%%%%%%%%%%%%%%%%%%%%%%%%  

We finally note that   possible  correlations between the
properties of the Einasto halos with those of the baryonic matter
(absolute magnitude, baryonic gas masses, disk and bulge characteristic scalelengths, etc.)  will be fully studied 
 into a forthcoming paper of our series (Chemin et al., in preparation). A preliminary analysis nevertheless shows Einasto indices scaling with (total) 
 stellar masses, but barely with e.g. the bulge-to-disk luminosity ratio.

\subsection{The influence of total gas densities on Einasto parameters}
\label{sec:molecgas}
\begin{figure}[b!]
 \begin{center}
\includegraphics[width=\columnwidth]{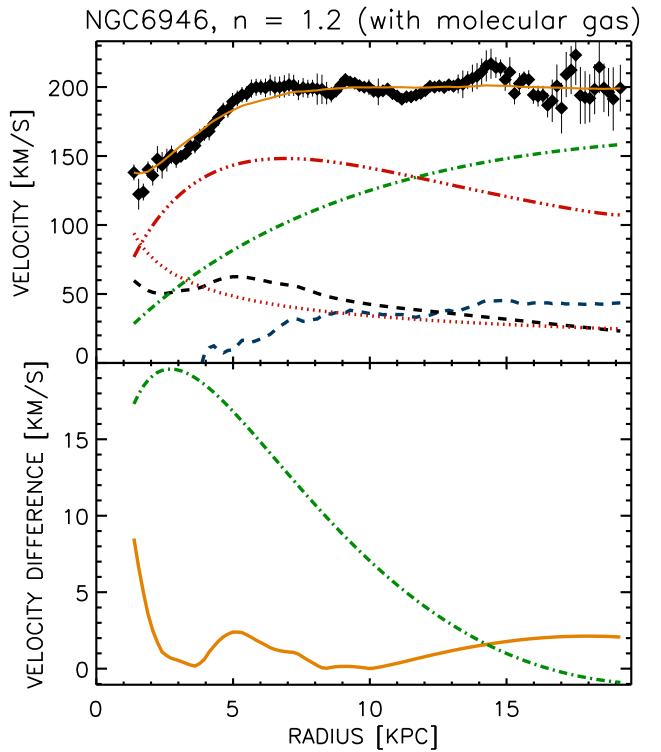}
 \caption{Mass model of NGC 6946 with the molecular gas contribution \emph(top panel). The molecular gas RC is shown by a black dashed line. 
 All other stellar and gas curves are from Fig.~\ref{fig:rc-kroupa}. 
 The bottom panel displays the differences between RCs of the total models of NGC 6946 done with and without the molecular gas (orange solid line, in absolute values),
  and between the corresponding Einasto halo RCs (green dashed-dotted line).}
 \label{fig:n6946molec}
 \end{center}
 \end{figure}

 The contribution of the molecular gas has been neglected in our study. This is  
justified since the total gas surface densities  are  dominated 
by those of the atomic gas for most of the sample.  Even when molecules dominate the total gas density in the innermost regions of 
a few galaxies from our sample \citep[NGC 3521, NGC 4736, NGC 5055, NGC 6946;][]{ler08}, 
the stellar surface densities completely dominate the mass budget of the baryons in those regions. They 
are generally more than ten times larger than the molecular densities,
with the exception of NGC 6946 where  the ratio of the surface density of molecules to that of the stars sometimes reaches 
a factor of 1/3 in its central kiloparsec \citep{ler08}.\footnote{Since the stellar masses we use are larger than those given by  \citet{ler08},  as explained in D08, 
that ratio of molecular gas to stellar surface densities is even smaller in our current analysis than in \citet{ler08}.}   
 We have thus fit Einasto models  by adding the molecular gas component to the mass distribution for NGC 6946 only,  
with $H_2$ surface densities from \citet{ler08}. The top panel of Fig.~\ref{fig:n6946molec}  shows the results of that mass distribution model.
As a result, the quality of fit remains unchanged with the Kroupa IMF ($\chi^2_r \sim 1.0$), as well as $\rho_{-2}$ and $r_{-2}$ within the uncertainties, with 
$\rho_{-2} = (2.7 \pm 1.0)\times$  \odensunit\ and $r_{-2} = (15.9 \pm 3.7)$ kpc. The Einasto index becomes $n=1.2 \pm 0.4$, which is about half the value derived 
with no molecular gas ($n=2.6 \pm 0.9$). 
 Such a difference of halo cuspiness is clearly seen in the bottom panel of Fig.~\ref{fig:n6946molec} (green curve), where the difference of the Einasto halo RCs reach 
20 \kms\ at 3 kpc. 
 Though NGC 6946 is probably an extreme case where the influence of the molecular gas is maximum on the halo shape, it is 
likely that Einasto indices would be even lower had we included the molecules in our fits.  
We also think   the total gas component could play a more important contribution to the total baryonic dynamical budget  than 
 in our current study if models were done under the assumption of \emph{free} stellar mass-to-light ratios. 
 However such an hypothesis of $\Upsilon_{3.6\mu m}$ is beyond the scope of the article and we defer a complete analysis of the influence 
 of the total/molecular gas contribution onto Einasto models with free  mass-to-light ratios to 
 our forthcoming article (Chemin et al., in preparation). 

 %%%%%%%%%%%%%%%%%%%% Fit of ngc 6946 RC with and without molecular gas contribution %%%%%%%%%%%%%%%%%%%%%%%
%%%%%%%%%%%%%%%%%%%%%%%%%%%%%%%%%%%%%%%%%%%%%%%%%%%%%%%%%%%  

%%%%%%%%%%%%%%%%%%%%% Table Einasto concentrations and masses%%%%%%%%%%%%%%%%%%%%%%%
\begin{deluxetable}{lcccccc}
\tablecaption{Einasto halo concentration $c_{200}$ and virial mass $M_{200}$}
\tablehead{Galaxy  & $\log(c_{200})$ & $\sigma_{\log}$ & $\log(M_{200})$ &   $\log(c_{200})$ & $\sigma_{\log}$  & $\log(M_{200})$  \\
           &  (1)   &  (2) & (3)    & (4)  & (5) & (6) }
\startdata
 NGC925   & 0.71  & 46.9  &  10.07 &  0.78 & 0.03 &  10.15   \\
 NGC2366  & 1.01  & 0.05  &  9.49  &  1.02 & 0.05 &   9.51   \\
 NGC2403  & 0.96  & 0.06 & 11.48   &  1.00 & 0.05 &  11.50   \\
 NGC2841  & 1.04  & 0.08  & 12.44  &  1.41 & 0.02 &  12.34   \\
 NGC2903  & 1.58  & 0.04  & 11.59  &  1.72 & 0.06 &  11.61    \\
 NGC3031  & 0.88  & 0.05  & 10.24  &  1.22 & 0.02 &  10.62   \\
 NGC3198  & 0.83  & 0.04  & 11.14  &  0.95 & 0.02 &  11.16  \\
 IC2574   & 0.70  & 0.19  & 10.66  &  0.72 & 0.14 &  10.66   \\
 NGC3521  & 0.61  & 1.73  &  11.15 &  1.04 & 0.07 &  10.99   \\
 NGC3621  & 0.78  & 0.11  & 11.59  &  0.69 & 0.13 &  11.83   \\
 NGC4736  & 0.77  & 1.35  &   9.48 &  1.40 & 0.07 &   9.92   \\
 DDO154   & 0.89  & 0.06  &  9.97  &  0.90 & 0.06 &   9.96     \\
 NGC5055  & 0.54  & 0.06  & 11.10  &  0.72 & 0.14 &  11.19     \\
 NGC6946  & 0.75  & 0.53  & 10.72  &  0.95 & 0.02 &  11.63   \\
 NGC7793  & 1.12  & 0.01  & 10.00  &  1.18 & 0.04 &  10.18	  
\enddata
\tablecomments{Columns (1)-(2)-(3) are for the diet-Salpeter IMF and (4)-(5)-(6) for the Kroupa IMF. 
$\sigma_{\log}$ is the $1\sigma-$error on $\log(c_{200})$. $M_{200}$ is in $h^{-1}$ \msol. }
\label{tab:concvirmass}
\end{deluxetable}
%%%%%%%%%%%%%%%%%%%%%%%%%%%%%%%%%%%%%%%%%%%%%%%%%%%%%%%%%%%  

\section{Strengths and weaknesses of  Einasto  $\Lambda$CDM halos} 
\label{sec:compsimu}

  We now compare more precisely the parameter   space of
observed Einasto halos with those  derived from $\Lambda$CDM dissipationless 
simulations
\citep[e.g.][]{nav04,nav10}. 

A first caveat to this comparison is that,
despite the wide global range of halo masses explored in these simulations,
the halo masses do not span the whole range of our galaxies, namely 
$10^{9.5-12.5}\ h^{-1} M_\odot$.
A larger mass range of simulated galaxy-sized halos is required for future comparisons with observations.  

Secondly, and more
importantly, these simulations  usually  ignore the gaseous component of galaxies,
whose dissipative nature leads them to concentrate in the cores of halos and
possibly force the dark matter to respond to the concentration of the
dominant central baryons.
For this reason, we have also
compared our results with those from \citet{tis10} who have included
the physics of baryons in the galactic halos of the Aquarius
simulations \citep{spr08} by means of hydrodynamical models.  

Finally,
the innermost regions of galactic halos are still hardly resolved by
 cosmological  numerical simulations (at scales $r<r_{200}/1000$), 
as  noted  in, e.g., \citet{nav10}.
 
We distinguish here   between   fits performed with a free index  
and fits done with a  fixed index in order to investigate whether the
typical Einasto index seen in $\Lambda$CDM simulations  matches  the
observations.  We have thus performed fits with a fixed index $n = 6.2$,
which is the average value of galaxy-sized halos presented in
\citet{nav04}.

\subsection{Models with a free index}

Figures~\ref{fig:haloparam1} and~\ref{fig:haloparam2} show that most
of  the observed  and simulated galaxy-sized halos share the same
characteristic scale densities and radii but not their indices.   The
indices of half of our observed halos  are usually less than half the mean
value of the simulated ones (i.e. $n < 3$).  The Einasto indices for baryons$+$dark matter
simulations from \citet{tis10} differ even more from our observations than do
the pure dark matter halo indices. The addition of hydrodynamics in the
$\Lambda$CDM simulations  drags more dark matter towards the centre of halos, as likely caused by gas dissipational effects, thus pushing  the index towards larger values ($n
\rightarrow 10$) than for the pure dark matter case ($n \sim 6$),
while   the observations go to an opposite way ($n
\rightarrow 1-2$). 
 Note though recent hydrodynamical cosmological simulations by \citet{gov10}
with strong SN feedback coupled with a high threshold for star formation lead to
 dark matter halos without cusps.  As theoretically explained by \citet{pon11}, episodic 
SN events in dwarf galaxies alter the dark matter distribution in non-adiabatic ways leading to cored profiles. 
 
Only a few galaxies have a  parameter  space in rough agreement with
simulations within the quoted uncertainties, though with some subtle
differences.     For instance, NGC~2903  can be
associated with a  dwarf-sized  simulated  halo in the index-radius
graph,  but with a halo more massive than that of a dwarf galaxy  
($\log(\rho_{-2}) \sim -1$ \msol\ pc$^{-3}$) in the index-density
parameter space.  For NGC~3621   any (dis)agreement depends critically on
the chosen IMF.    Only NGC~2841 and NGC~2403 seem to be galaxies
whose halo parameters are in full agreement with cuspy $\Lambda$CDM Einasto
halos.

% GRAPHS OF RC fits  

%%%%%%%%%%%%%%%%%%%%% kroupa imf - fixed index %%%%%%%%%%%%%%%%%%%%%%%
 \begin{figure*}[ht] 
 \begin{center}
\includegraphics[width=0.25\textwidth]{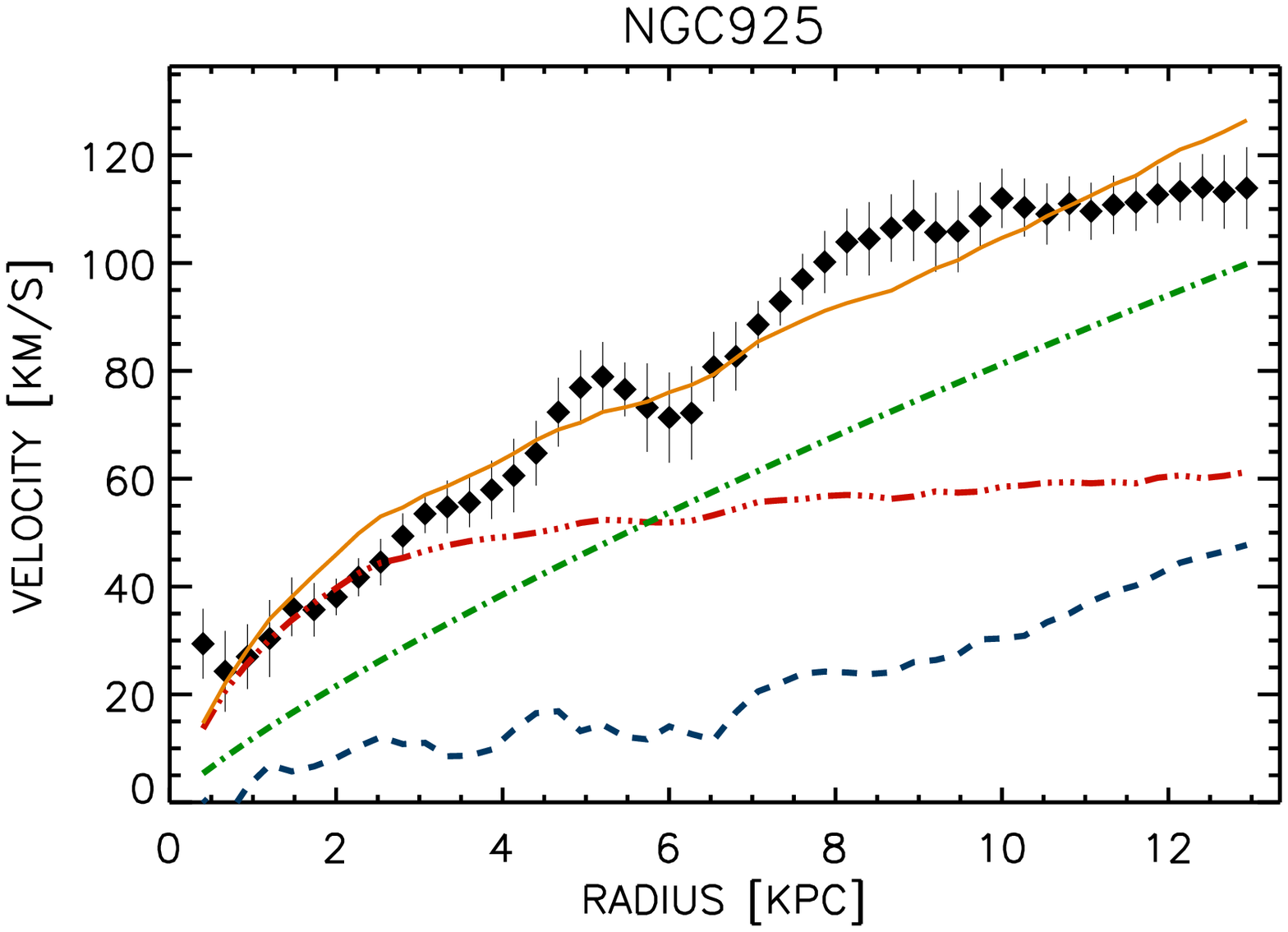}\includegraphics[width=0.25\textwidth]{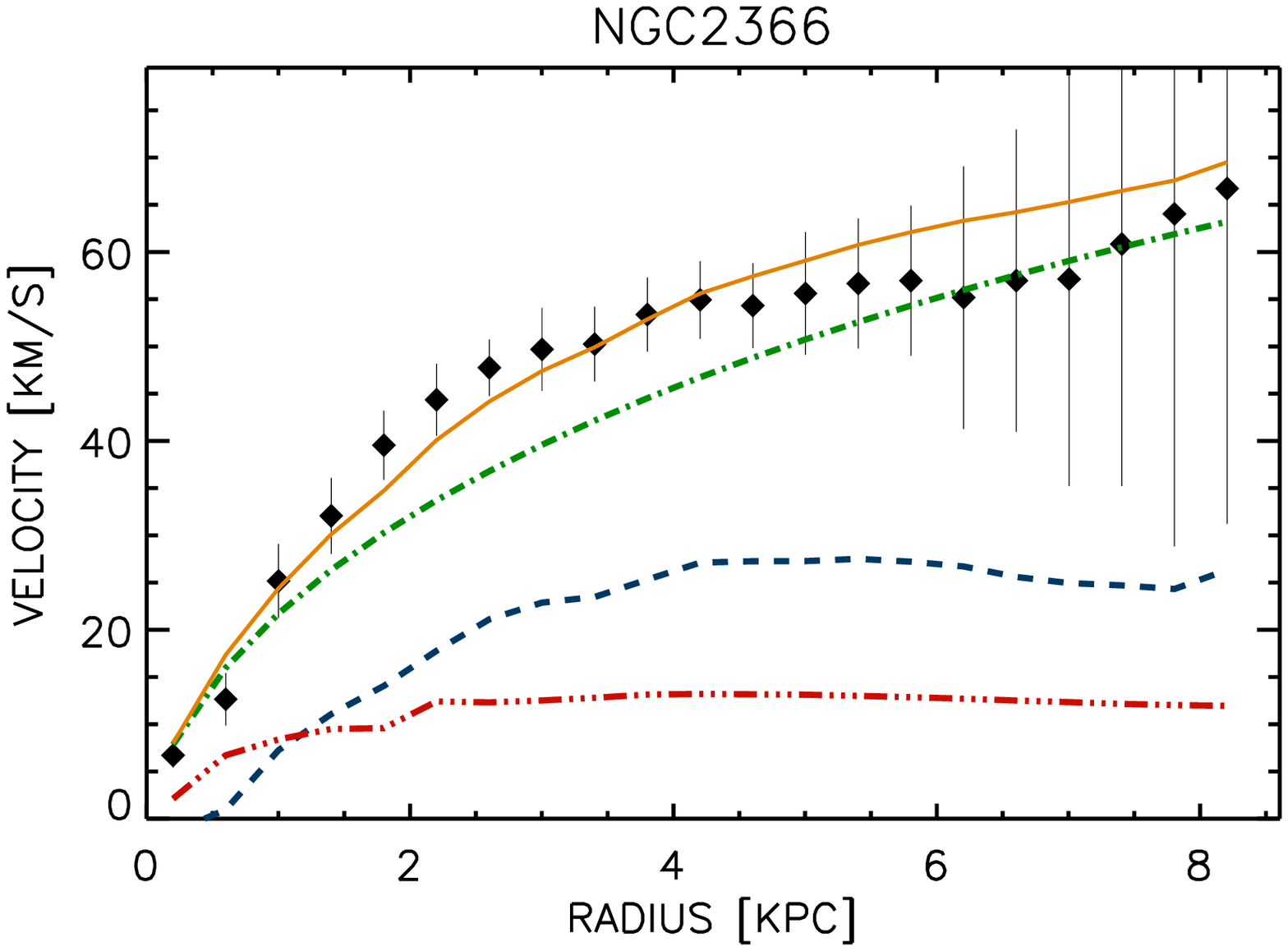}\includegraphics[width=0.25\textwidth]{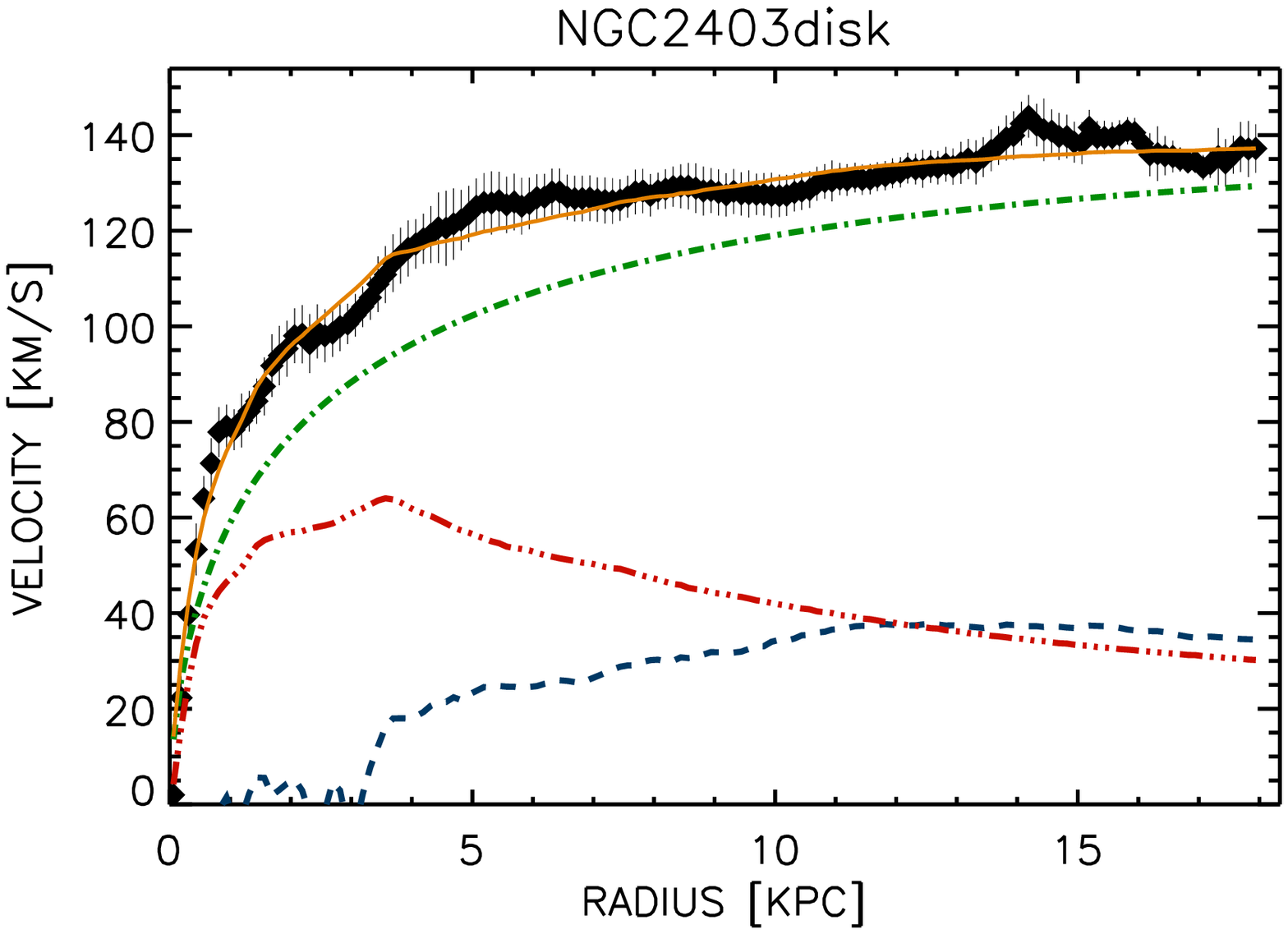}\includegraphics[width=0.25\textwidth]{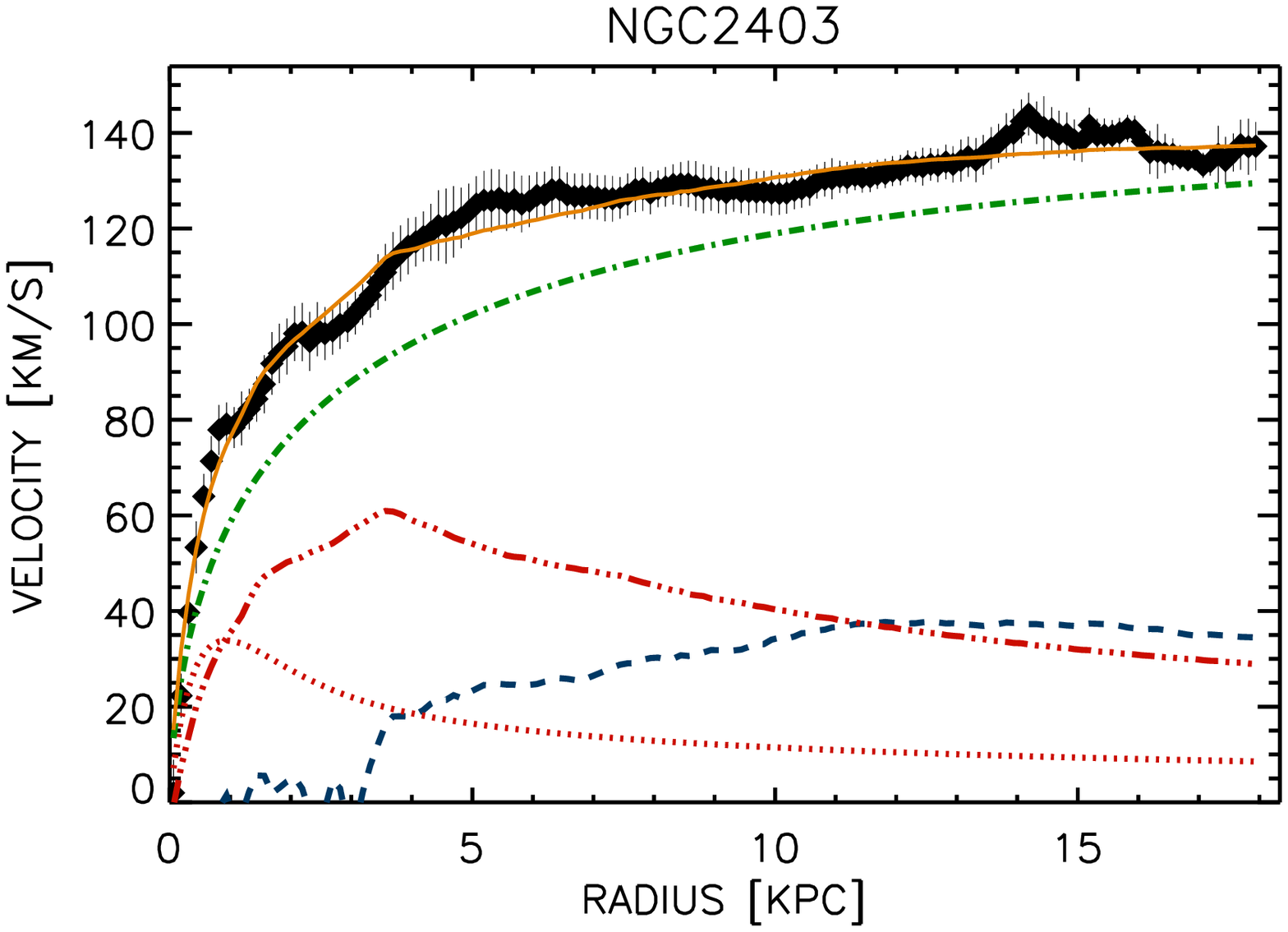}
\includegraphics[width=0.25\textwidth]{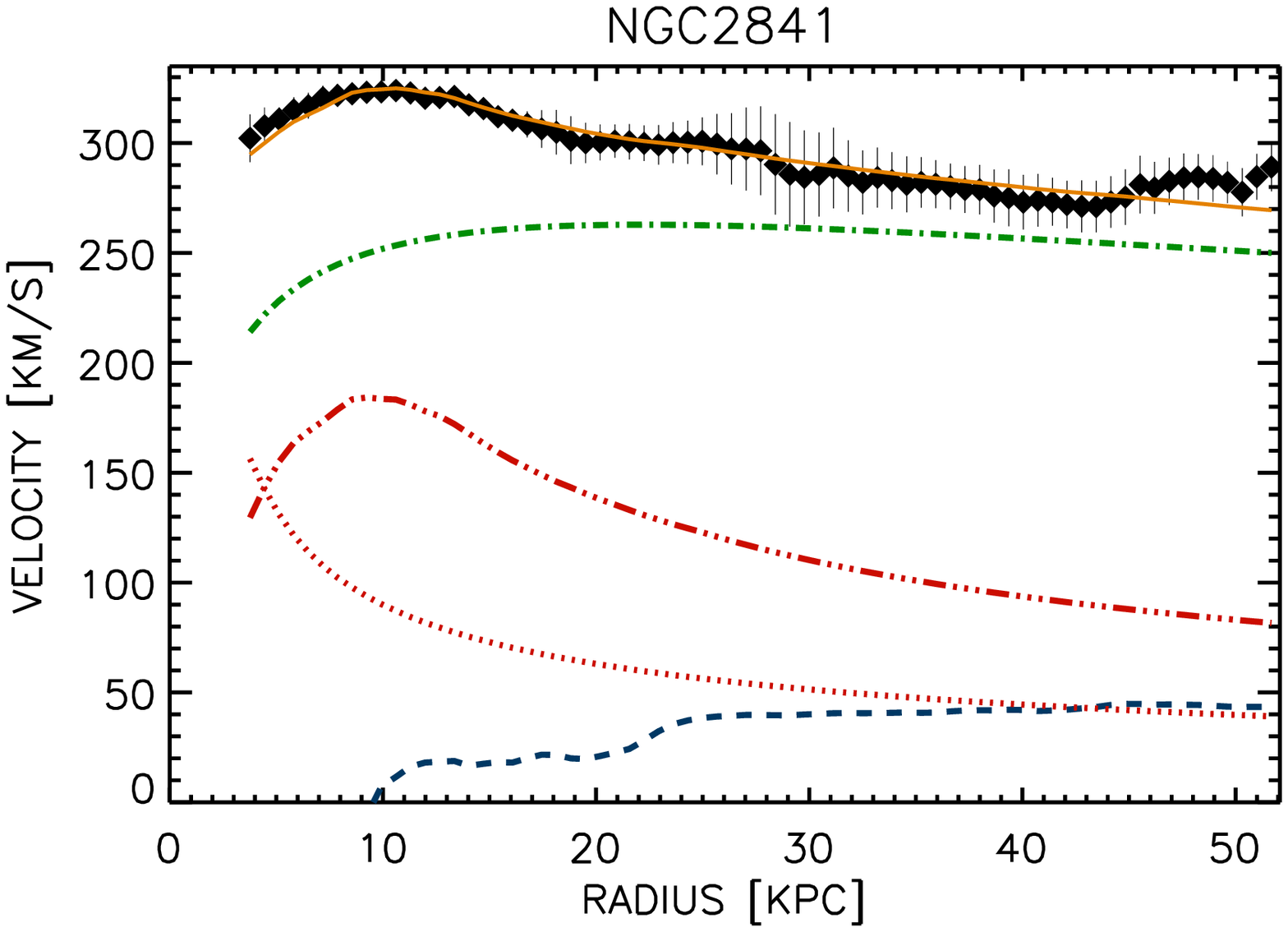}\includegraphics[width=0.25\textwidth]{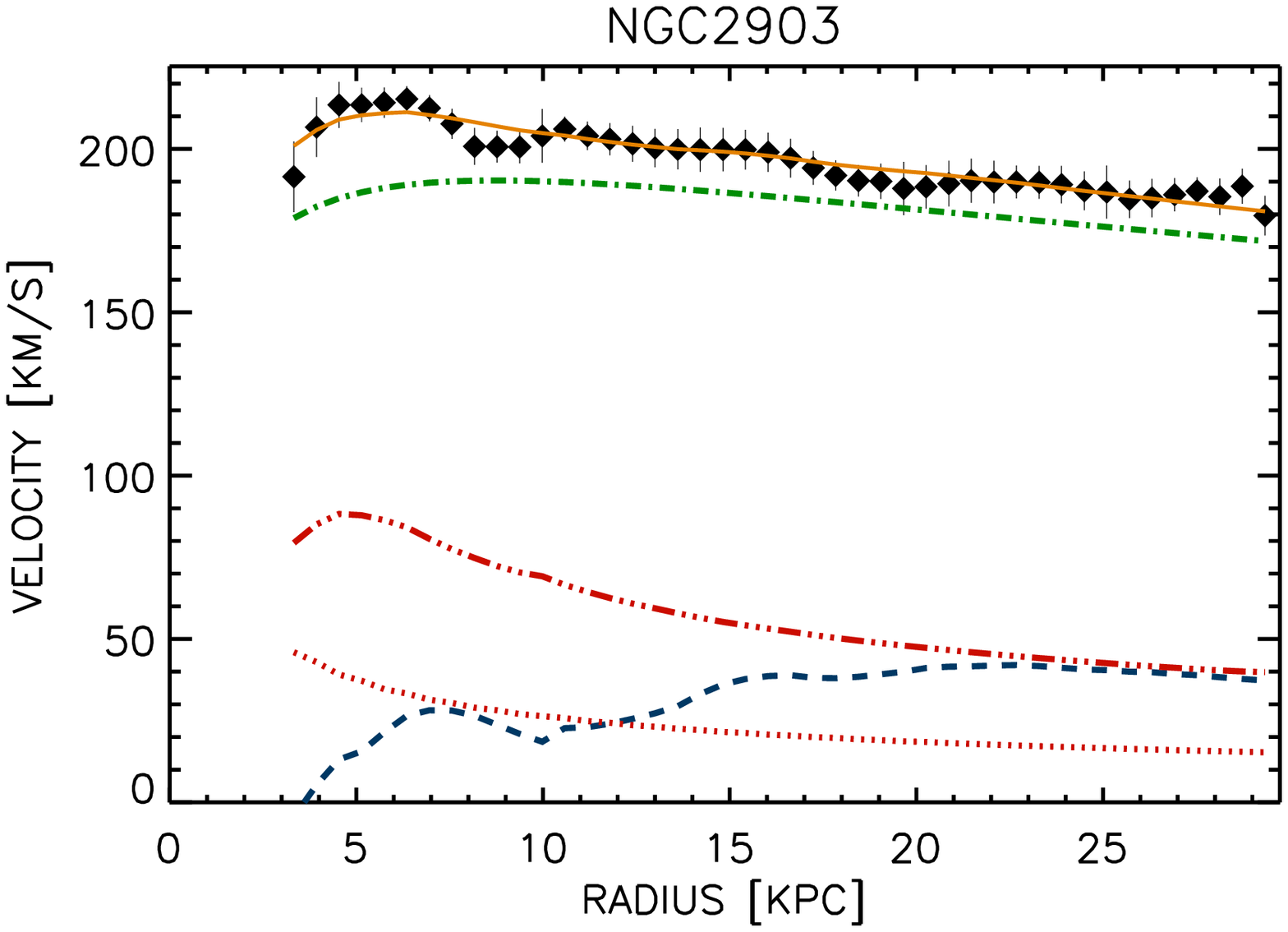}\includegraphics[width=0.25\textwidth]{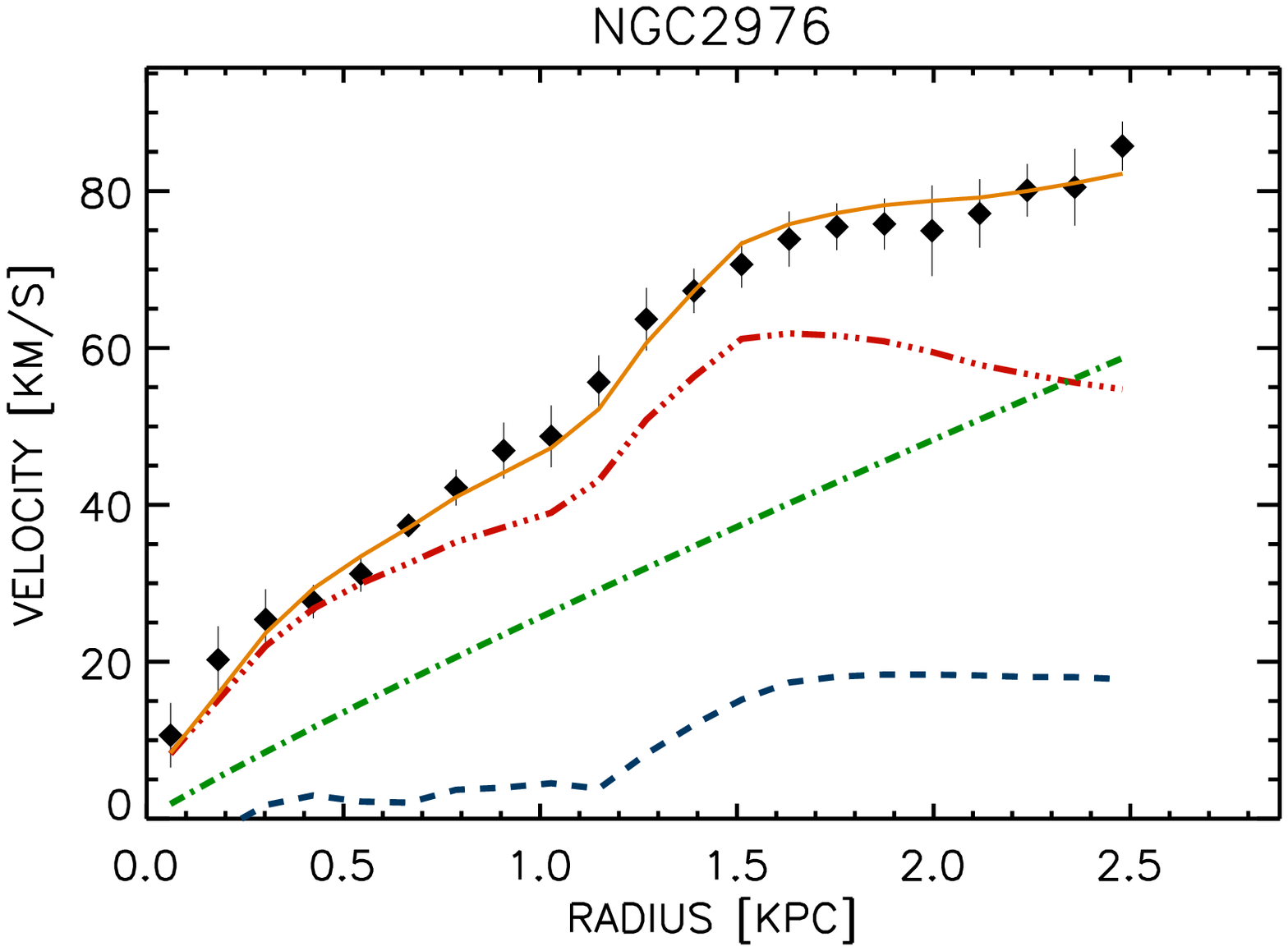}\includegraphics[width=0.25\textwidth]{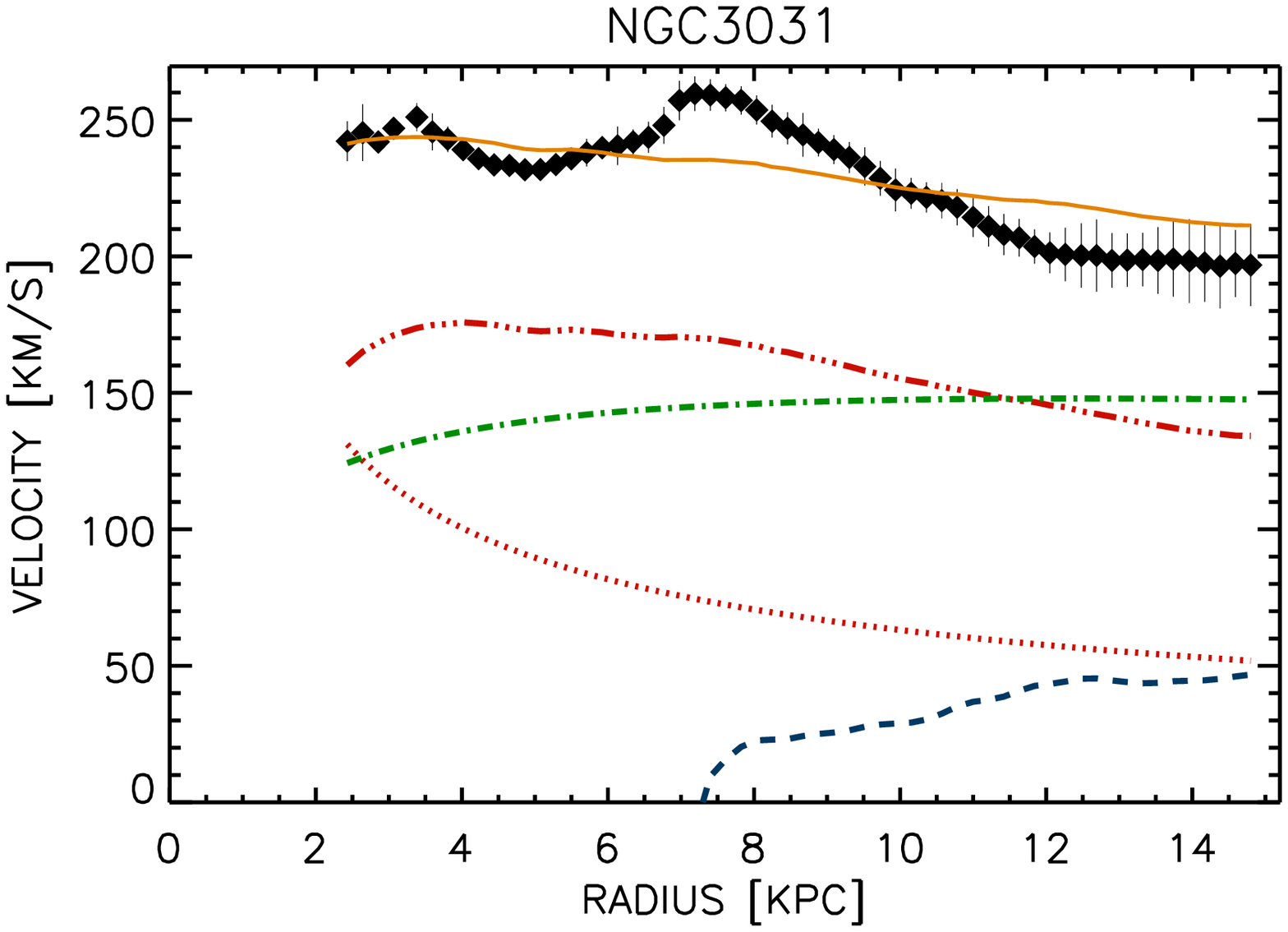}
\includegraphics[width=0.25\textwidth]{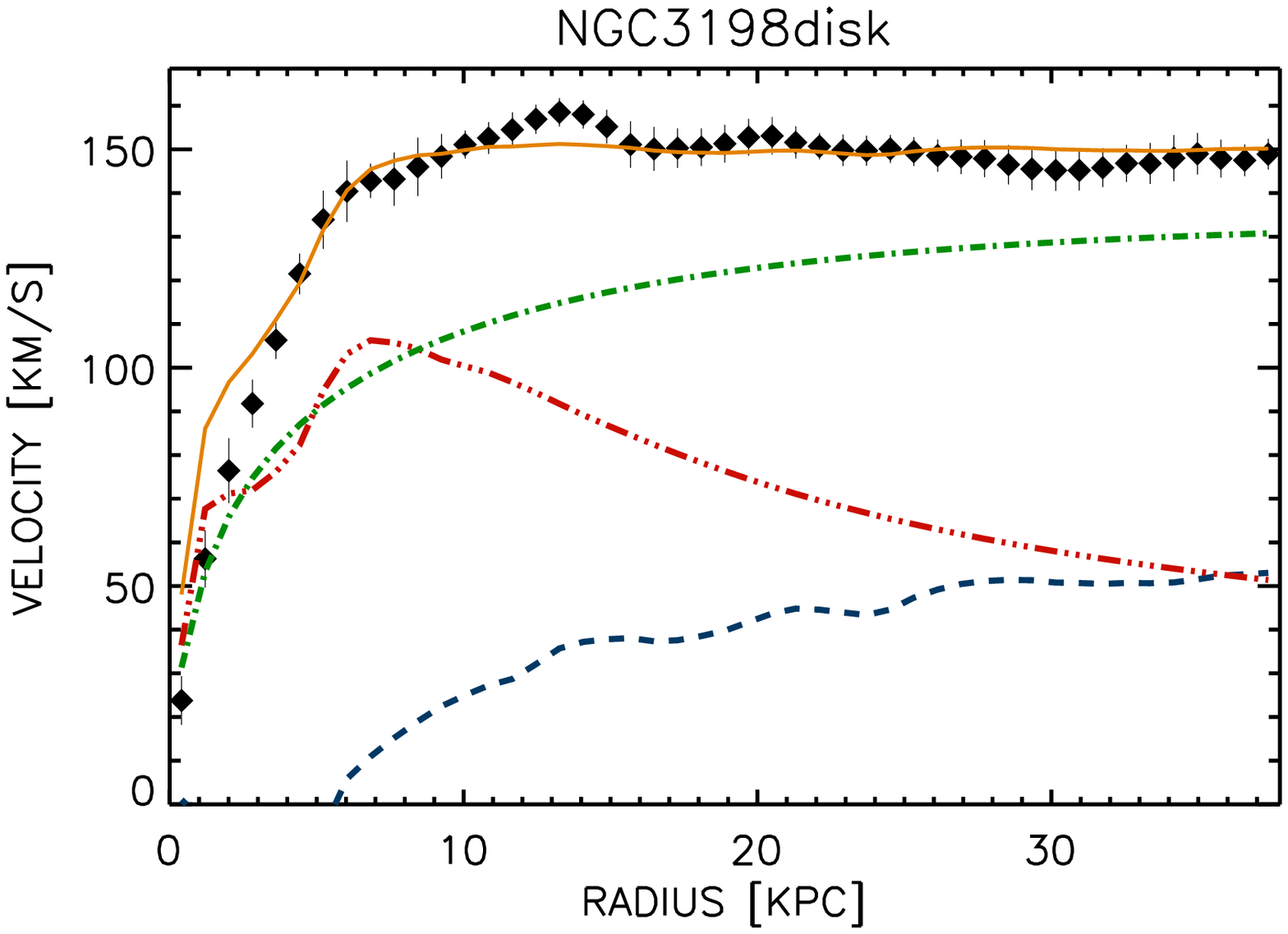}\includegraphics[width=0.25\textwidth]{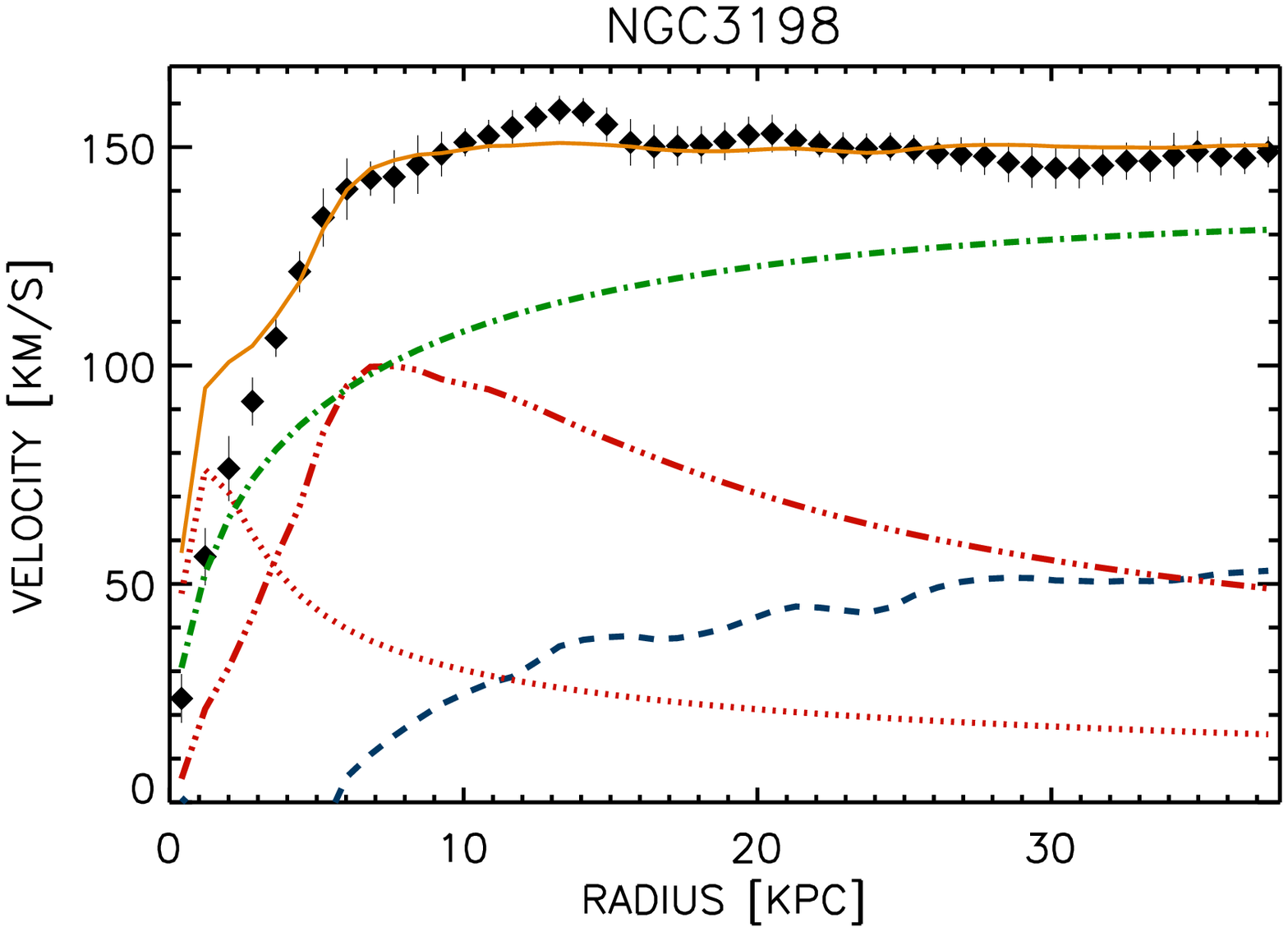}\includegraphics[width=0.25\textwidth]{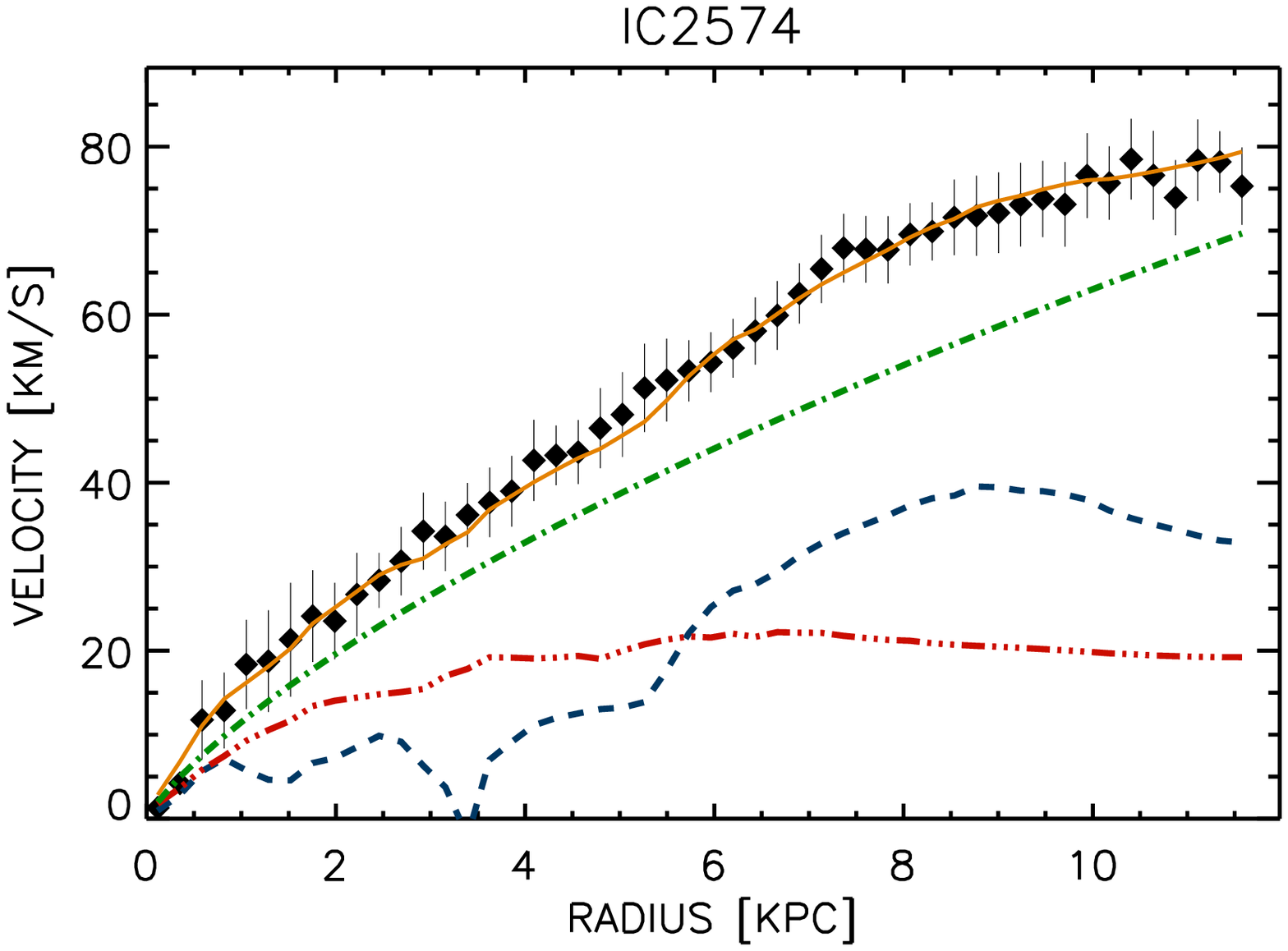}\includegraphics[width=0.25\textwidth]{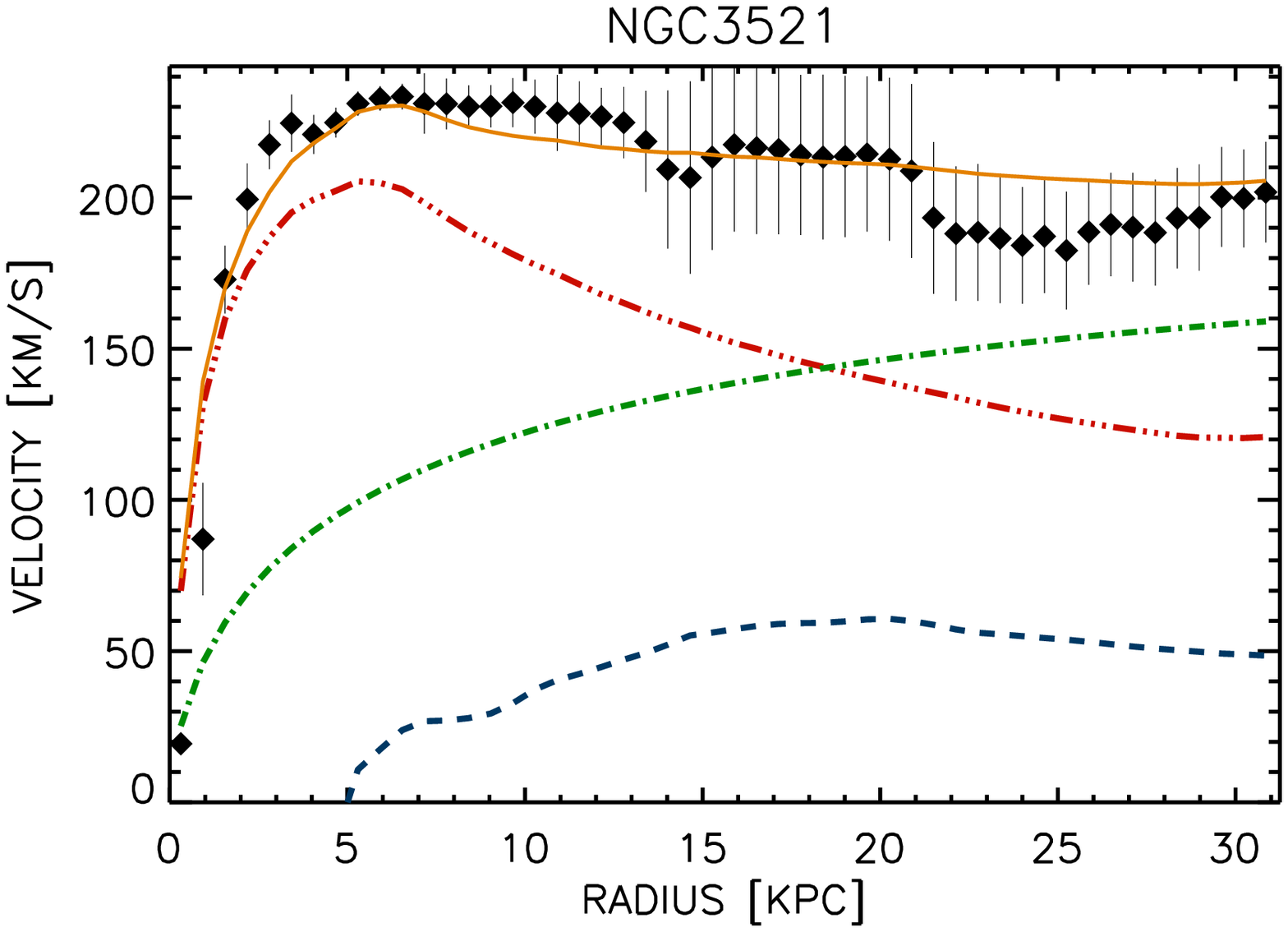}
\includegraphics[width=0.25\textwidth]{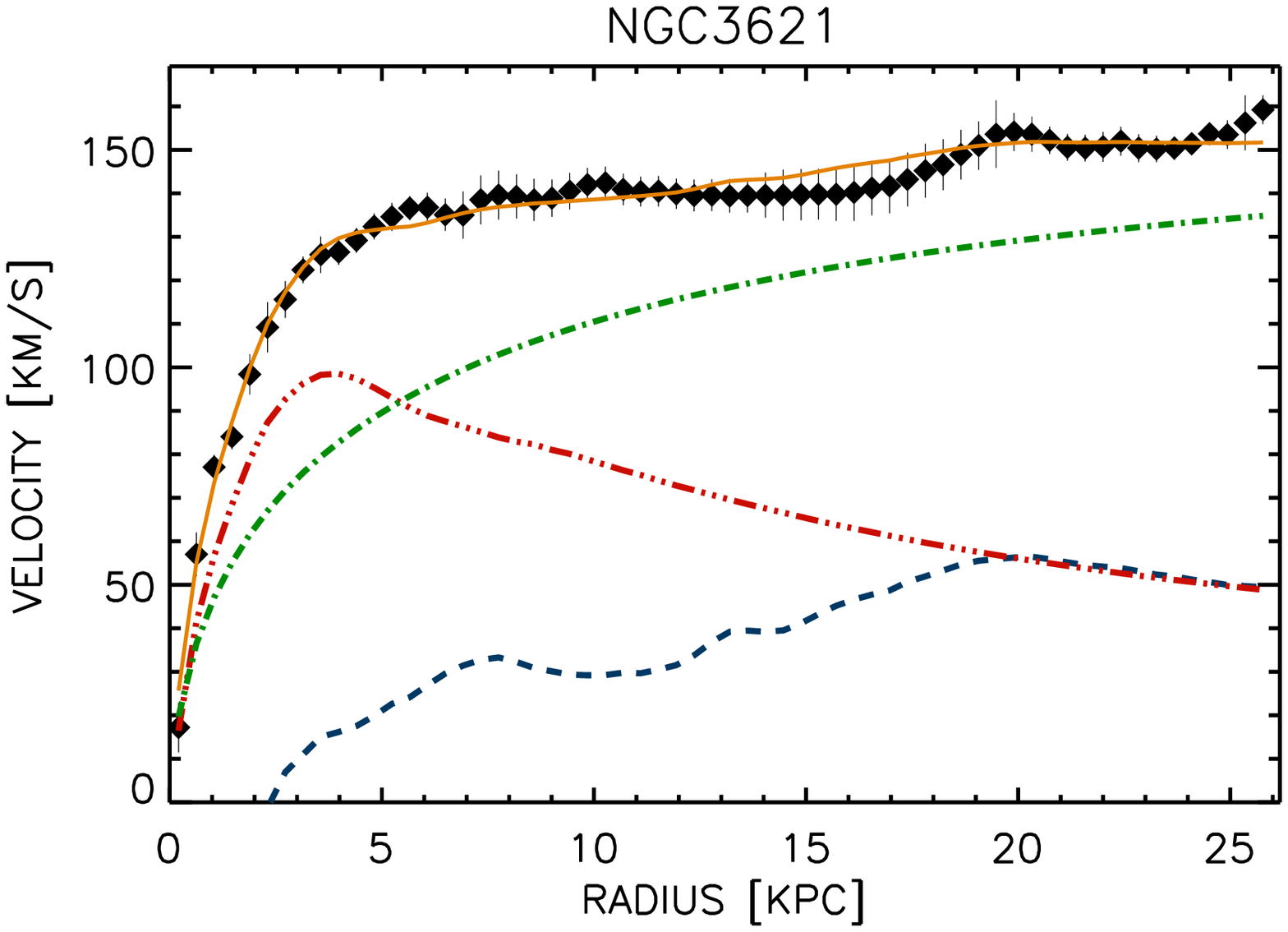}\includegraphics[width=0.25\textwidth]{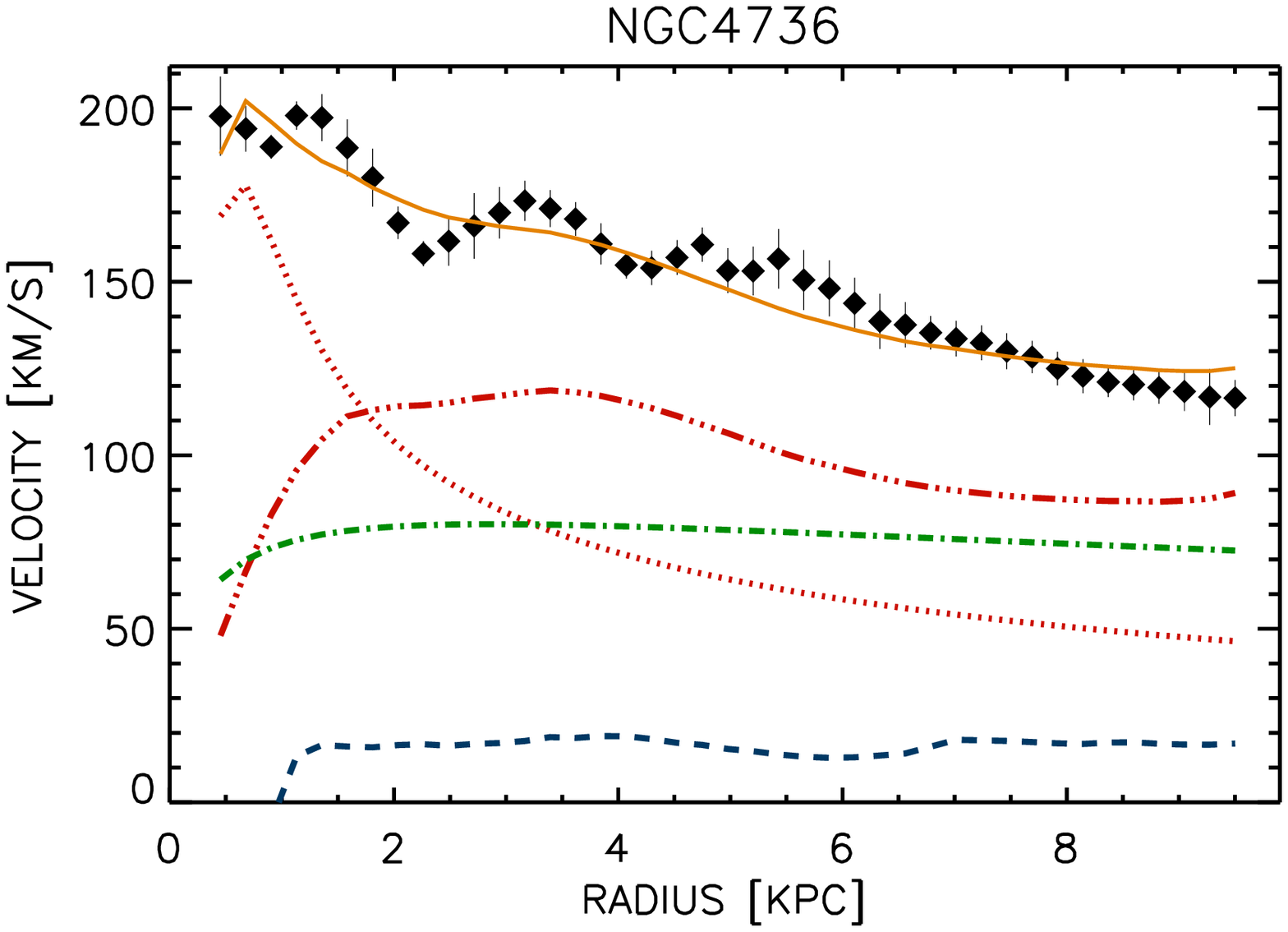}\includegraphics[width=0.25\textwidth]{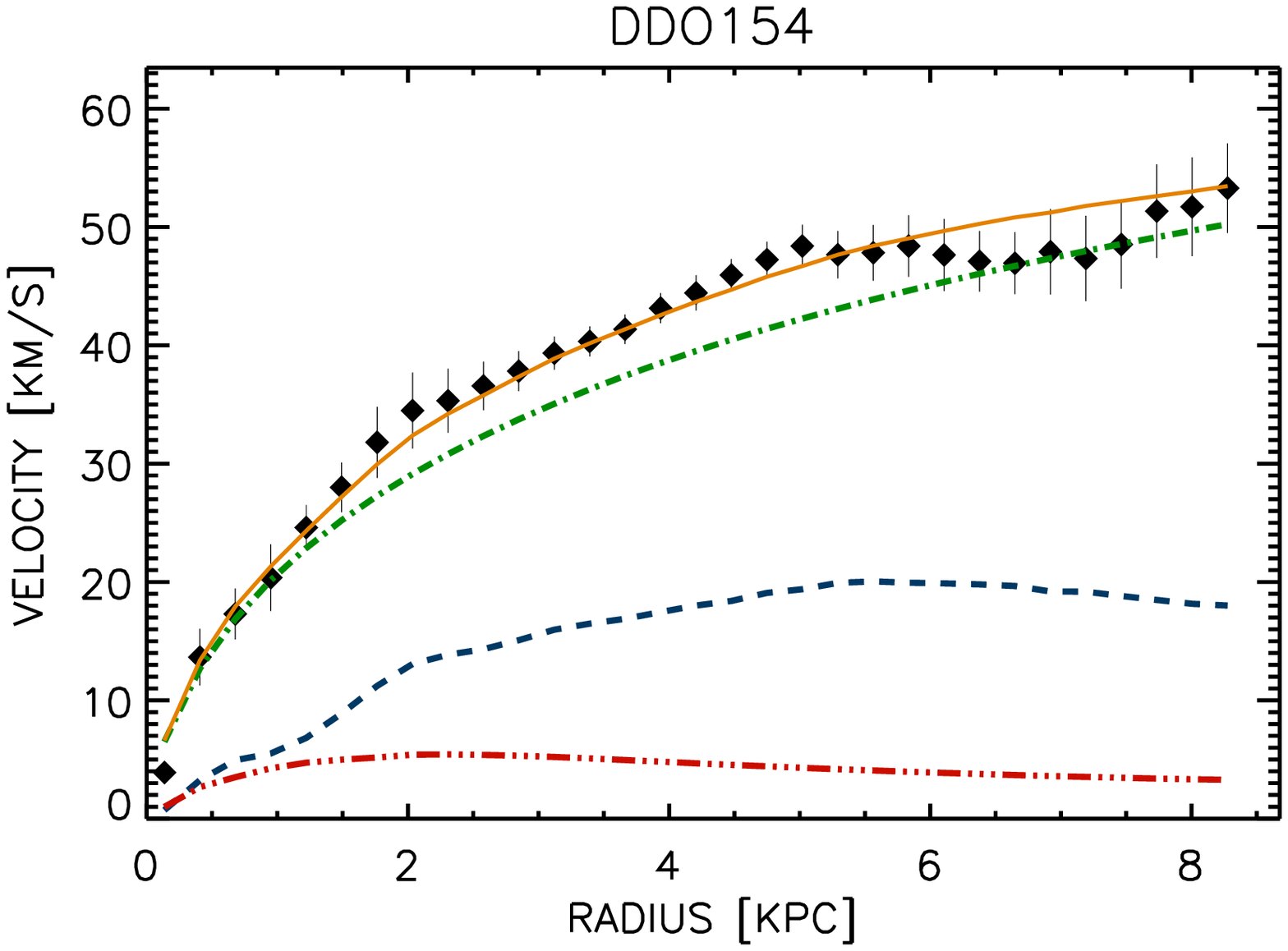}\includegraphics[width=0.25\textwidth]{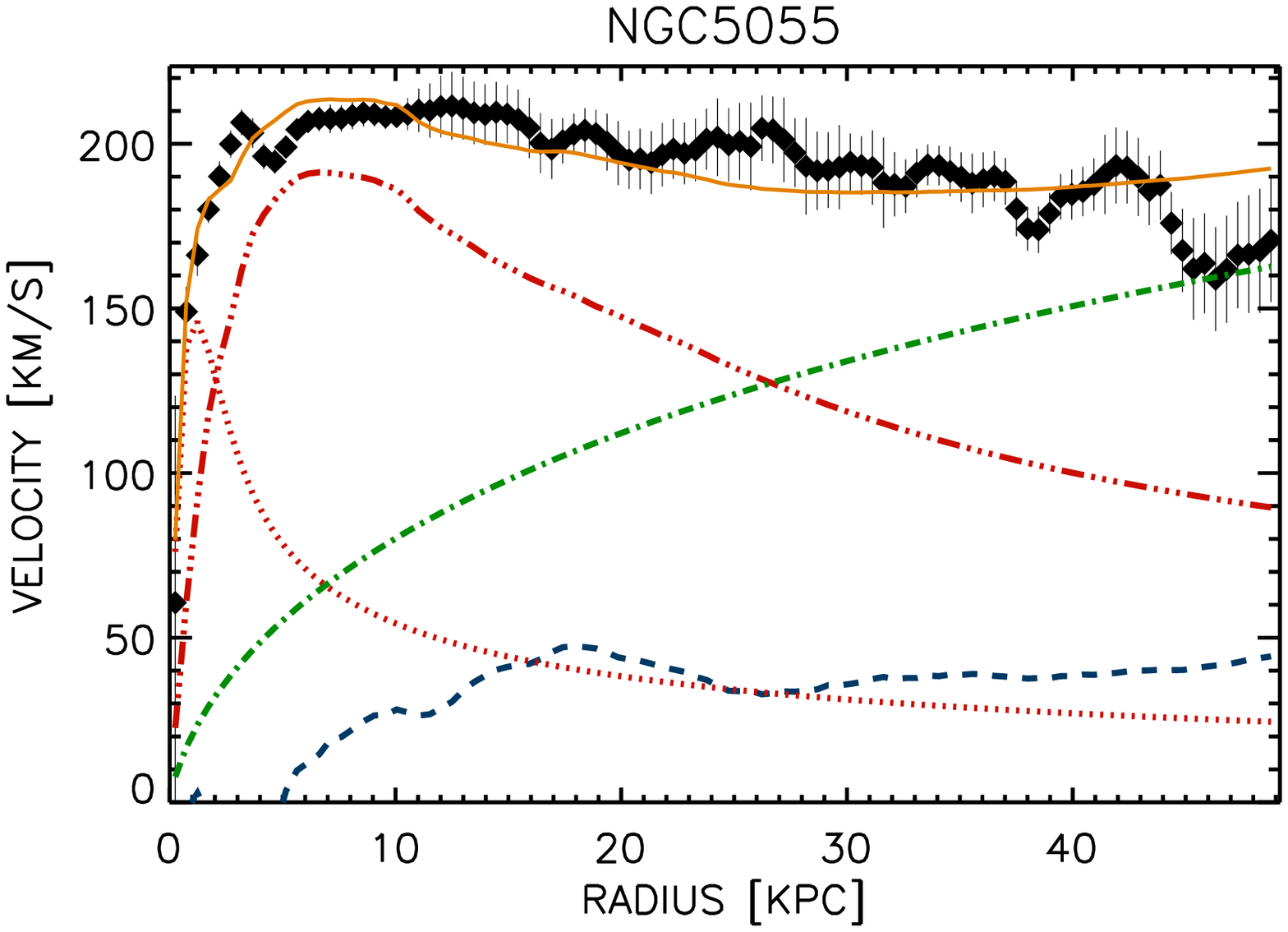}
\includegraphics[width=0.25\textwidth]{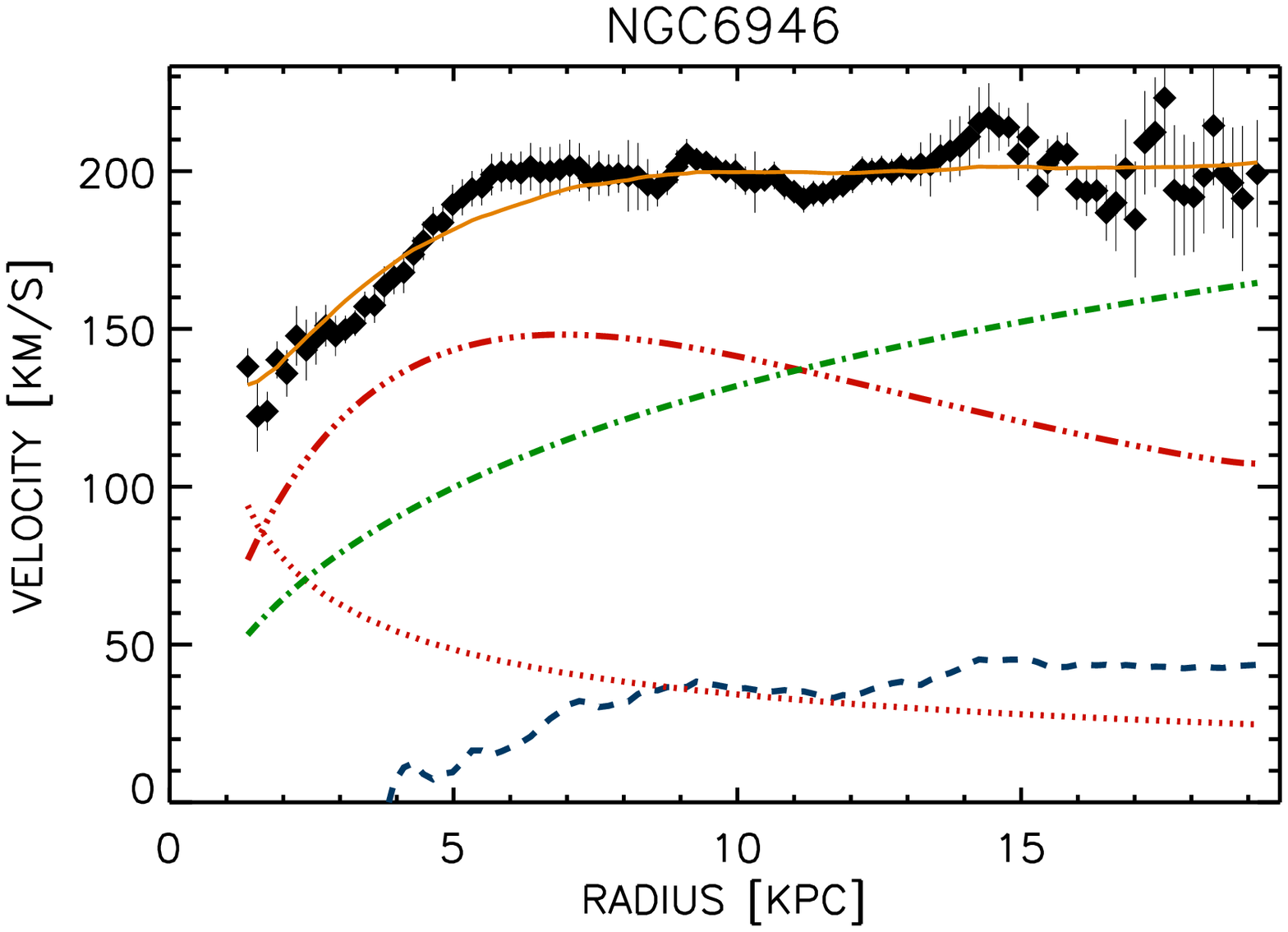}\includegraphics[width=0.25\textwidth]{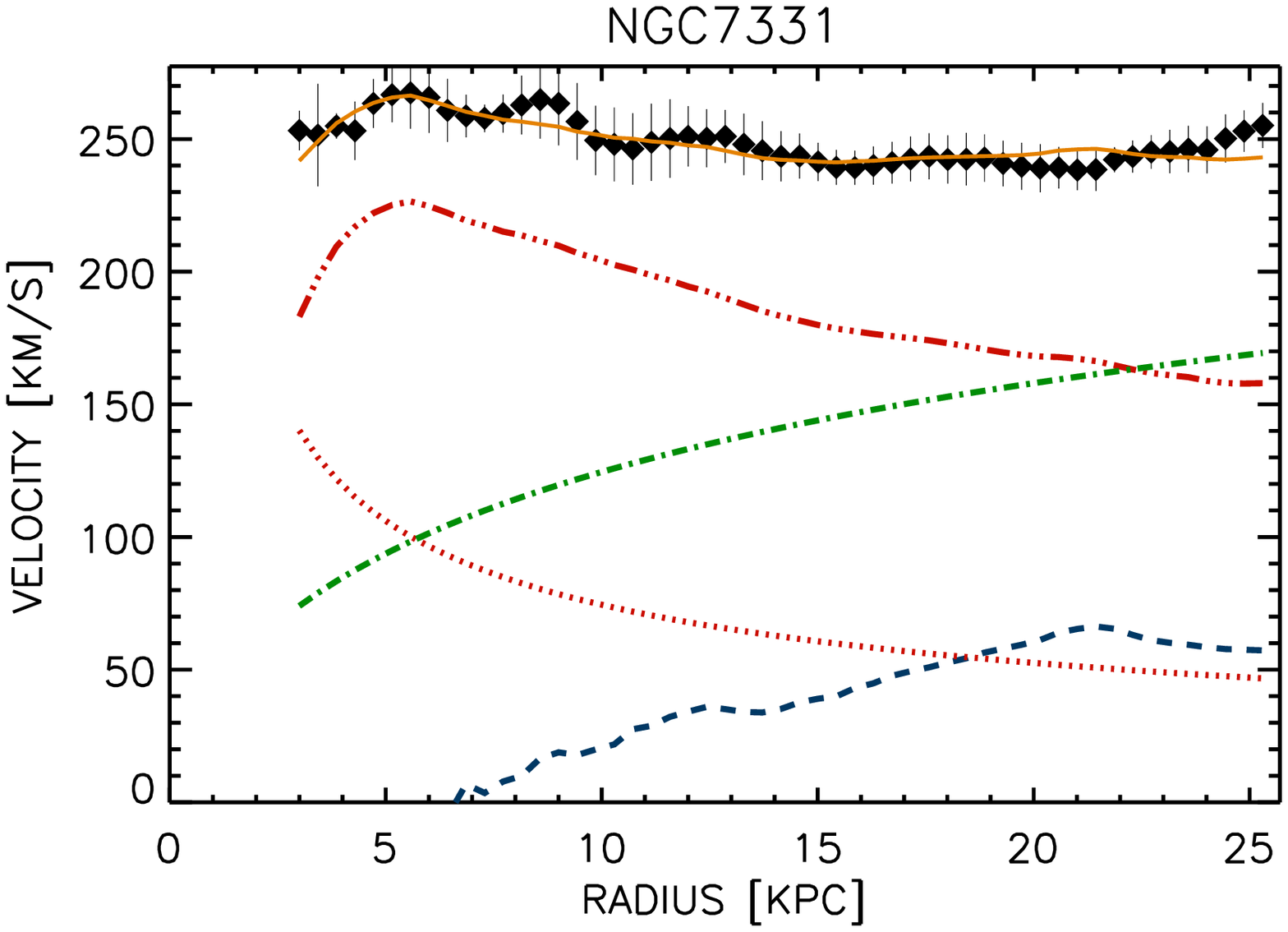}\includegraphics[width=0.25\textwidth]{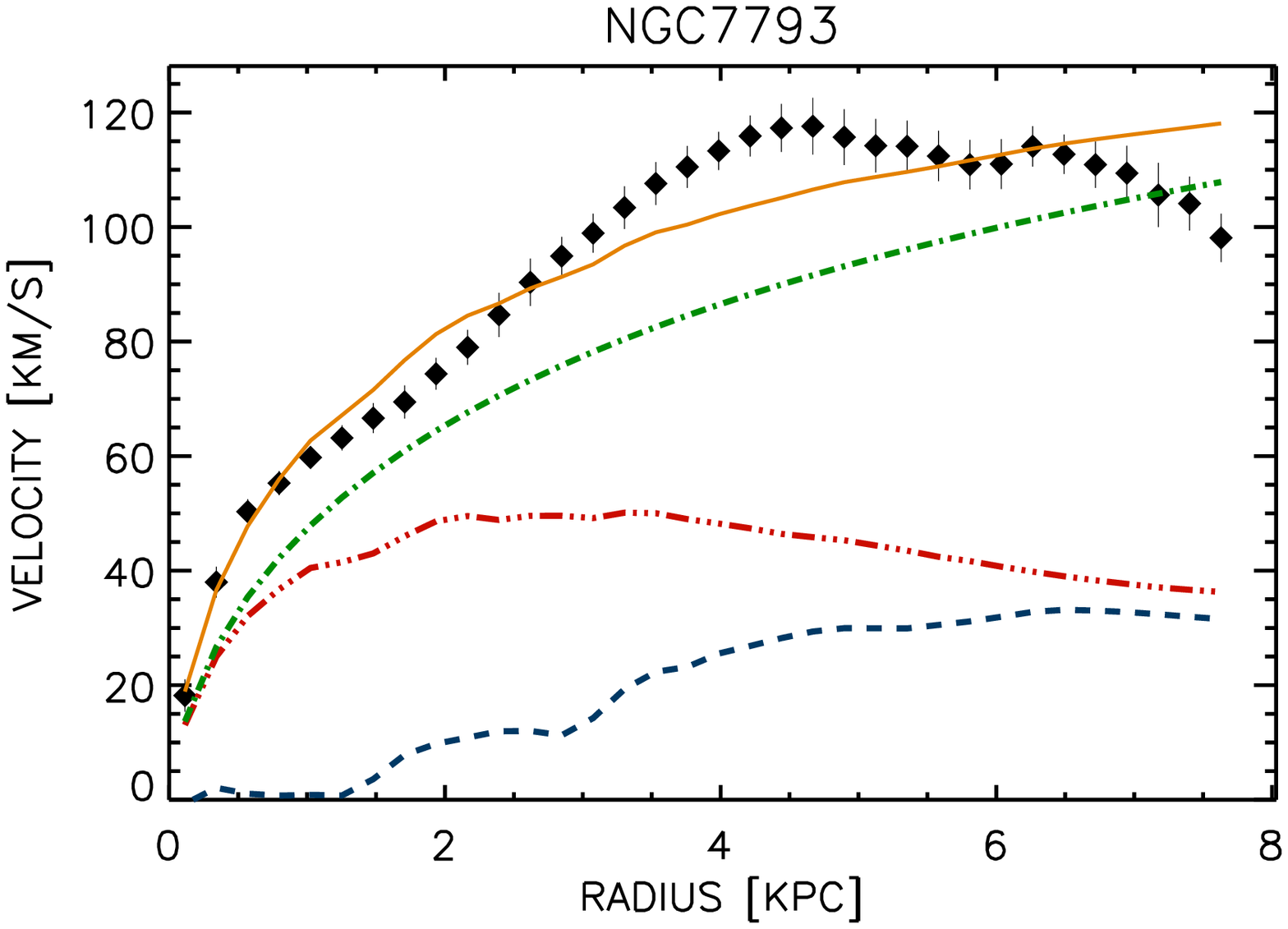}\includegraphics[width=0.25\textwidth]{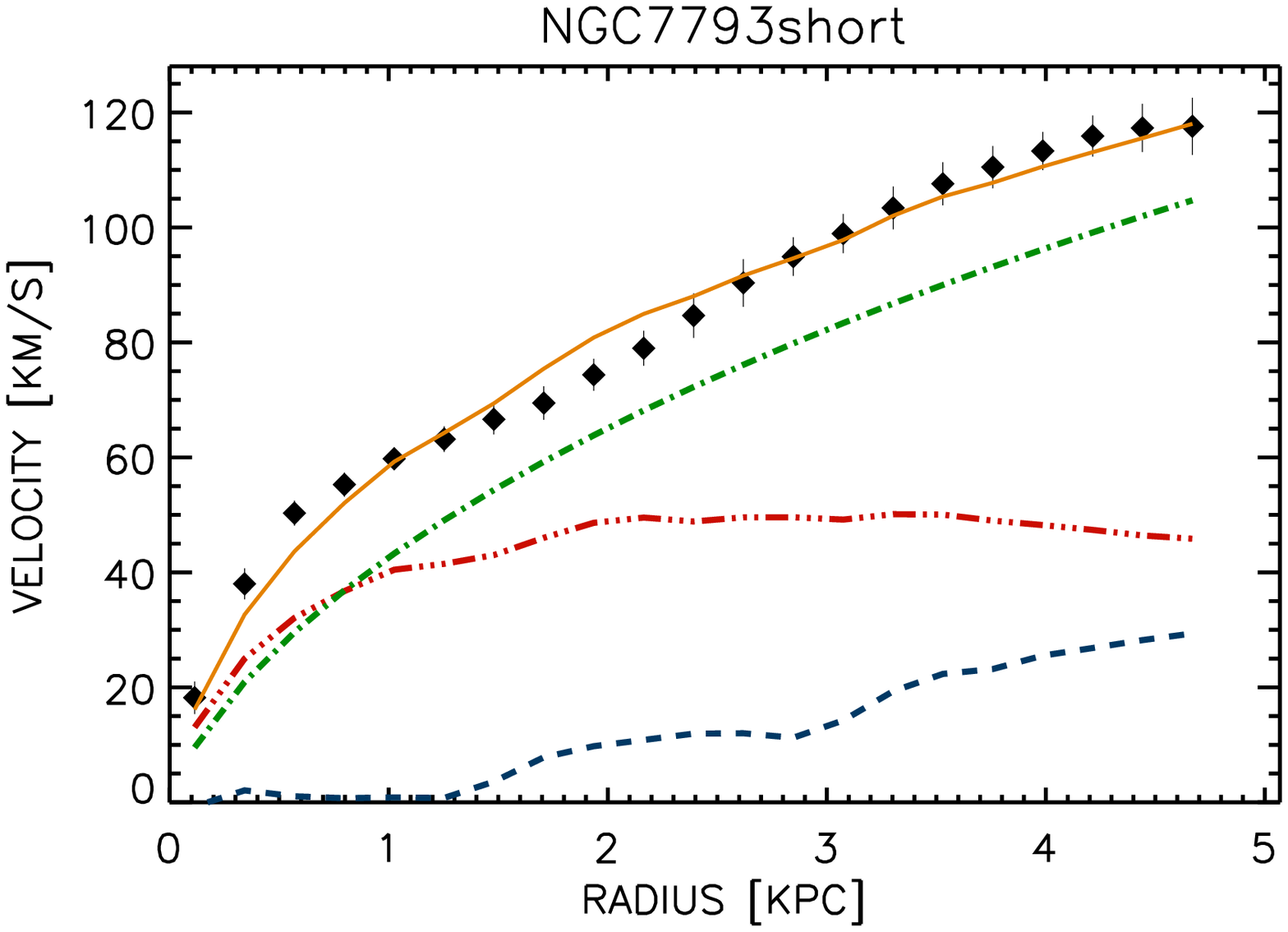}
 \caption{Same as Fig.~\ref{fig:rc-kroupa} but with a fixed  index $n=6.2$.  }
 \label{fig:rc-kroupa-fix}
 \end{center}
 \end{figure*}
%%%%%%%%%%%%%%%%%%%%%%%%%%%%%%%%%%%%%%%%%%%%%%%%%%%%%%%%%%%
 
Another interesting point  concerns the results for the few dwarf and low surface
brightness galaxies DDO 154 and NGC 2366.  Due to their  dark matter dominance  they are ideal
candidates for a direct comparison with $\Lambda$CDM low mass halos
($10^{9}\ h^{-1}$ \msol). They perfectly agree with  simulated
dwarf galaxy halos in the density-radius parameter space   (the two
closest symbols to the red triangles in Figs~\ref{fig:haloparam1} and
~\ref{fig:haloparam2}) but they  have   a low Einasto index, contrary
to all simulated  dwarf-galaxy-sized  halos.  

\subsection{Cuspy Einasto models with a fixed index}
\label{sec:cuspyeinasto}

\begin{figure}[ht]
 \begin{center}
\includegraphics[width=0.5\columnwidth]{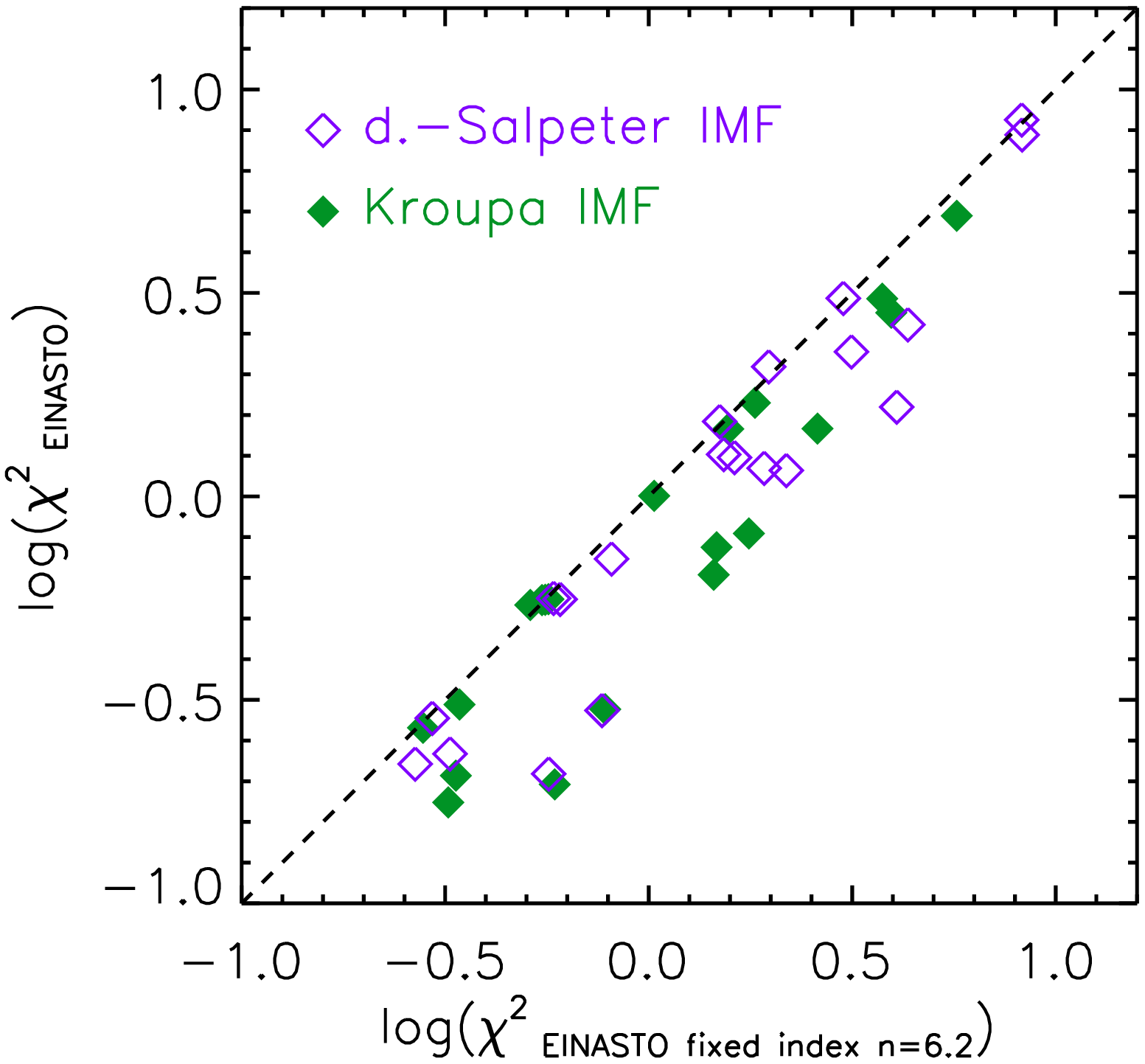}\includegraphics[width=0.5\columnwidth]{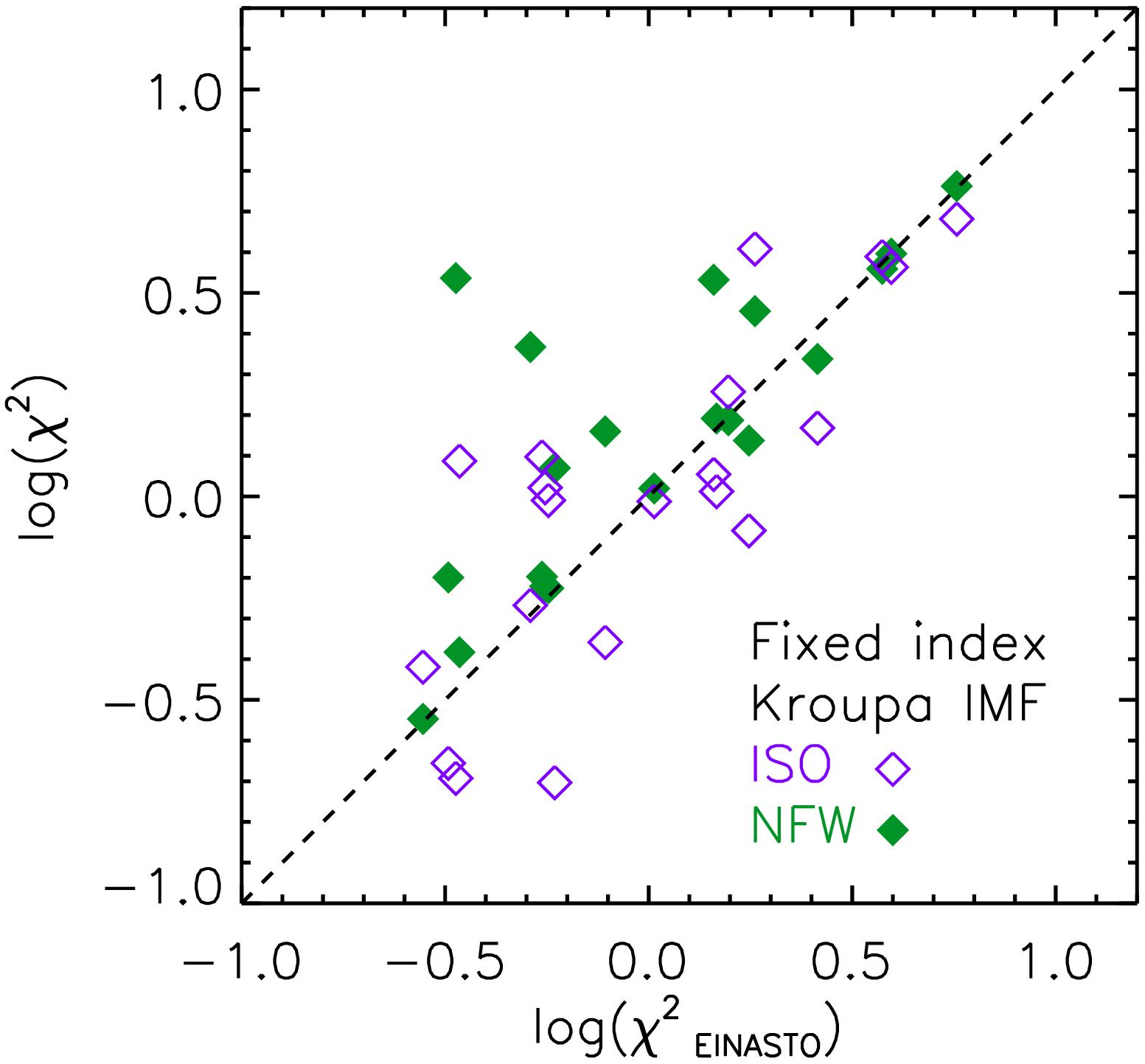}
 \caption{Comparison of reduced $\chi^2$ for
   Einasto halo models done at  fixed index ($n=6.2$) with Einasto halo models done at free index (\emph{left}). Open and filled symbols 
   are for the diet-Salpeter and Kroupa IMFs, respectively.   
   Comparison of reduced $\chi^2$ for
   $n=6.2$ Einasto halo models with the NFW and Iso halos
   (respectively filled and open symbols)  with the Kroupa IMF (\emph{right}). See Tab.~\ref{tab:resfit3} for results obtained 
   with the diet-Salpeter IMF.}
 \label{fig:compredchi2fixfree}
 \end{center}
 \end{figure}

\begin{figure}[!t]
 \begin{center}
\includegraphics[width=0.75\columnwidth]{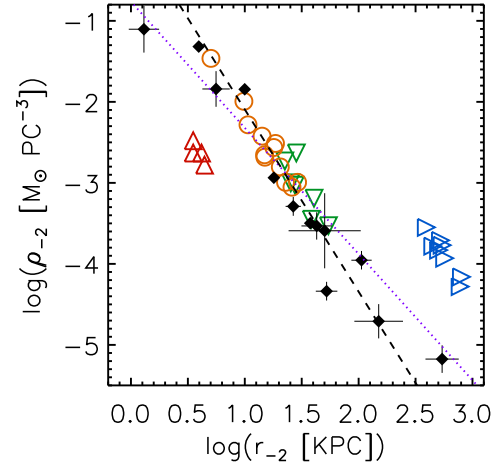}
 \caption{Parameter space of Einasto halos with a fixed index ($n=6.2$) for the Kroupa IMF. See Tab.~\ref{tab:resfit3} for results obtained 
   with the diet-Salpeter IMF.   Results for the few halos having an 
 extremely large scale radius are not shown for clarity. Colored symbols correspond to the same cosmological halos 
 as in Figs.~\ref{fig:haloparam1} and~\ref{fig:haloparam2}. Additional blue, open triangles correspond to simulated 
 cluster-sized halos from \citet{nav04}. A dashed line is a power law fit    to the relationship $\log(\rho_{-2}) = (-2.25 \pm 0.04)\log(r_{-2})+(0.16 \pm 0.05)$. 
 As a comparison, a dotted line represents the  linear fit to the relationship obtained 
 from Einasto models done at free Einasto index (see \S\ref{sec:nonuniversality} and Fig.~\ref{fig:haloparam2}).}
 \label{fig:compredchi2fix}
 \end{center}
 \end{figure}

We now  investigate whether Einasto halos with an index close to the
typical index seen in cosmological simulations agree better with the
observations than the NFW   and Iso  halos.  The
\hi\ RCs have therefore been fit with a Einasto fixed
value ($n = 6.2$)  model.  These fits have two free
parameters, like   the NFW and core halos.  The results for both
IMFs are reported in Tab.~\ref{tab:resfit3} and the RC
decompositions are displayed in Fig.~\ref{fig:rc-kroupa-fix} (Kroupa
IMF only).   Figure~\ref{fig:compredchi2fixfree} compares the quality of fits 
between the fixed Einasto index  models to the free Einasto index models (left-hand panel), and 
 to the NFW and Iso models (right-hand panel), whereas 
 Figure~\ref{fig:compredchi2fix} displays  the  parameter  space $[\rho_{-2},r_{-2}]$ obtained for $n = 6.2$.

 %%%%%%%%%%%%%%%  table EINASTO  FIXED INDEX -diet Salpeter - Kroupa %%%%%%%%%%%%%%%%%%%%
\begin{deluxetable*}{l||cll|cll}
\tablecaption{Fit parameters of the Einasto halo to THINGS galaxies for a fixed index $n=6.2$ and mass-to-light ratios derived using a diet-Salpeter (left panel) and Kroupa (right
panel) IMFs.}
\tablehead{
   &    &   diet-Salpeter  &   &   & Kroupa  &   \\ 
 Galaxy  &  $\chi^2_r$ &  $ \rho_{-2}$ & $r_{-2}$ &  $\chi^2_r$ &  $ \rho_{-2}$ &  $r_{-2}$ \\
  &    & \densunit& (kpc)&       &  \densunit &  (kpc)  }
\startdata
    NGC925 &  2.2 &  1.2$\times 10^{-5}$ $\pm$ 8.0$\times 10^{-7}$ & $>1000$  &  1.4 & 1.8$\times 10^{-4}$ $\pm$ 1.5$\times 10^{-4}$ &  1.6$\times 10^5$ $\pm$  3.3$\times 10^5$ \\ 
   NGC2366 &  0.6 &  1.8$\times 10^{-2}$ $\pm$ 8.9$\times 10^{-3}$ &  159.0 $\pm$   80.9 &  0.6 & 2.0$\times 10^{-2}$ $\pm$ 9.5$\times 10^{-3}$ &  150.0 $\pm$   73.8 \\ 
  NGC2403d &  0.6 & 5.8$\times 10^{-1}$ $\pm$ 5.0$\times 10^{-2}$ &   26.7 $\pm$    1.5 &  0.6 & 1.2    $\pm$ 8.7$\times 10^{-2}$ &   17.8 $\pm$    0.8 \\ 
   NGC2403 &  0.6 & 5.5$\times 10^{-1}$ $\pm$ 4.9$\times 10^{-2}$ &   27.7 $\pm$    1.6 &  0.6 & 1.1    $\pm$ 8.5$\times 10^{-2}$ &   18.2 $\pm$    0.8 \\ 
   NGC2841 &  0.3 & 3.6 $\pm$ 2.1$\times 10^{-1}$ &   18.9 $\pm$    0.6 &  0.3 & 14.3  $\pm$ 7.7$\times 10^{-1}$ &   10.0 $\pm$    0.3 \\ 
   NGC2903 &  0.3 &  31.6 $\pm$  2.5 &    4.7 $\pm$    0.2 &  0.3 & 48.1  $\pm$    3.9 &	3.9 $\pm$    0.1 \\ 
   NGC2976 &  2.0 & 4.7$\times 10^{-5}$ $\pm$ 1.3$\times 10^{-5}$ & $>1000$ &  0.5 & 4.2$\times 10^{-4}$ $\pm$ 3.5$\times 10^{-4}$ &  3.6$\times 10^{6}$ $\pm$  1.7$\times 10^{7}$ \\ 
   NGC3031 &  4.3 & 1.2$\times 10^{-2}$ $\pm$ 2.2$\times 10^{-2}$ &  417.0 $\pm$  760.0 &  3.7 & 14.4  $\pm$    7.3 &	5.6 $\pm$    1.5 \\ 
  NGC3198d &  1.9 & 1.1$\times 10^{-1}$ $\pm$ 3.6$\times 10^{-2}$ &   62.9 $\pm$   13.6 &  1.8 & 5.3$\times 10^{-1}$ $\pm$ 1.3$\times 10^{-1}$ &   26.1 $\pm$    3.6 \\ 
   NGC3198 &  3.2 & 9.7$\times 10^{-2}$ $\pm$ 4.2$\times 10^{-2}$ &   66.9 $\pm$   18.9 &  2.6 & 4.9$\times 10^{-1}$ $\pm$ 1.5$\times 10^{-1}$ &   27.2 $\pm$    4.7 \\ 
    IC2574 &  0.3 & 2.4$\times 10^{-4}$ $\pm$ 8.0$\times 10^{-5}$ &  $>1000$ &  0.3 & 4.5$\times 10^{-4}$ $\pm$ 1.3$\times 10^{-4}$ &   9170 $\pm$   4520 \\ 
   NGC3521 &  8.3 & 4.2$\times 10^{-6}$ $\pm$ 1.5$\times 10^{-6}$ & $>1000$ &  5.7 & 2.6$\times 10^{-1}$ $\pm$ 2.7$\times 10^{-1}$ &   50.3 $\pm$   36.7 \\ 
   NGC3621 &  0.8 & 3.9$\times 10^{-2}$ $\pm$ 6.7$\times 10^{-3}$ &  147.0 $\pm$   19.4 &  0.5 & 3.2$\times 10^{-1}$ $\pm$ 3.1$\times 10^{-2}$ &   37.6 $\pm$    2.3 \\ 
   NGC4736 &  1.6 & 3.3$\times 10^{-1}$ $\pm$    1.1 &   12.1 $\pm$   24.8 &  1.6 & 78.8  $\pm$  52.2 &	1.3 $\pm$    0.4 \\ 
    DDO154 &  0.8 & 4.4$\times 10^{-2}$ $\pm$ 1.2$\times 10^{-2}$ &   54.1 $\pm$   11.8 &  0.8 & 4.6$\times 10^{-2}$ $\pm$ 1.2$\times 10^{-2}$ &   52.3 $\pm$   11.4 \\ 
   NGC5055 &  8.3 & 3.7$\times 10^{-6}$ $\pm$ 3.2$\times 10^{-7}$ & $>1000$ &  1.5 & 6.7$\times 10^{-3}$ $\pm$ 2.6$\times 10^{-3}$ &  541.0 $\pm$  182.0 \\ 
   NGC6946 &  1.5 & 6.2$\times 10^{-5}$ $\pm$ 5.0$\times 10^{-5}$ & $>1000$ &  1.0 & 1.1$\times 10^{-1}$ $\pm$ 2.9$\times 10^{-2}$ &  106.0 $\pm$   21.8 \\ 
   NGC7331 &  3.0 & 6.4$\times 10^{-6}$ $\pm$ 6.7$\times 10^{-7}$ & $>1000$ & 0.3 & 9.2$\times 10^{-2}$  $\pm$ 2.7$\times 10^{-2}$   & 112.0	  $\pm$  24.4    \\ 
   NGC7793 &  4.1 & 1.1$\times 10^{-1}$ $\pm$ 5.3$\times 10^{-2}$ &   89.2 $\pm$   39.1 &  3.9 & 2.9$\times 10^{-1}$ $\pm$ 1.1$\times 10^{-1}$ &   42.8 $\pm$   13.2 \\ 
  NGC7793s &  1.5 & 5.4$\times 10^{-3}$ $\pm$ 3.0$\times 10^{-3}$ &   $>1000$ &  1.8 & 4.1$\times 10^{-2}$ $\pm$ 1.8$\times 10^{-2}$ &  324.0 $\pm$  162.0 
 \enddata
 \label{tab:resfit3}
 \end{deluxetable*}
 %%%%%%%%%%%%%%% end table EINASTO  FIXED INDEX -diet Salpeter - Kroupa %%%%%%%%%%%%%%%%%%%%

 %%%%%%%%%%%%%%%%%%%%% Figure Comparison of reduced chi2 Fixed vs Free index (left panel) - Fixed index vs NFW-Iso Kroupa only (right) %%%%%%%%%%%%%%%%%%%%%%%
%%%%%%%%%%%%%%%%%%%%%%%%%%%%%%%%%%%%%%%%%%%%%%%%%%%%%%%%%%%

 As seen in Fig.~\ref{fig:compredchi2fixfree}, the fits performed with a   fixed index  
$n = 6.2$ are  significantly  worse than those with a free index, regardless of the IMF
($84\%$ and $89\%$ of our sample for the diet-Salpeter and Kroupa IMFs,
respectively).   They also provide  worse  results than the
Iso  model, regardless of the IMF ($65\%$ and $55\%$
for the diet-Salpeter and Kroupa IMFs, respectively),
 and better results than the NFW  model fits,
regardless of the IMF ($75\%$ and $80\%$ for the diet-Salpeter and
Kroupa IMFs, respectively).

The  parameter  space of the characteristic scale densities and
radii still shows a  correlation (Fig.~\ref{fig:compredchi2fix}), 
but with more scatter than seen in Figs.~\ref{fig:haloparam1} and~\ref{fig:haloparam2} for fits done with a free index.  
 The striking feature is the larger range of densities and radii than for the fits performed with free indices.
 The relation between $r_{_2}$ and $\rho(r_{-2})$ is steeper than the previous one given in
  \S\ref{sec:nonuniversality} and Fig.~\ref{fig:haloparam2}. Besides the fact that
extreme halos of low density and large extent appear unrealistic for
galaxies, the agreement with the expected galaxy-size range seen in
cosmological simulations becomes worse and several fit halos
exceed the cluster-sized regime.  We therefore conclude that Einasto
halos with a fixed index $n \sim 6$ are ruled out by the analysis of
RCs of galaxies.

 %%%%%%%%%%%%%%%%%%%%% Figure Einasto Parameter space fixed index Kroupa IMF %%%%%%%%%%%%%%%%%%%%%%%
%%%%%%%%%%%%%%%%%%%%%%%%%%%%%%%%%%%%%%%%%%%%%%%%%%%%%%%%%%%

 \section{Two-parameter Einasto models}
\label{sec:twoparam}
 In the previous section we have studied two-parameter, fixed index
Einasto halos that were consistent with CDM, but noted they were 
not consistent with the observations. In this section, we try to identify
other two-parameter Einasto halos and ask whether there are any
Einasto models that can describe the observed halos better than the
Iso or  NFW halo models.  Before answering these
questions, we  emphasize  that our galaxy sample 
does not pretend to be representative of the whole diversity of
spiral galaxies in the local Universe.  Nevertheless, in this section,
we aim at finding a family of generic Einasto halos that reproduce
fairly well the RCs of THINGS spiral galaxies, at least
as good as the usual core or cosmological cuspy models.
 
 \subsection{Cuspy Einasto halos versus the NFW halo}
 \label{sec:twoparam1}
We estimate the critical index value where the NFW halo becomes better
than the Einasto halo for the majority of the current sample.  We find
$n_{\rm up} \sim 10-11$ for the diet-Salpeter IMF and $n_{\rm up} \sim
11-12$ for the Kroupa IMF. Those values have been derived by
increasing progressively the (fixed) index in the fits and by
 identifying  the turnover index where more than half of the NFW
fits from the sample becomes better than the Einasto fits.  
 Note that those models represent halos which are less and less dense, and larger as
the index increases.    Another advantage of  these  models with
respect to the NFW model is the better agreement with $\Lambda$CDM
simulations. 

We repeat the same procedure  but
towards smaller  indices and find $n_{\rm low} \sim 0.5$ (diet-Salpeter IMF) 
and $n_{\rm low}\sim 1$ (Kroupa IMF) as the  values where the majority of
constrained Einasto fits turn better than NFW ones. This result is completely expected 
because small indices provide core halos, whose halo shape is naturally better for the sample 
than any cusp.

 \subsection{Core Einasto halos versus the pseudo-isothermal sphere} 
 \label{sec:twoparam2}
 In contrast with the   cuspy  halos,  we find  the cored ($n<4$) halos at
the low end of the Einasto index distribution.  We investigate the
goodness of fit of two-parameter Einasto models performed with a fixed, small 
index   to find a critical index where the
majority of such constrained Einasto fits become worse than those
performed with the Iso model.  It turns out a low limit cannot be obtained for the diet-Salpeter IMF; the 
Iso model is always a better fit than the constrained Einasto model at a level of $\sim$70\%, which is likely 
explained by the relatively scattered index distribution, as seen  in \S\ref{sec:nonuniversality}. 
In other words it is almost impossible to find an index that would satisfy most of all halos at the same 
time for that IMF. It is not the case for the Kroupa IMF for which $n_{\rm low} \sim 2$ indicates
the turnover index. Indeed, 65\%, 55\% and 45\% of all the sample 
is better fit by the Iso model than $n=1$, $n=2$ and $n=3$ (respectively) Einasto fits.  
Notice this value appears  consistent with the mean index $\bar{n} = 1.4 \pm 0.8$ of non-cuspy
halos fit under the Kroupa IMF assumption
(\S\ref{sec:nonuniversality}). 

For larger indices $n_{\rm up} \sim 5$ is the limit  where most of
constrained Einasto fits become worse than Iso ones. Past this value,  
the core nature of the Iso model starts to dominate the cuspy nature of these Einasto
 halos.  The index range $2  \leq  n  \leq 4$ of constrained Einasto core
models is therefore  small compared with that of
constrained Einasto cusps. 

\subsection{Towards a new two-parameter density profile}
Both results of \S\ref{sec:twoparam1} and \S\ref{sec:twoparam2} make us conclude that 
the family of two-parameter Einasto models with $4 < n \leq 12$ is more suited 
for the modeling of observational cuspy halos  than is the NFW model, 
whereas the Einasto model with $2 \leq n \leq 4$ is more suited for 
the modeling of cored halos than is the isothermal model.
While allowing the variation of the Einasto index is essential in finding the best fit models to the THINGS RCs 
(\S\ref{sec:comparisoneinastoisonfw}, \S\ref{sec:halofamily} and \S\ref{sec:compsimu}), those  
fits at fixed indices point out the exponential shape of the density profile 
also contributes in better modeling the mass distribution of the galaxies. 
We thus suggest that a new  shape of dark matter density, where the volumic density profile 
 decreases exponentially as a function of radius following Eq.~\ref{eq:rhoeinasto},  may be a good 
  alternative to the usual mass divergent, pseudo-isothermal and NFW models, 
  with Einasto indices chosen to match the particular type of halo (core or cusp) one wants to model. 
   
\section{Conclusions}
\label{conclusion}
For the first time, the Einasto dark matter halo has been applied to model 
the mass distribution of  nearby and undisturbed galactic disks. 
Our results show  that the Einasto model is a good fit of the dark matter contribution to  highly-resolved \hi\ rotation curves of galaxies, better
than other models like the (cored) Iso model or the (cuspy) NFW model. 
The significant  improvement  is caused by the
parameter that shapes the radial behaviour of the mass 
density profile, the Einasto index. 
The big advantage of the Einasto formalism  over the NFW cusp or the pseudo-isothermal sphere 
lies on its third parameter, the index, which allows for extra freedom in describing different shapes of density profiles. 
 The larger (respectively, smaller) is the Einasto index, the cuspier (shallower) is the halo in its central parts.
 Since there is  no  unique index
describing all halos, it is not possible to scale the density profile from one
galaxy to another with a  different mass. In other words, we find no
evidence for a universal dark matter Einasto halo, at least in the
galaxy sample we have analyzed.  This result corroborates the one
found in dissipationless $\Lambda$CDM simulations. Noteworthy however is the fact that 
the index appears to scale with the halo mass  for halos more  
massive than $M_{200} \sim  2 \times 10^{11}\ h^{-1}$ \msol. 

We  find  that the Einasto index is generally smaller than the one
inferred from cosmological simulations, by a factor of two or more.  
This discrepancy between observed and simulated Einasto indices  extends  the core-cusp
controversy (Iso versus NFW) to the Einasto formalism.  A $\Lambda$CDM
Einasto halo is cuspier than an observed galactic Einasto halo. In 
our sample, galaxies with a cuspy (i.e. with $n > 4$) Einasto halo are not frequent.
 It is interesting to note that our sample consists of both
normal, bright galaxies and of a few dwarf, low surface density
galaxies, while the  cusp-core controversy usually deals with small
and low surface density galaxies only.   Cored dark matter halos are thus found
 in galaxies whose  baryonic content significantly differs, for 
halos less massive than $M_{200} \sim  2 \times 10^{11}\ h^{-1}$ \msol.

 The RC decomposition has also shown that  
cuspy Einasto halos with fixed index $4 < n \leq 12$ give a better description of RCs than
does the NFW  model.  Conversely core Einasto models with fixed index $2  \leq  n  \leq 4$  give
  better RC fits than the Iso model. It implies that the exponential decrease of the Einasto dark matter density  is 
  also responsible for the improvement in modeling the mass distribution of galaxies, in addition to the effect of the Einasto index. 
   Those constrained Einasto models can be a good alternative to
   Iso or NFW models for galactic dynamics, 
   but we re-iterate here that Einasto models with free index are preferable over any of those two-parameter models.

The analysis has  shown that the stellar masses derived with the \citet{bel01} prescription under a diet-Salpeter IMF hypothesis 
are in conflict with the Einasto model. This is not the case for the Kroupa IMF. In future papers from this series, we will perform mass models
 with free stellar masses and with other fixed stellar mass scalings, like the one described in \citet{cha03}. 
 Additional improvements of dynamical models will take into account 
  the molecular gas content, where available. 
     
 Furthermore we will investigate possible correlations between the properties of Einasto dark matter halos and baryons. 
 The mass distribution for a large sample of low surface density disks will also be studied. 
 We expect that those galaxies exhibit low Einasto indices because of the 
shallow nature of the density profile in their core, but to which extent  
their values compare with those of normal, higher surface brightness disks like most 
of our current THINGS subsample is not  known yet. Both these samples will help us to measure the variation of dark halo inner shape  
as a function of galaxy mass.

We finally emphasize the importance of
testing the Einasto halo model for objects at extreme ends of the
galaxy luminosity function (dwarf spheroidals and giant ellipticals).
This will be crucial for getting a complete description of the
observational behaviour of Einasto models over several decades of halo mass and comparing with 
expectations from $\Lambda$CDM simulations at similar cosmological scales. 

\acknowledgments  We are very grateful to the referee, Matt Bershady, for helpful comments and suggestions.  Laurent Chemin acknowledges 
financial support from RadioNet-FP7 and R\'egion Aquitaine. The work of WJGdB is based upon research supported by
the South African Research Chairs Initiative of the Department of
Science and Technology and National Research Foundation.


\begin{thebibliography}{}
  \bibitem[Akaike(1974)]{aka74} Akaike, H.,  1974, IEEE Trans. Automatic Control,  19, 716
  \bibitem[Barnes et al.(2004)]{bar04} Barnes, E. I., Sellwood, J. A., Kosowsky, A., 2004, \aj, 128, 2724
  \bibitem[Bell \& de Jong(2001)]{bel01} Bell, E. F., \& de Jong, R. S., 2001, \apj, 550, 212
  \bibitem[Cardone, Piedipalumbo \& Tortora(2005)]{car05} Cardone, V.F., Piedipalumbo, E., \& Tortora, C. 2005, \mnras, 358, 1325
  \bibitem[Chabrier(2003)]{cha03} Chabrier, G., 2003, \pasp, 115, 763
  \bibitem[Chemin et al.(2009)]{che09} Chemin, L., Carignan, C. \& Foster, T. 2009, \apj, 705, 1395
   \bibitem[de Blok \& Bosma(2002)]{deb02} de Blok, W. J. G., \& Bosma, A. 2002, \aap, 385, 816
  \bibitem[de Blok(2004)]{deb04} de Blok, W. J. G., 2004, IAU Symp. 220, Dark Matter in Galaxies, eds. S. D. Ryder et al. (San Francisco: ASP), 69
  \bibitem[de Blok et al.(2008)]{deb08} de Blok, W. J. G.,  Walter, F., Brinks, E., Trachternach, C., Oh, S.-H., \&  Kennicutt, R. C., 2008, \aj, 136, 2648
  \bibitem[Diemand et al.(2004)]{die04} Diemand, J., Moore, B., Stadel, J., 2004, \mnras, 353, 624
  \bibitem[Donato et al.(2009)]{don09} Donato, F., Gentile, G., Salucci, P., Frigerio Martins, C., Wilkinson, M. I., 
  Gilmore, G., Grebel, E. K., Koch, A., et al., 2009, \mnras, 397, 1169 
  \bibitem[Einasto(1965)]{ein65} Einasto, J., 1965,  Trudy Inst. Astrofiz. Alma-Ata, 51, 87 
  \bibitem[Einasto(1968)]{ein68} Einasto, J., 1968, PTarO, 36, 414
  \bibitem[Einasto(1969)]{ein69} Einasto, J., 1969, Astronomische Nachrichten 291, 97
  \bibitem[Gao et al.(2008)]{gao08} Gao, L., Navarro, J. F., Cole S., Frenk, C. S., White, S. D. M., Springel, V., Jenkins, A., Neto, A. F., 2008, \mnras, 387, 536
  \bibitem[Governato et al.(2010)]{gov10} Governato, F., Brook, C., Mayer, L., Brooks, A., Rhee, G., Wadsley, J., Jonsson, P., Willman, B., et al., 2010, Nature, 463, 203 
  \bibitem[Graham et al.(2006)]{gra06} Graham, A. W., Merritt, D., Moore, B., Diemand, J., \& Terzi\'c, B., 2006, \aj, 132, 2701
  \bibitem[Jing \& Suto(2000)]{jin00} Jing, Y. P., \& Suto, Y., 2000, \apj, 529, L69
  \bibitem[Kennicutt et al.(2003)]{ken03} Kennicutt, R. C., et al. 2003, PASP, 115, 928
  \bibitem[Kormendy \& Freeman(2004)]{kor04} Kormendy, J., \& Freeman, K. C., 2004, IAU Symp. 220, 
  Dark Matter in Galaxies, eds. S. D. Ryder et al. (San Francisco: ASP), 377 
  \bibitem[Kormendy et al.(2009)]{kor09} Kormendy, J., Fisher, D. B., Cornell, M. E., Bender, R., 2009, \apjs, 182, 216
  \bibitem[Kregel et al.(2002)]{kre02} Kregel, M., van der Kruit, P. C., \& de Grijs, R. 2002, \mnras, 334, 646
  \bibitem[Kroupa(2001)]{kro01} Kroupa, P., 2001, \mnras, 322, 231
  \bibitem[Kuzio de Naray et al.(2006)]{kuz06} Kuzio de Naray, R., McGaugh, S. S., de Blok, W. J. G., \& 
  Bosma, A., 2006, \apjs, 165, 461
  \bibitem[Kuzio de Naray et al.(2008)]{kuz08} Kuzio de Naray, R., McGaugh, S. S., \& de Blok, W. J. G., 2008, \apj, 676, 920
  \bibitem[Leroy et al.(2008)]{ler08}  Leroy, A. K., Walter, F., Brinks, E., Bigiel, F., de Blok, W. J. G., 
  Madore, B., \& Thornley, M. D., 2008, \aj, 136, 2782
  \bibitem[{\L}okas \& Mamon(2001)]{lok01}  {\L}okas, E. L., \& Mamon, G. A., 2001, \mnras, 321, 155
  \bibitem[Macci\`o et al.(2008)]{mac08} Macci\`o, A. V., Dutton, A. A., \& van den Bosch, F. C., 2008, \mnras, 391, 1940
  \bibitem[Mamon \& {\L}okas(2005)]{mam05} Mamon, G.A., \& {\L}okas, E.L. 2005, \mnras, 362, 95
  \bibitem[Merritt et al.(2005)]{mer05} Merritt, D., Navarro, J. F., Ludlow, A., \& Jenkins, A. 2005, \apj, 624, 85
  \bibitem[Merritt et al.(2006)]{mer06} Merritt, D., Graham, A. W., Moore, B., Diemand, J., \& Terzi\'c, B. 2006, \aj, 132, 2685
  \bibitem[Moore et al.(1999)]{moo99} Moore, B., Quinn, T., Governato, F., Stadel, J., Lake, G., 1999, \mnras, 310, 1147
  \bibitem[Navarro, Frenk \& White(1996)]{nfw96} Navarro, J. F., Frenk, C. S., \& White, S. D. M. 1996, \apj, 462, 563
  \bibitem[Navarro, Frenk \& White(1997)]{nfw97} Navarro, J. F., Frenk, C. S., \& White, S. D. M. 1997, \apj, 490, 493
  \bibitem[Navarro et al.(2004)]{nav04} Navarro, J. F., Hayashi, E., Power, C.,
   Jenkins, A. R., Frenk, C. S., White, S. D. M., Springel, V., Stadel, J., et al. 2004, \mnras, 349, 1039
  \bibitem[Navarro et al.(2010)]{nav10} Navarro, J. F., Ludlow, A., Springel, V., Wang, J., Vogelsberger,  M., 
   White, S. D. M., Jenkins, A. R., Frenk, C. S.,  et al. 2010, \mnras, 402, 21
  \bibitem[Neto et al.(2007)]{net07} Neto, A. F., Gao, L., Bett, P., Cole, S., Navarro, J. F., Frenk, C. S., White, S. D. M., 
  Springel, V., et al., 2007, \mnras, 381, 1450
  \bibitem[Oh et al.(2008)]{oh08} Oh, S., de Blok, W. J. G., Walter, F., Brinks, E., \& Kennicutt, R. C., 2008, AJ 136, 2761
  \bibitem[Pontzen \& Governato(2011)]{pon11} Pontzen, A., \& Governato, F., 2011,  submitted to ApJ, arXiv:1106.0499
  \bibitem[S\'ersic(1968)]{ser68} S\'ersic, J.-L.\ 1968, Atlas de galaxias australes
  \bibitem[Spano et al.(2008)]{spa08} Spano, M., Marcelin, M., 
  Amram, P., Carignan, C., \'Epinat, B., \& Hernandez, O., 2008, \mnras, 383, 297
  \bibitem[Spergel et al.(2007)]{spe07} Spergel, D. N. et al. 2007, \apjs, 170, 377
  \bibitem[Springel et al.(2008)]{spr08} Springel, V., Wang, J., 
  Vogelsberger, M., Ludlow, A., Jenkins, A., Helmi, A., J. F. Navarro, J. F. N., Frenk, C. S., et al., MNRAS, 391, 1685 
  \bibitem[Stoehr(2006)]{sto06} Stoehr, F., 2006, \mnras, 365, 147
  \bibitem[Tissera et al.(2010)]{tis10} Tissera, P. B., White, S. D. M., Pedrosa, S., \& Scannapieco, C., 2010, \mnras, 406, 922
  \bibitem[van der Kruit \& Searle(1981)]{vdk81} van der Kruit, P. C., \& Searle, L. 1981, \aap, 95, 105
  \bibitem[Walter et al.(2008)]{wal08} Walter, F., Brinks, E., de Blok, W. J. G., Bigiel, F., Kennicutt, R. C.,  Thornley, M. D., \& Leroy, A. 2008, \aj, 136, 2563
  \end{thebibliography}
\end{document}